\documentclass[%
 reprint,
superscriptaddress,
 amsmath,amssymb,
 aps,
floatfix,
]{revtex4-2}

\usepackage[T1]{fontenc}
\usepackage{comment}
\usepackage{multirow}
\usepackage{graphicx}% Include figure files
\usepackage{dcolumn}% Align table columns on decimal point
\usepackage{bm}% bold math
\usepackage{braket}  % Dirac symbols
\usepackage{hyperref}% add hypertext capabilities
\hypersetup{hypertex=true,
            colorlinks=true,
            linkcolor=blue,
            anchorcolor=blue,
            citecolor=blue}
\preprint{APS/123-QED}
\begin{document}
%\linenumbers\relax % Commence numbering lines
\title{Universal control of four singlet-triplet qubits
%Universal control of a 4 singlet-triplet-qubit quantum processor in a germanium quantum dot ladder
}

\author{Xin Zhang}
\email{These authors contributed equally to this work}
\affiliation{QuTech, Delft University of Technology, Delft, The Netherlands.}%
\affiliation{Kavli Institute of Nanoscience, Delft University of Technology, Delft, The Netherlands.}%

\author{Elizaveta Morozova}
\email{These authors contributed equally to this work}
\affiliation{QuTech, Delft University of Technology, Delft, The Netherlands.}%
\affiliation{Kavli Institute of Nanoscience, Delft University of Technology, Delft, The Netherlands.}%

\author{Maximilian Rimbach-Russ}
\affiliation{QuTech, Delft University of Technology, Delft, The Netherlands.}%
\affiliation{Kavli Institute of Nanoscience, Delft University of Technology, Delft, The Netherlands.}%

\author{Daniel Jirovec}
\affiliation{QuTech, Delft University of Technology, Delft, The Netherlands.}%
\affiliation{Kavli Institute of Nanoscience, Delft University of Technology, Delft, The Netherlands.}%

\author{Tzu-Kan Hsiao}
\affiliation{QuTech, Delft University of Technology, Delft, The Netherlands.}%
\affiliation{Kavli Institute of Nanoscience, Delft University of Technology, Delft, The Netherlands.}%

\author{Pablo Cova Fari$\tilde{\text{n}}$a}
\affiliation{QuTech, Delft University of Technology, Delft, The Netherlands.}%
\affiliation{Kavli Institute of Nanoscience, Delft University of Technology, Delft, The Netherlands.}%

\author{Chien-An Wang}
\affiliation{QuTech, Delft University of Technology, Delft, The Netherlands.}%
\affiliation{Kavli Institute of Nanoscience, Delft University of Technology, Delft, The Netherlands.}%

\author{Stefan D. Oosterhout}
\affiliation{QuTech, Delft University of Technology, Delft, The Netherlands.}%
\affiliation{Netherlands Organisation for Applied Scientific Research (TNO), Delft, The Netherlands.}%

\author{Amir Sammak}
\affiliation{QuTech, Delft University of Technology, Delft, The Netherlands.}%
\affiliation{Netherlands Organisation for Applied Scientific Research (TNO), Delft, The Netherlands.}%

\author{Giordano Scappucci}
\affiliation{QuTech, Delft University of Technology, Delft, The Netherlands.}%
\affiliation{Kavli Institute of Nanoscience, Delft University of Technology, Delft, The Netherlands.}%

\author{Menno Veldhorst}
\affiliation{QuTech, Delft University of Technology, Delft, The Netherlands.}%
\affiliation{Kavli Institute of Nanoscience, Delft University of Technology, Delft, The Netherlands.}%

\author{Lieven M. K. Vandersypen}
\email{L.M.K.Vandersypen@tudelft.nl}
\affiliation{QuTech, Delft University of Technology, Delft, The Netherlands.}%
\affiliation{Kavli Institute of Nanoscience, Delft University of Technology, Delft, The Netherlands.}%

\date{\today}

\begin{abstract}
The coherent control of interacting spins in semiconductor quantum dots is of strong interest for quantum information processing as well as for studying quantum magnetism from the bottom up. 
\begin{comment}
On paper, individual spin-spin couplings can be independently controlled through gate voltages, but nonlinearities and crosstalk introduce significant complexity that has slowed down progress in past years.  
\end{comment} 
Here, we present a $2\times4$ germanium quantum dot array with full and controllable interactions between nearest-neighbor spins. As a demonstration of the level of control, we define four singlet-triplet qubits in this system and show two-axis single-qubit control of each qubit and SWAP-style two-qubit gates between all neighbouring qubit pairs, yielding average single-qubit gate fidelities of 99.49(8)-99.84(1)\% and Bell state fidelities of 73(1)-90(1)\%. Combining these operations, we experimentally implement a circuit designed to generate and distribute entanglement across the array. A remote Bell state with a fidelity of 75(2)\% and concurrence of 22(4)\% is achieved. These results highlight the potential of singlet-triplet qubits as a competing platform for quantum computing and indicate that scaling up the control of quantum dot spins in extended bilinear arrays can be feasible.
\end{abstract}

\keywords{Suggested keywords}
\maketitle 

%\section{\label{sec:level1}Introduction} 
The coherent control of a large-scale array of spins in the solid state represents a major challenge in the field of quantum-coherent nanoscience~\cite{vandersypen2017interfacing, heinrich2021quantum, gonzalez2021scaling, chatterjee2021semiconductor}. As a quintessential platform for quantum spin control, the lithographically-defined semiconductor quantum dot has shown great promise both for fault-tolerant digital quantum computation~\cite{stano2022review, burkard2023semiconductor, xue2022quantum, noiri2022fast, mills2022two, hendrickx2021four, philips2022universal, lawrie2023simultaneous, shulman2012demonstration, nichol2017high, weinstein2023universal} and for analog quantum simulation of emergent quantum phenomena~\cite{hensgens2017quantum, dehollain2020nagaoka, van2021quantum, wang2023probing}. Nevertheless, the inherent nanoscale dimensions of the devices, the geometrical constraints in integrating all the required components, and the necessity of employing high-frequency electromagnetic fields in cryogenic environments present important challenges for the integration and control of a large number of spins.

Already, significant efforts have been undertaken to tackle these challenges. For single-spin qubits, the number of coherently controlled interacting spins has been scaled up to six in a one-dimensional array~\cite{philips2022universal} and four in a two-dimensional array~\cite{hendrickx2021four}. A six-dot linear array was also used to achieve universal control of two qubits that are each encoded in a subspace of three electron spins distributed over three dots~\cite{weinstein2023universal}. For singlet-triplet qubits defined in a subspace of two spins across two dots, recent progress includes the individual control of three to four uncoupled qubits~\cite{jang2020individual, fedele2021simultaneous} and the operation of a single qubit in a $3\times3$ quantum dot array~\cite{mortemousque2021coherent}.

Similar to exchange-only qubits, singlet-triplet qubits~\cite{burkard2023semiconductor, Levy2002pair, petta2005coherent, maune2012coherent, shulman2012demonstration, wu2014two, nichol2017high, jock2018silicon, cerfontaine2020closed, cerfontaine2020high, jirovec2021singlet, fedele2021simultaneous} allow fully electrical qubit control using baseband voltage pulses. The use of baseband-only control signals can avoid commonly encountered problems of single-spin qubits such as microwave heating effects~\cite{takeda2018optimized, philips2022universal, undseth2023hotter} and may furthermore alleviate crosstalk effects~\cite{PhysRevApplied.19.044078}. Singlet-triplet qubits also map naturally to the spin-readout basis in Pauli spin blockade (PSB), which is a common method for spin-to-charge conversion in quantum dots~\cite{ono2002science,petta2005coherent}. By using pulse optimization, single-qubit control fidelities of singlet-triplet qubits have exceeded $99\%$~\cite{cerfontaine2020closed}, whereas two-qubit gate fidelities relying on the relatively weak capacitive (Coulomb) interaction reached 72-90\%~\cite{shulman2012demonstration, nichol2017high}. In theory, the two-qubit gate fidelity can be further improved by replacing the capacitive coupling with the stronger exchange coupling~\cite{cerfontaine2020high}, although it has been little investigated in experiments~\cite{qiao2021floquet}. Despite this progress, universal control of more than two interacting singlet-triplet qubits remains yet to be achieved. Recently, a controlled number of charge carriers were loaded in $2\times4$ arrays, a $4\times4$ array and a $1\times12$ array~\cite{chanrion2020charge, duan2020remote, hsiao2023exciton, borsoi2022shared, neyens2023probing}. These advances set the stage for exploring the operation of three or more interacting singlet-triplet qubits experimentally.

\begin{figure*}[htb]
\includegraphics{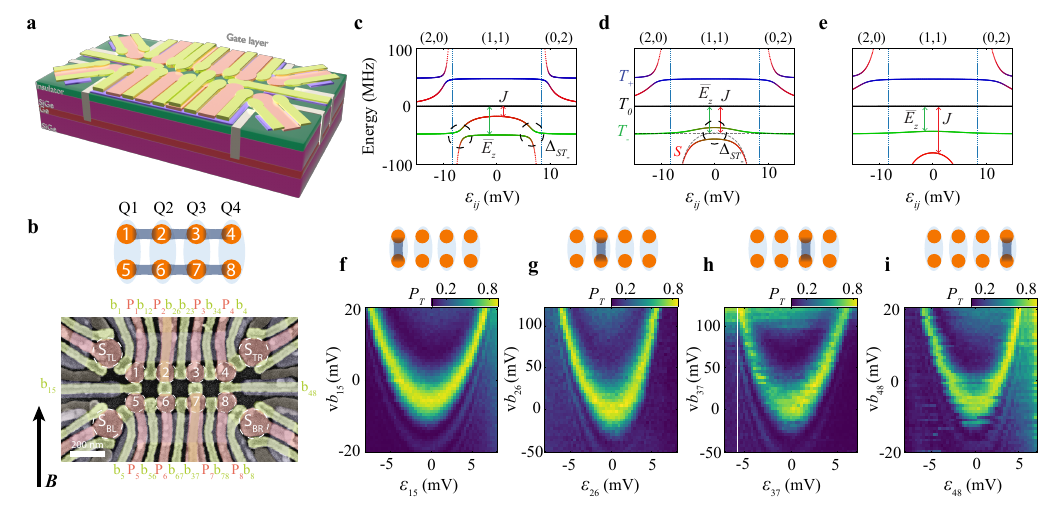}
%\internallinenumbers
\caption{\label{Device} \textbf{Device and energy spectroscopy.} \textbf{a}, Schematic drawing showing the Ge/SiGe heterostructure and three layers of gate electrodes on top to define the quantum dot ladder and sensing dots: screening gates (purple), plunger gates (red), barrier gates (green). Ohmic contacts ({gray}) extend towards the Ge quantum well in which the holes are confined. The aluminum oxide dielectric between different gate layers is omitted for clarity. \textbf{b}, False-colored scanning electron microscope image of a device nominally identical to that used in the measurements, where the in-plane position of the 8 quantum dots is indicated with numbers 1-8 in circles. Charge sensors close to the ladder corners are labeled within larger circles. The plunger (red) and barrier (green) gates of the quantum dots are labeled outside the image. A schematic of the ladder structure of the quantum dots is shown on top, with Q1-Q4 formed by vertical double quantum dots (DQD). \textbf{c-e}, The energy levels of two-spin states in a DQD as a function of energy detuning $\varepsilon_{ij}$ between dot $i$ and $j$ with from left to right the case $J(\varepsilon_{ij} = 0) < \overline{E}_z$, $J(\varepsilon_{ij} = 0) = \overline{E}_z$, $J(\varepsilon_{ij} = 0) > \overline{E}_z$. The dashed black circles denote the positions of $S-T_-$ anticrossings. \textbf{f-i}, The measured energy spectra that probe the positions of the $S-T_-$ anticrossings as a function of the detuning and the barrier gate voltage for each vertical DQD {at $B=5$ mT}. The color scale shows the measured spin triplet probability $P_T$ after initializing a vertical DQD in a singlet state (in (0,2) or (2,0)) and applying a gate voltage pulse (20 ns ramp in, 40 ns wait time, 0 ns ramp out) to the detuning shown on the horizontal axis, for different $\text{v}b_{ij}$. {The cartoons on top of panels \textbf{f-i} represent the eight dots, and the dark grey line indicates which exchange coupling is active in the panel below.}}
\end{figure*}

Here we demonstrate coherent control of four interacting singlet-triplet qubits in a 2x4 germanium quantum dot array, which forms a quantum dot ladder. Taking advantage of the strong intrinsic spin-orbit coupling and small in-plane g-factors of holes in strained germanium quantum wells~\cite{giordano2021germanium}, we encode the qubit in the singlet ($\ket{S}$) and the lowest triplet ($\ket{T_-}$) of two exchange-coupled spins, a variant of the originally proposed singlet-triplet qubit~\cite{petta2010coherent, ribeiro2010harnessing, nichol2015quenching, PhysRevLett.128.126803, PhysRevB.104.195421, PhysRevApplied.18.054090, cai2023coherent, wang2023probing, rooney2023gate}. By controlling the exchange interaction inside each spin pair along the ladder rungs, we first map out the qubit energy spectrum. Then we show universal control of each qubit by pulsing both the detuning and tunneling barrier of the corresponding double quantum dot (DQD). With proper simultaneous control of detunings and tunneling barriers of neighbouring $S-T_-$ qubits, we achieve a two-qubit SWAP-style gate induced by exchange interactions for each pair of neighbouring qubits in the ladder. Finally, with the demonstrated single- and two-qubit control, we implement a quantum circuit for quantum state transfer across the ladder.

\section{\label{sec:level2}Germanium quantum dot ladder}

As shown in Fig.~\ref{Device}a and b, the 2x4 quantum dot ladder is fabricated in a germanium quantum-well heterostructure~\cite{lodari2021low}. The gate pattern and substrate have the same design as that in ref.~\cite{hsiao2023exciton}. The eight quantum dots are labeled with numbers 1 to 8 and the four charge sensors to measure the charge states in the quantum dots are labeled $\text{S}_{\text{TL}}$, $\text{S}_{\text{TR}}$, $\text{S}_{\text{BL}}$ and $\text{S}_{\text{BR}}$, respectively. The quantum dot potentials are controlled by plunger gates $\text{P}_i$, and the interdot or dot-sensor tunnel couplings are controlled by barrier gates $\text{b}_{ij}$ or $\text{b}_{i}$, with $i$ or $j$ denoting the corresponding quantum dot number. Linear combinations of plunger gate voltages $\{P_i\}$ allow us to set the overall electrochemical potential of each DQD $\mu_{ij}= (\text{v}P_i + \text{v}P_j)/2 $ and the interdot detuning $\varepsilon_{ij} = (\text{v}P_i - \text{v}P_j)/2 $. The prefix ``$\text{v}$'' indicates that the physical gate voltages are virtualized to compensate the crosstalk on the dot potentials~\cite{hensgens2017quantum} (see Supplementary Information section II for the virtual gate matrix). Single-hole occupation of each quantum dot in the array is confirmed by measuring the charge stability diagrams using sensors $\text{S}_{\text{BL}}$ and $\text{S}_{\text{BR}}$ (see Extended Data Fig.~\ref{CSD_ED}). All plunger and interdot barrier gates are connected to a bias tee to allow both DC voltages and voltage pulses to be applied.

\section{\label{sec:level3}Singlet-triplet qubit and energy spectroscopy}
We encode the qubit into the two-spin singlet-triplet states, $\ket{S}$ and $\ket{T_-}$, of the DQDs along the rungs of the quantum dot ladder, with the singlet $\ket{S}=(\ket{\uparrow\downarrow}-\ket{\downarrow\uparrow})/\sqrt{2}$ and the lowest-energy triplet $\ket{T_-}=\ket{\downarrow\downarrow}$. Thus Q1, Q2, Q3 and Q4 are formed using DQD 1-5, 2-6, 3-7 and 4-8. Qubit readout is achieved by pulsing the corresponding DQD to the PSB regime{, i.e. in the (0,2) or (2,0) regime but close to (1,1)}. This regime converts the singlet and triplet states into distinct charge states, which are then measured through the charge sensor {(see Extended Data Fig.~\ref{CSD_ED} for details)}. The single-qubit Hamiltonian can be written as
\begin{equation} \label{ST-}
H_{ST_-} = \frac{\overline{E}_z-J}{2}\sigma_z + \frac{\Delta_{ST_-}}{2}\sigma_x,
\end{equation}
where $\sigma_x$ and $\sigma_z$ are the Pauli matrices, $J = J(\varepsilon_{ij}, \text{v}b_{ij})$ is a function of both the detuning $\varepsilon_{ij}$ and the barrier gate voltage $\text{v}b_{ij}$, and $\overline{E}_z = \overline{g}_{ij} \mu_B B$ is the average Zeeman energy of the two hole spins in the DQD, with $\overline{g}_{ij}$ the average $g$-factor, $\mu_B$ the Bohr magneton, and $B$ the magnetic field strength. Unless indicated otherwise, an in-plane magnetic field (up to alignment precision) of {$B = 5$ or $10$ mT} is applied to the device. The intrinsic spin-orbit interaction for holes in germanium couples the states $\ket{S}$ and $\ket{T_-}$ with an energy $\Delta_{ST_-}$.

Figs.~\ref{Device}c-e show the energy levels of the two-spin $\ket{S}$ and $\ket{T_-}$ states in a DQD with $J(\varepsilon_{ij} = 0) < \overline{E}_z$, $J(\varepsilon_{ij} = 0) = \overline{E}_z$, $J(\varepsilon_{ij} = 0) > \overline{E}_z$, respectively. The other two-spin states are $\ket{T_0}=(\ket{\uparrow\downarrow}+\ket{\downarrow\uparrow})/\sqrt{2}$ and $\ket{T_+}=\ket{\uparrow\uparrow}$. In a DQD, we can describe the charge states as ($n_L$,$n_R$) to denote the charge number distribution in the left ($n_L$) and right ($n_R$) quantum dot. By adjusting the detuning $\varepsilon_{ij}$ of the DQD from negative to positive, we can change the charge state from (2,0) to (1,1) and then to (0,2), as indicated by the labels on top of each diagram, and the energy levels of the two-spin states in the DQD will change accordingly. As shown in Fig.~\ref{Device}c, when $J(\varepsilon_{ij} = 0)$ is smaller than $\overline{E}_z$, the singlet $\ket{S}$ crosses the triplet $\ket{T_-}$ twice in the (1,1) regime. Due to intrinsic spin-orbit coupling, these are in fact avoided crossings with a gap $\Delta_{ST_-}$, where the states $\ket{S}$ and $\ket{T_-}$ are admixed. As $J(\varepsilon_{ij} = 0)$ increases, the two anticrossings approach each other and eventually merge into one, as shown in Fig.~\ref{Device}d. When $J(\varepsilon_{ij} = 0)$ increases even further, see Fig.~\ref{Device}e, $\ket{S}$ and $\ket{T_-}$ no longer exhibit an anvoided crossing. 

Experimentally, we probe the position of the avoided crossings as follows. First, we initialize one of the qubits to a singlet by pulsing from (0,2) or (2,0) to the detuning $\varepsilon_{ij}$ in (1,1). {A 20 ns ramp-in time is used to ensure adiabaticity with respect to the tunnel coupling, which is around 2 GHz, and diabaticity with respect to the $S-T_-$ anticrossing.} After waiting for a certain time, we pulse the qubit back to the PSB regime to record the triplet probability {(see Extended Figure~\ref{CSD_ED} for full details)}. When the pulse takes the system to an anticrossing, the singlet will evolve into a triplet during the waiting time (of 40 ns duration, {close to a $\pi$ rotation} is chosen to obtain a sizable triplet probability). Performing such measurements as a function of the barrier gate voltage $\text{v}b_{ij}$ that controls $J$ for each qubit, results in the parabola-like patterns, also called spin mixing maps~\cite{PhysRevLett.115.096801, mortemousque2021coherent}, in Fig.~\ref{Device}f-i. As expected, when $\text{v}b_{ij}$ is tuned from positive to negative, $J$ increases and the positions of the $S-T_-$ anticrossings move inwards before disappearing. The asymmetry visible in {these panels} can arise from imperfect virtualization of the barrier gates or from a detuning-dependent Zeeman energy~\cite{hendrickx2020single} (see Supplementary Information section III).

\section{\label{sec:level31}Universal single-qubit control}
\begin{figure*}[htb]
\includegraphics{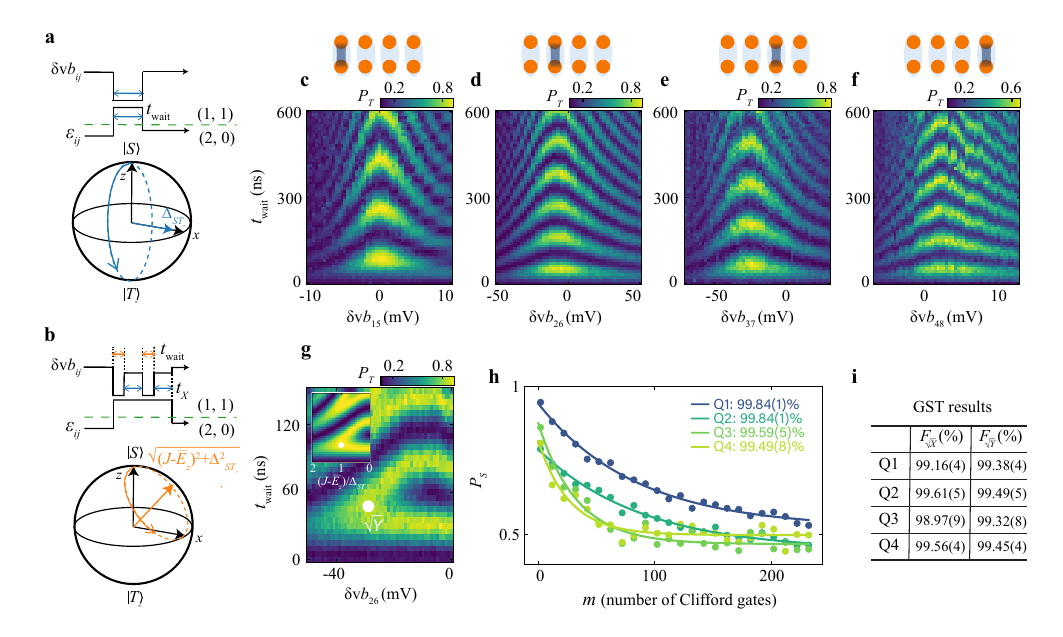}
%\internallinenumbers
\caption{\label{single qubit gate} {\textbf{Universal single-qubit control of four singlet-triplet qubits.} \textbf{a},\textbf{b}, The pulse schemes used for $x$-axis control (\textbf{a}) and $y$-axis control (\textbf{b}). In the experiments, the detuning pulse in \textbf{a} and \textbf{b} has a 20 ns ramp (not shown) from (2,0) to (1,1), similar to the pulse used for the energy spectroscopy. \textbf{c-f}, Experimental results for $x$-axis rotations of each qubit, showing measured triplet probabilities $P_T$ as a function of $t_\text{wait}$ and the corresponding barrier voltage $\delta \text{v}b_{ij}$. \textbf{g}, Measured $P_T$ for the sequence shown in panel \textbf{b} as a function of $t_\text{wait}$ and the barrier voltage change $\delta \text{v}b_{26}$. The inset shows the numerically computed $P_T$ as a function of $t_\text{wait}$ and the ratio of $z$-axis component to the $x$-axis component, $(J-\overline{E}_z) / \Delta_{ST_-}$. The position where$\sqrt{Y}$ is properly calibrated is indicated by a white dot. \textbf{h}, Single-qubit randomized benchmarking data for Q1-Q4. The numbers in the legend are the extracted average gate fidelities, which are obtained from the Clifford gate fidelities using a ratio of 3.625. \textbf{i}, Table showing the single-qubit gate fidelities of Q1-Q4 measured by gate set tomography (GST). All the data above are measured at $B=5$ mT.}}
\end{figure*}

With the knowledge of the energy spectrum of the four $S-T_-$ qubits, we next implement the two-axis control of each qubit, which is necessary and sufficient for universal single-qubit control. By operating the qubit in the regime where $J = \overline{E}_z$, the first term of Eq.~\ref{ST-} goes to zero and $\Delta_{ST_-}$ rotates the qubit around the $x$-axis in the Bloch sphere, as shown in Fig.~\ref{single qubit gate}a. Furthermore, we tune the barrier voltage to obtain $J = \overline{E}_z$ at zero detuning, which is a symmetry point where the effect of detuning noise is strongly suppressed~\cite{PhysRevLett.116.116801, PhysRevLett.116.110402}. The pulse scheme for testing $x$-axis control is shown in Fig.~\ref{single qubit gate}k: first we initialize the qubit into a singlet by starting in the (2,0) (or (0,2)) regime, then pulse the detuning to the center of the (1,1) regime where $J(\varepsilon_{ij} = 0) = \overline{E}_z$, next allow the qubit to evolve for a variable time $t_\text{wait}$, and finally pulse the detuning back to a point in the (2,0) (or (0,2)) regime for spin readout via PSB. The measured rotations of Q1-Q4 as a function of the {corresponding barrier gate voltage are shown in Fig.~\ref{single qubit gate}c-f. By choosing the point where the oscillation speed is the slowest, i.e., at the $S-T_-$ anticrossing,  the qubits rotate around the $x$-axis. Long-timescale $x$-rotations are shown in Extended Data Fig.~\ref{Additional single-qubit results2}. At $B=5$ mT, the dephasing times are in the range of 1.5-2.2 $\mu$s for Q1-Q4, mostly limited by low-frequency or quasi-static noise.} 

{To realize $y$-axis control, we use the relationship $\sqrt{Y}=XH$, where $\sqrt{Y}$ stands for a $\pi/2$ rotation around the $y$-axis, $X$ for a $\pi$ rotation around the $x$-axis, and $H$ refers to the Hadamard gate (for direct $z$-axis control, see Extended Data Fig.~\ref{Additional single-qubit results}). To obtain a Hadamard gate, we slightly increase $J$ to a value where $(J-\overline{E}_z) = \Delta_{ST_-}$. The qubit then rotates around an axis halfway between the $x$ and $z$ axes. A rotation angle of $\pi$ then corresponds to a Hadamard gate. In order to calibrate the rotation axis and rotation angle of the $H$ gate, we concatenate two $\sqrt{Y}$ gates and evaluate the probability of having flipped the qubit (Fig.~\ref{single qubit gate}b). Specifically, we first initialize the qubit into a singlet, then change $J$ diabatically by pulsing the corresponding barrier gate by an amount $\delta\text{v}b_{ij}$ for a time $t_\text{wait}$ to implement the $H$ gate, and finally apply a $X$ gate (we aim to stay at the detuning symmetry point when pulsing the barrier to minimize the sensitivity to charge noise~\cite{PhysRevLett.116.116801, PhysRevLett.116.110402}). This combination is repeated such that a $Y$ gate is expected for the right choice of $\delta\text{v}b_{ij}$ and $t_\text{wait}$. This procedure is illustrated for Q2 in Fig.~\ref{single qubit gate}g. The white dot shows the position where the sequence of Fig.~\ref{single qubit gate}b produces a $Y$ gate. Running only the first half of the sequence implements a $\sqrt{Y}$ gate. A corresponding numerical simulation result is shown in the inset of Fig.~\ref{single qubit gate}g, showing the same pattern as the experimental result.} 

\begin{figure*}[htb]
\includegraphics{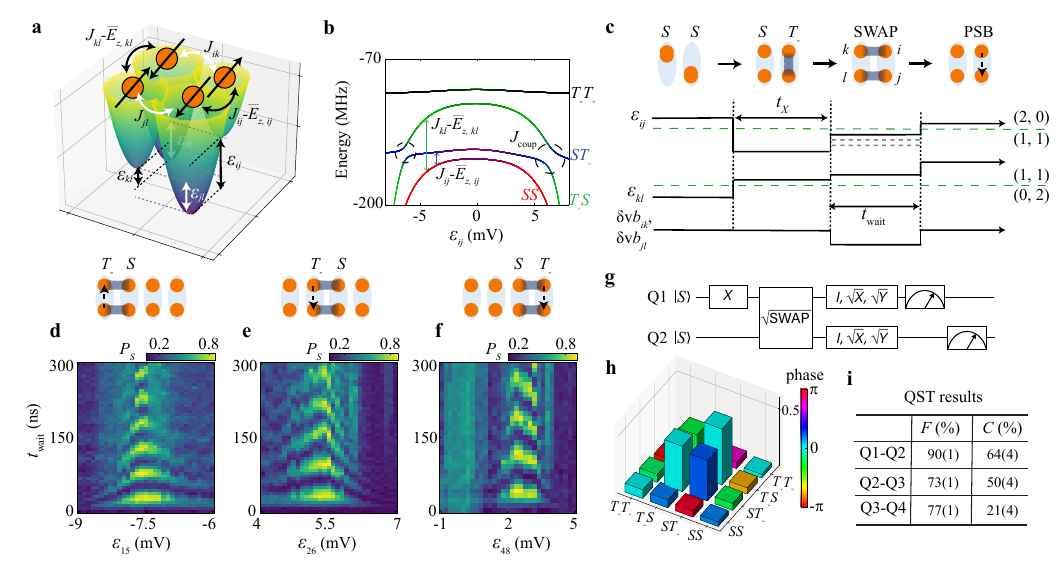}
%\internallinenumbers
\caption{\label{two qubit gate} \textbf{Two-qubit interactions across the quantum dot ladder.} \textbf{a}, A plaquette of two connected DQDs. The $S-T_-$ qubits have a splitting of $J_{ij}-\overline{E}_{z, ij}$ and $J_{kl}-\overline{E}_{z, kl}$ (neglecting $\Delta_{ST_-}$), which are controlled by the detunings $\varepsilon_{ij}$ and $\varepsilon_{kl}$, respectivily. The qubit-qubit coupling $J_\text{coup}$ is an average of $J_{ik}$ and $J_{jl}$ between the corresponding dots, which are controlled by $\varepsilon_{ik}$ and $\varepsilon_{jl}$. \textbf{b}, The energy levels of two-qubit states, where we fix $\varepsilon_{kl}$ to be positively large and scan $\varepsilon_{ij}$. At the positions where $J_{ij}-\overline{E}_{z, ij}$ equals $J_{kl}-\overline{E}_{z, kl}$, an anticrossing with a gap $J_\text{coup}$ forms (black dashed circles), which can be used to induce SWAP oscillations between $\ket{ST_-}$ and $\ket{T_-S}$. {The parameters used in this calculation are based on the experimental results for Q3-Q4 shown in Supplementary Fig. 4c.} \textbf{c}, The pulse scheme for SWAP operations. We start in (0,2) or (2,0), at large positive or negative detuning, and diabatically pulse one qubit to (1,1) at modest detuning such that it remains in $\ket{S}$, and pulse the other qubit to zero detuning where a $\pi$ rotation for a time $t_X$ takes it to $\ket{T_-}$. At this point, the qubits are set to $\ket{ST_-}$ or $\ket{T_-S}$. Next, we pulse the detunings of both qubits to make their energies resonant, while at the same time activating $J_{ik}$ and $J_{jl}$. This will kickstart SWAP oscillations between the two qubits. The dashed lines in the pulse of $\varepsilon_{ij}$ show that we scan the detuning of one qubit to find the condition for SWAP operations. After an evolution time $t_\text{wait}$, we pulse the detunings to the PSB readout configuration for one of the qubits. \textbf{d-f}, The experimental results of SWAP oscillations, showing measured triplet probabilities $P_T$ or singlet probabilities $P_S$ as a function of operation time $t_\text{wait}$ and the detuning voltage for Q1-Q2 (\textbf{d}), Q2-Q3 (\textbf{e}) and Q3-Q4 (\textbf{f}). The initial states of two qubits (before the SWAP oscillations) are denoted on the top, and the qubit pair that is read out is indicated by the dashed arrow showing the readout pulse direction. \textbf{g}, {The quantum circuit used to create a generalized Bell state between Q1 and Q2 and to characterize it via quantum state tomography (QST). \textbf{h}, Measured two-qubit density matrix of Q1-Q2, after removal of SPAM errors and using maximum-likelihood estimation (MLE). \textbf{i}, State fidelities and concurrence estimated from the density matrices of the Bell states of Q1-Q2, Q2-Q3 and Q3-Q4. The data of panels \textbf{d-f} and \textbf{h-i} is measured at $B=5$ mT.} }
\end{figure*}

{Using the $\sqrt{X}$ and $\sqrt{Y}$ gates, we perform randomized benchmarking (RB) to obtain the average gate fidelities (see Fig.~\ref{single qubit gate}h). All 4 qubits yield average gate fidelities at or above 99\%, extracted from the Clifford gate fidelities. We also measure the $\sqrt{X}$ and $\sqrt{Y}$ gate fidelities with gate set tomography (GST), of which the results are summarized in Fig.~\ref{single qubit gate}i. Overall, most of the fidelities in GST results are slightly lower than those from RB. These fidelity differences may stem from the presence of low-frequency noise in our system, which causes different uncertainties in the fidelity estimates between GST and RB ~\cite{wang2024operating}. Full details on the quantum process for those two gates, derived from GST, are given in Extended Data Fig.~\ref{Single qubit GST restuls} and Supplementary Information section VIII.}  

\section{\label{sec:level3X}Two-qubit gate}

In order to realize universal control of the full four-qubit register, we need to complement single-qubit gates with two-qubit entangling gates. Assuming isotropic exchange interactions between adjacent $S-T_-$ qubits, the two-qubit Hamiltonian in the basis of $\lbrace \ket{SS}$, $\ket{ST_-}$, $\ket{T_-S}$ and $\ket{T_-T_-}\rbrace$ can be written as:
\begin{equation} \label{SWAP}
\begin{split}
%H_\text{2Q}
% &=\frac{(\overline{E}_{z, ij}-J_{ij})\sigma_z +\Delta_{ST_-, ij}\sigma_x }{2}\otimes I +\\
% &I \otimes \frac{(\overline{E}_{z, kl}-J_{kl})\sigma_z + \Delta_{ST_-, kl}\sigma_x}{2}+\\
% &\frac{J_\text{coup}}{4}[\sigma_x \otimes \sigma_x + \sigma_y \otimes \sigma_y + \frac{1}{2}(\sigma_z - I) \otimes (\sigma_z - %I)],
H_\text{2Q}
 &=\frac{(\overline{E}_{z, ij}-J_{ij})\sigma_z^{ij} +\Delta_{ST_-, ij}\sigma_x^{ij}}{2} +\\
 &\frac{(\overline{E}_{z, kl}-J_{kl})\sigma_z^{kl} + \Delta_{ST_-, kl}\sigma_x^{kl}}{2}+\\
 &\frac{J_\text{coup}}{4}[\sigma_x^{ij}\sigma_x^{kl} + \sigma_y^{ij}\sigma_y^{kl} + \frac{1}{2}(\sigma_z^{ij} - I)(\sigma_z^{kl} - I)],
\end{split}
\end{equation}
where $ij$ and $kl$ refer to the respective qubit dot pair, and the interqubit coupling $J_\text{coup}=(J_{ik}+J_{jl})/2$. The coupling term is reminiscent of two well-known interaction Hamiltonians. If the factor $1/2$ of the $\sigma_z\sigma_z$ coupling term were 1 instead, we recover the exchange Hamiltonian that generates the SWAP gate and the universal $\sqrt{\text{SWAP}}$ gate. If that factor were zero, only the flip-flop terms would survive, which generate the iSWAP and $\sqrt{\text{iSWAP}}$ gate. The coupling Hamiltonian in Eq.~\ref{SWAP} thus generates a SWAP-style gate that is not a standard two-qubit gate but is also universal from the perspective of quantum computing (see Supplementary Information section VII). For simplicity, we call it a SWAP gate in the remainder of this work. 

To activate the SWAP gate, we equalize the energy splittings of two qubits and turn on $J_{ik}$ and $J_{jl}$ such that the flip-flop terms can exchange the qubit populations (note that if the qubit energies were set very different from each other, a CZ gate would result instead). Our strategy for meeting both requirements at the same time is to use the interdot detuning of both qubits~\cite{van2021quantum, wang2023probing}. A typical potential landscape for the two qubits in DQD $ij$ and $kl$ is shown in Fig.~\ref{two qubit gate}a, where we pulse $\varepsilon_{ij}$ to large positive and $\varepsilon_{kl}$ to large negative detuning. The detunings $\varepsilon_{ik}$ and $\varepsilon_{jl}$, which control the interactions between the qubits, are then automatically increased as well. Therefore, all the exchange interactions involved are enhanced simultaneously and the effect of the single-qubit terms $\sigma_x$ is made negligible. In practice, we fix the (large) detuning of one qubit and fine-tune that of the other to find the position where two qubits have equal energy splittings. This is illustrated by the energy spectrum in Fig.~\ref{two qubit gate}b, where we fix the detuning $\varepsilon_{kl}$ to a large negative value and scan the detuning $\varepsilon_{ij}$. We see that the states $\ket{ST_-}$ and $\ket{T_-S}$ anticross at the two positions where $J_{ij}-\overline{E}_{z, ij}$ is equal to $J_{kl}-\overline{E}_{z, kl}$. The gap size is given by $J_\text{coup}$. Since $\varepsilon_{ik}$ and $\varepsilon_{jl}$, which control $J_\text{coup}$ via $J_{ik}$ and $J_{jl}$, are also dependent on $\varepsilon_{ij}$, the sizes of the two gaps are not necessarily the same. 

Fig.~\ref{two qubit gate}c shows an example of the pulse scheme we use in the experiment to observe two-qubit SWAP oscillations. Starting from both qubits in (0,2) or (2,0), we initialize one qubit to $\ket{T_-}$ using single-qubit control by pulsing $\varepsilon_{ij}$ to zero and waiting for a $\pi$ rotation, and we initialize the other qubit to $\ket{S}$ by pulsing $\varepsilon_{kl}$ to a large value in (1,1) (other qubits are either initialized to singlets by pulsing back and forth to (0,2) or (2,0), or remain in the (1,1) regime all the time). Then we pulse the detuning of one qubit such that the energies of the two qubits match, and SWAP oscillations are initiated. Simultaneously, several barrier voltages are pulsed to help set the respective exchange-interaction strengths to appropriate values (details of these pulses vary). 

Fig.~\ref{two qubit gate}d-f show the resulting SWAP oscillations for Q1-Q2, Q2-Q3, and Q3-Q4. Chevron-style patterns are observed with the energies of the two qubits aligned in the middle of the patterns. Moving away from the middle, the energy of one qubit is shifted with respect to that of the other. This qubit-qubit energy detuning tilts the rotation axis and accelerates the rotation. Looking closely, the chevron patterns are not symmetrical. This can be understood by the fact that the qubit energy does not vary linearly with detuning. {In some panels, single-qubit oscillations around an axis close to the $x$-axis are also observed, such as the data at $\varepsilon_{26} = 4$ mV in Fig.~\ref{two qubit gate}e.} These $\varepsilon_{ij}$ values are close to zero interdot detuning, and when $J(\varepsilon_{ij} = 0)$ is not much larger than $\overline{E}_z$, such rotations are expected. We note that SWAP oscillations between $\ket{ST_-}$ and $\ket{T_-S}$ were also observed in previous research on simulating the dynamics of an antiferromagnetic spin chain and resonating-valence-bond states based on the Heisenberg model in four-quantum-dot systems~\cite{van2021quantum, wang2023probing}.  

{With a combination of single-qubit gates and the SWAP-style gate, we prepare a Bell state and characterize it using quantum state tomography (QST). The pulse sequence shown in Fig.~\ref{two qubit gate}g is expected to produce a generalized Bell state $\ket{\psi}=\frac{1}{\sqrt{2}}(\ket{ST_-}+e^{i\theta}\ket{T_-S})$ between the two qubits, where $\theta$ is a single-qubit phase term, followed by single-qubit gates ($I$, $\sqrt{X}$, $\sqrt{Y}$) applied to both qubits to achieve basis changes before measurement along the $z$-axis. The density matrix of the Bell state formed by Q1-Q2 is shown in Fig.~\ref{two qubit gate}h, and the acquired fidelity and concurrence of all the neighbouring qubit pairs are shown in Fig.~\ref{two qubit gate}i (see Extended Data Fig.~\ref{Bell state} for details). The fidelities are in the range of 73(1)-90(1)\% and the concurrence ranges from 21(4)\% to 64(4)\%. Concurrence is a measure of the entanglement between two qubits, which ranges from 0 (no entanglement) to 1 (maximal entanglement). Therefore, the measured concurrence for all the qubit pairs demonstrates the implemented $\sqrt{\text{SWAP}}$ gate can generate entanglement between qubits. To evaluate the performance of the $\sqrt{\text{SWAP}}$ gate, we perform two-qubit GST on Q1 and Q2. By fitting the result to a theoretical model, a gate fidelity of $\sim$80\% is obtained (see Extended Data Fig.~\ref{Bell state} and Supplementary Information section VIII for details).} 

\begin{figure}[htb]
\includegraphics{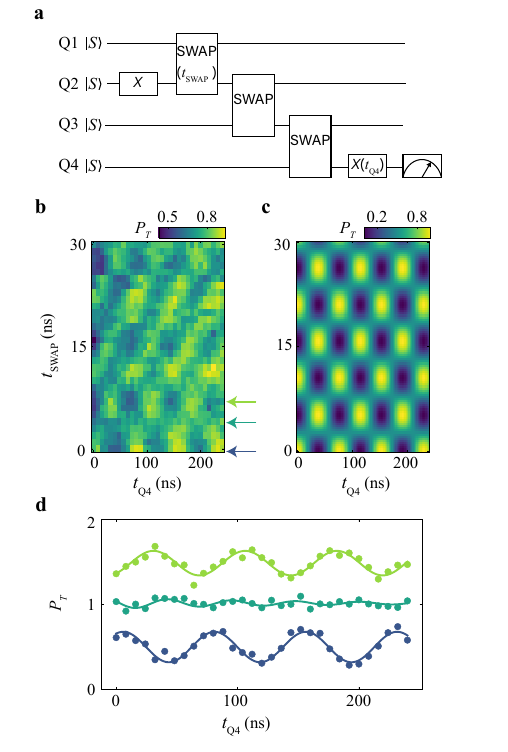}
%\internallinenumbers
\caption{\label{state transfer} \textbf{Implementation of a quantum circuit for entanglement generation and quantum state transfer.} \textbf{a}, Quantum circuit with all the qubits initialized into the singlet. {An $X$ gate (32 ns) rotates Q2 into a triplet state, and a SWAP interaction for a variable time $t_\text{SWAP}$ periodically produces entanglement between Q1 and Q2. Two subsequent SWAP gates (30 ns and 24 ns) transfer the state of Q2 to Q4 and a final single-qubit rotation of Q4 for a variable time $t_\text{Q4}$ is followed by Q4 readout.} The delay time between each quantum gate is set to zero. \textbf{b},\textbf{c}, Experimental (\textbf{b}) and numerical (\textbf{c}) results after running the quantum circuit of panel \textbf{a}, with triplet probabilities $P_T$ of Q4 shown as a function of $t_\text{SWAP}$ and $t_\text{Q4}$. \textbf{d}, Linecuts from \textbf{a} showing triplet probabilities $P_T$ of Q4 as a function of control time $t_\text{Q4}$. The data is vertically shifted by 0.5 for clarity. All the data above are measured at $B=10$ mT.}
\end{figure}

\section{\label{sec:level3XX} Quantum circuit implementation}

Finally, using a combination of the single- and two-qubit gates demonstrated above, we aim to implement a quantum circuit designed to create and distribute an entangled state {across the array}. As shown in Fig.~\ref{state transfer}a, we first initialize Q1 and Q2 into $\ket{ST_-}$ by applying a $\pi$ rotation on Q2 and then activate a SWAP interaction between Q1 and Q2 for a duration $t_\text{SWAP}$. This interaction is expected to generate entanglement when $t_\text{SWAP}$ corresponds to a quarter period, i.e. for a $\sqrt{\text{SWAP}}$ gate. Next, we apply consecutive half-period SWAP gates of Q2-Q3 and Q3-Q4 to transfer the state of Q2 to Q4 via Q3. Finally, we perform a single-qubit $x$-axis rotation of Q4 for a time $t_{\text{Q4}}$ and measure Q4. 

The experimental results are shown in Fig.~\ref{state transfer}b, where the single-qubit oscillations of Q4 as a function of $t_{\text{Q4}}$ are modulated in phase by $t_{\text{SWAP}}$, resulting in a checkerboard pattern. The underlying mechanism is that the state of Q2 oscillates as a function of $t_{\text{SWAP}}$, as quantum information is periodically exchanged between Q1 and Q2. Therefore the state of Q4 following the quantum state transfer also oscillates with $t_{\text{SWAP}}$. Where the evolution of Q4 changes phase, $t_{\text{SWAP}}$ corresponds to the duration of a $\sqrt{\text{SWAP}}$ operation (modulo an integer number of SWAP operations), at which point maximal entanglement between Q1 and Q2 is expected. When two qubits are maximally entangled, the density matrix of each qubit by itself is fully mixed. At this point, the measured $P_T$ of Q4 should not oscillate as a function of $t_{\text{Q4}}$. This is indeed what we observe, see Fig.~\ref{state transfer}d, where we show the linecuts from Fig.~\ref{state transfer}b. A trace without {apparent} oscillations is observed between two sets of out-of-phase oscillations of Q4, as expected. The same features are seen in Fig.~\ref{state transfer}c, which shows the ideally expected checkerboard pattern obtained from numerical simulations of the protocol of Fig.~\ref{state transfer}a, assuming perfect initialization, operations and readout. 

We note the checkerboard pattern is quite robust to errors in the SWAP gates. Small errors will merely change the contrast of the pattern; for large SWAP errors, the alternating rows are no longer equal in height. However, when the initialization of Q1 or Q2 leads to superposition states with a $y$-axis component (and assuming perfect SWAP gates), the pattern acquires a tilt. In this case, the rotation angle of the final $x$-axis rotation needed to maximize or minimize $P_T$ is no longer exactly $0$ or $\pi$ but depends on the $y$-axis component of Q4 (and hence also on $t_{\text{SWAP}}$) after the sequence of SWAP operations. Looking closely, the blue and green oscillations in Fig.~\ref{state transfer}d are not perfectly out of phase with each other, and the data in Fig.~\ref{state transfer}b shows weak diagonal features not seen in the numerical simulations. These point at the imperfect initialization of Q1 or Q2.

{We also characterize the remote Bell state of Q1 and Q4 by performing QST. The experiment was performed at $B=5$ mT and the quantum circuit is similar except Q1 was initialized into a triplet instead of Q2. The resulting Bell state fidelity is 75(2)\% and the concurrence is 22(4)\%. Compared to the concurrence of the Bell state of Q1-Q2 before state transfer, which is 90(1)\%, the remote entanglement is reduced by the transfer process of two consecutive SWAP gates.}

\section{\label{sec:level4}Conclusion}
In conclusion, we have experimentally demonstrated initialization, readout, and universal control of four singlet-triplet ($S-T_-$) qubits in a 2x4 germanium quantum dot array. {By using randomized benchmarking and quantum state tomography, we obtain average single-qubit gate fidelities of 99.49(8)-99.84(1)\% and Bell state fidelities of 73(1)-90(1)\% for all the nearest qubit pairs. For the $\sqrt{\text{SWAP}}$ gate, we estimate a gate fidelity of $\sim$80\% by fitting the GST result to a theoretical model.} Furthermore, through independent control of the exchange interactions between any pair of neighbouring spins across the device, we are able to implement a quantum circuit that spans the entire array{, yielding remote entanglement of two singlet-triplet qubits with a Bell state fidelity of 75(2)\% and a concurrence of 22(4)\%}. 

With four universally controlled qubits in a bilinear array, these results put baseband-controlled singlet-triplet spin qubits in germanium firmly on the map as a potential candidate for large-scale quantum computing. In future experiments, {the two-qubit gate fidelity must be increased in order to allow fault-tolerant quantum computation.} {The gate fidelities can be potentially improved by suppressing low-frequency noise using feedback control or pulse optimization~\cite{cerfontaine2020closed, PhysRevApplied.18.054090, berritta2024real} and by a more detailed modeling of the effects of spin-orbit interaction. Additionally, other types of two-qubit gates like the CZ gate for $S-T_-$ qubits can be investigated. Additional improvements can be reached if the tunnel barriers are more tunable, which can be achieved by depositing the barrier gates either before~\cite{hendrickx2021four, wang2023probing} or together with ~\cite{ha2021flexible} the plunger gates. Moreover, with programmable control of exchange interactions in the array, this spin ladder can also be used for analog simulation of a wealth of rich physical phenomena such as quantum magnetism~\cite{dagotto1996surprises}.} 

\begin{comment}
$\left( \begin{smallmatrix} 1 & 1 & 1 & 1 \\ 0 & 0 & 0 & 0 \end{smallmatrix} \right)$.
\end{comment}

\begin{acknowledgments}
We thank C. Déprez, {I. Fernandez De Fuentes, B. Undseth, Y. Matsumoto and X.X. Yang} for insightful discussions and kind help. We also thank other members of the Vandersypen group and Veldhorst group for stimulating discussions. We acknowledge S. L. de Snoo's help in software development and technical support by O. Benningshof, J. D. Mensingh, R. Schouten, R. Vermeulen, R. Birnholtz, E. Van der Wiel, and N. P. Alberts. This work was funded by an Advanced Grant from the European Research Council (ERC) under the European Union’s Horizon 2020 research (882848). C.-A.W. and M.V. acknowledge support by the European Union through an ERC Starting Grant QUIST (850641). M.R.-R. acknowledges support from the Netherlands Organization of Scientific Research (NWO) under Veni grant VI.Veni.212.223.\\

\textbf{Author contributions} X.Z. and E.M. performed the experiment and analyzed the data with help from M.R.-R. and D.J.. X.Z. performed the numerical simulations with help from M.R.-R. and D.J.. M.R-R developed the theory model. T.-K.H., P.C.F. and C.-A.W. contributed to the preparation of the experiments. S.D.O. fabricated the device with inputs from T.-K.H., P.C.F. and X.Z.. A.S. grew the Ge/SiGe heterostructure, M.V. supervised the device fabrication and G.S. supervised the heterostructure growth and development. X.Z. and L.M.K.V. conceived the project and L.M.K.V. supervised the project. X.Z., M.R.-R., E.M. and L.M.K.V. wrote the manuscript with inputs from all authors.\\

\textbf{Data and Code availability} The data supporting this work and codes used for data processing and numerical simulation are archived on a Zenodo data repository at https://doi.org/10.5281/zenodo.12801188 %https://doi.org/10.5281/zenodo.10431402
\end{acknowledgments}

% \nocite{*}
\bibliography{main}

\providecommand{\noopsort}[1]{}\providecommand{\singleletter}[1]{#1}%
\begin{thebibliography}{10}
\expandafter\ifx\csname url\endcsname\relax
  \def\url#1{\texttt{#1}}\fi
\expandafter\ifx\csname urlprefix\endcsname\relax\def\urlprefix{URL }\fi
\providecommand{\bibinfo}[2]{#2}
\providecommand{\eprint}[2][]{\url{#2}}

\bibitem{vandersypen2017interfacing}
\bibinfo{author}{Vandersypen, L.} \emph{et~al.}
\newblock \bibinfo{title}{Interfacing spin qubits in quantum dots and
  donors—hot, dense, and coherent}.
\newblock \emph{\bibinfo{journal}{npj Quantum Inf.}}
  \textbf{\bibinfo{volume}{3}}, \bibinfo{pages}{34} (\bibinfo{year}{2017}).

\bibitem{heinrich2021quantum}
\bibinfo{author}{Heinrich, A.~J.} \emph{et~al.}
\newblock \bibinfo{title}{Quantum-coherent nanoscience}.
\newblock \emph{\bibinfo{journal}{Nat. Nanotechnol.}}
  \textbf{\bibinfo{volume}{16}}, \bibinfo{pages}{1318--1329}
  (\bibinfo{year}{2021}).

\bibitem{gonzalez2021scaling}
\bibinfo{author}{Gonzalez-Zalba, M.} \emph{et~al.}
\newblock \bibinfo{title}{Scaling silicon-based quantum computing using {CMOS}
  technology}.
\newblock \emph{\bibinfo{journal}{Nat. Electron.}}
  \textbf{\bibinfo{volume}{4}}, \bibinfo{pages}{872--884}
  (\bibinfo{year}{2021}).

\bibitem{chatterjee2021semiconductor}
\bibinfo{author}{Chatterjee, A.} \emph{et~al.}
\newblock \bibinfo{title}{Semiconductor qubits in practice}.
\newblock \emph{\bibinfo{journal}{Nat. Rev. Phys.}}
  \textbf{\bibinfo{volume}{3}}, \bibinfo{pages}{157--177}
  (\bibinfo{year}{2021}).

\bibitem{stano2022review}
\bibinfo{author}{Stano, P.} \& \bibinfo{author}{Loss, D.}
\newblock \bibinfo{title}{Review of performance metrics of spin qubits in gated
  semiconducting nanostructures}.
\newblock \emph{\bibinfo{journal}{Nat. Rev. Phys.}}
  \textbf{\bibinfo{volume}{4}}, \bibinfo{pages}{672--688}
  (\bibinfo{year}{2022}).

\bibitem{burkard2023semiconductor}
\bibinfo{author}{Burkard, G.}, \bibinfo{author}{Ladd, T.~D.},
  \bibinfo{author}{Pan, A.}, \bibinfo{author}{Nichol, J.~M.} \&
  \bibinfo{author}{Petta, J.~R.}
\newblock \bibinfo{title}{Semiconductor spin qubits}.
\newblock \emph{\bibinfo{journal}{Rev. Mod. Phys.}}
  \textbf{\bibinfo{volume}{95}}, \bibinfo{pages}{025003}
  (\bibinfo{year}{2023}).

\bibitem{xue2022quantum}
\bibinfo{author}{Xue, X.} \emph{et~al.}
\newblock \bibinfo{title}{Quantum logic with spin qubits crossing the surface
  code threshold}.
\newblock \emph{\bibinfo{journal}{Nature}} \textbf{\bibinfo{volume}{601}},
  \bibinfo{pages}{343--347} (\bibinfo{year}{2022}).

\bibitem{noiri2022fast}
\bibinfo{author}{Noiri, A.} \emph{et~al.}
\newblock \bibinfo{title}{Fast universal quantum gate above the fault-tolerance
  threshold in silicon}.
\newblock \emph{\bibinfo{journal}{Nature}} \textbf{\bibinfo{volume}{601}},
  \bibinfo{pages}{338--342} (\bibinfo{year}{2022}).

\bibitem{mills2022two}
\bibinfo{author}{Mills, A.~R.} \emph{et~al.}
\newblock \bibinfo{title}{Two-qubit silicon quantum processor with operation
  fidelity exceeding 99\%}.
\newblock \emph{\bibinfo{journal}{Sci. Adv.}} \textbf{\bibinfo{volume}{8}},
  \bibinfo{pages}{eabn5130} (\bibinfo{year}{2022}).

\bibitem{hendrickx2021four}
\bibinfo{author}{Hendrickx, N.~W.} \emph{et~al.}
\newblock \bibinfo{title}{A four-qubit germanium quantum processor}.
\newblock \emph{\bibinfo{journal}{Nature}} \textbf{\bibinfo{volume}{591}},
  \bibinfo{pages}{580--585} (\bibinfo{year}{2021}).

\bibitem{philips2022universal}
\bibinfo{author}{Philips, S.~G.} \emph{et~al.}
\newblock \bibinfo{title}{Universal control of a six-qubit quantum processor in
  silicon}.
\newblock \emph{\bibinfo{journal}{Nature}} \textbf{\bibinfo{volume}{609}},
  \bibinfo{pages}{919--924} (\bibinfo{year}{2022}).

\bibitem{lawrie2023simultaneous}
\bibinfo{author}{Lawrie, W.} \emph{et~al.}
\newblock \bibinfo{title}{Simultaneous single-qubit driving of semiconductor
  spin qubits at the fault-tolerant threshold}.
\newblock \emph{\bibinfo{journal}{Nat. Commun.}} \textbf{\bibinfo{volume}{14}},
  \bibinfo{pages}{3617} (\bibinfo{year}{2023}).

\bibitem{shulman2012demonstration}
\bibinfo{author}{Shulman, M.~D.} \emph{et~al.}
\newblock \bibinfo{title}{Demonstration of entanglement of electrostatically
  coupled singlet-triplet qubits}.
\newblock \emph{\bibinfo{journal}{Science}} \textbf{\bibinfo{volume}{336}},
  \bibinfo{pages}{202--205} (\bibinfo{year}{2012}).

\bibitem{nichol2017high}
\bibinfo{author}{Nichol, J.~M.} \emph{et~al.}
\newblock \bibinfo{title}{High-fidelity entangling gate for double-quantum-dot
  spin qubits}.
\newblock \emph{\bibinfo{journal}{npj Quantum Inf.}}
  \textbf{\bibinfo{volume}{3}}, \bibinfo{pages}{3} (\bibinfo{year}{2017}).

\bibitem{weinstein2023universal}
\bibinfo{author}{Weinstein, A.~J.} \emph{et~al.}
\newblock \bibinfo{title}{Universal logic with encoded spin qubits in silicon}.
\newblock \emph{\bibinfo{journal}{Nature}} \textbf{\bibinfo{volume}{615}},
  \bibinfo{pages}{817--822} (\bibinfo{year}{2023}).

\bibitem{hensgens2017quantum}
\bibinfo{author}{Hensgens, T.} \emph{et~al.}
\newblock \bibinfo{title}{Quantum simulation of a {F}ermi--{H}ubbard model
  using a semiconductor quantum dot array}.
\newblock \emph{\bibinfo{journal}{Nature}} \textbf{\bibinfo{volume}{548}},
  \bibinfo{pages}{70--73} (\bibinfo{year}{2017}).

\bibitem{dehollain2020nagaoka}
\bibinfo{author}{Dehollain, J.~P.} \emph{et~al.}
\newblock \bibinfo{title}{Nagaoka ferromagnetism observed in a quantum dot
  plaquette}.
\newblock \emph{\bibinfo{journal}{Nature}} \textbf{\bibinfo{volume}{579}},
  \bibinfo{pages}{528--533} (\bibinfo{year}{2020}).

\bibitem{van2021quantum}
\bibinfo{author}{van Diepen, C.~J.} \emph{et~al.}
\newblock \bibinfo{title}{Quantum simulation of antiferromagnetic {H}eisenberg
  chain with gate-defined quantum dots}.
\newblock \emph{\bibinfo{journal}{Phys. Rev. X.}}
  \textbf{\bibinfo{volume}{11}}, \bibinfo{pages}{041025}
  (\bibinfo{year}{2021}).

\bibitem{wang2023probing}
\bibinfo{author}{Wang, C.-A.} \emph{et~al.}
\newblock \bibinfo{title}{Probing resonating valence bonds on a programmable
  germanium quantum simulator}.
\newblock \emph{\bibinfo{journal}{npj Quantum Inf.}}
  \textbf{\bibinfo{volume}{9}}, \bibinfo{pages}{58} (\bibinfo{year}{2023}).

\bibitem{jang2020individual}
\bibinfo{author}{Jang, W.} \emph{et~al.}
\newblock \bibinfo{title}{Individual two-axis control of three singlet-triplet
  qubits in a micromagnet integrated quantum dot array}.
\newblock \emph{\bibinfo{journal}{Appl. Phys. Lett.}}
  \textbf{\bibinfo{volume}{117}}, \bibinfo{pages}{234001}
  (\bibinfo{year}{2020}).

\bibitem{fedele2021simultaneous}
\bibinfo{author}{Fedele, F.} \emph{et~al.}
\newblock \bibinfo{title}{Simultaneous operations in a two-dimensional array of
  singlet-triplet qubits}.
\newblock \emph{\bibinfo{journal}{PRX Quantum}} \textbf{\bibinfo{volume}{2}},
  \bibinfo{pages}{040306} (\bibinfo{year}{2021}).

\bibitem{mortemousque2021coherent}
\bibinfo{author}{Mortemousque, P.-A.} \emph{et~al.}
\newblock \bibinfo{title}{Coherent control of individual electron spins in a
  two-dimensional quantum dot array}.
\newblock \emph{\bibinfo{journal}{Nat. Nanotechnol.}}
  \textbf{\bibinfo{volume}{16}}, \bibinfo{pages}{296--301}
  (\bibinfo{year}{2021}).

\bibitem{Levy2002pair}
\bibinfo{author}{Levy, J.}
\newblock \bibinfo{title}{Universal quantum computation with spin-$1/2$ pairs
  and {H}eisenberg exchange}.
\newblock \emph{\bibinfo{journal}{Phys. Rev. Lett.}}
  \textbf{\bibinfo{volume}{89}}, \bibinfo{pages}{147902}
  (\bibinfo{year}{2002}).

\bibitem{petta2005coherent}
\bibinfo{author}{Petta, J.~R.} \emph{et~al.}
\newblock \bibinfo{title}{Coherent manipulation of coupled electron spins in
  semiconductor quantum dots}.
\newblock \emph{\bibinfo{journal}{Science}} \textbf{\bibinfo{volume}{309}},
  \bibinfo{pages}{2180--2184} (\bibinfo{year}{2005}).

\bibitem{maune2012coherent}
\bibinfo{author}{Maune, B.~M.} \emph{et~al.}
\newblock \bibinfo{title}{Coherent singlet-triplet oscillations in a
  silicon-based double quantum dot}.
\newblock \emph{\bibinfo{journal}{Nature}} \textbf{\bibinfo{volume}{481}},
  \bibinfo{pages}{344--347} (\bibinfo{year}{2012}).

\bibitem{wu2014two}
\bibinfo{author}{Wu, X.} \emph{et~al.}
\newblock \bibinfo{title}{Two-axis control of a singlet--triplet qubit with an
  integrated micromagnet}.
\newblock \emph{\bibinfo{journal}{Proc. Natl. Acad. Sci. U.S.A.}}
  \textbf{\bibinfo{volume}{111}}, \bibinfo{pages}{11938--11942}
  (\bibinfo{year}{2014}).

\bibitem{jock2018silicon}
\bibinfo{author}{Jock, R.~M.} \emph{et~al.}
\newblock \bibinfo{title}{A silicon metal-oxide-semiconductor electron
  spin-orbit qubit}.
\newblock \emph{\bibinfo{journal}{Nat. Commun.}} \textbf{\bibinfo{volume}{9}},
  \bibinfo{pages}{1768} (\bibinfo{year}{2018}).

\bibitem{cerfontaine2020closed}
\bibinfo{author}{Cerfontaine, P.} \emph{et~al.}
\newblock \bibinfo{title}{Closed-loop control of a {G}a{A}s-based
  singlet-triplet spin qubit with 99.5\% gate fidelity and low leakage}.
\newblock \emph{\bibinfo{journal}{Nat. Commun.}} \textbf{\bibinfo{volume}{11}},
  \bibinfo{pages}{4144} (\bibinfo{year}{2020}).

\bibitem{cerfontaine2020high}
\bibinfo{author}{Cerfontaine, P.}, \bibinfo{author}{Otten, R.},
  \bibinfo{author}{Wolfe, M.}, \bibinfo{author}{Bethke, P.} \&
  \bibinfo{author}{Bluhm, H.}
\newblock \bibinfo{title}{High-fidelity gate set for exchange-coupled
  singlet-triplet qubits}.
\newblock \emph{\bibinfo{journal}{Phys. Rev. B}}
  \textbf{\bibinfo{volume}{101}}, \bibinfo{pages}{155311}
  (\bibinfo{year}{2020}).

\bibitem{jirovec2021singlet}
\bibinfo{author}{Jirovec, D.} \emph{et~al.}
\newblock \bibinfo{title}{A singlet-triplet hole spin qubit in planar {G}e}.
\newblock \emph{\bibinfo{journal}{Nat. Mater.}} \textbf{\bibinfo{volume}{20}},
  \bibinfo{pages}{1106--1112} (\bibinfo{year}{2021}).

\bibitem{takeda2018optimized}
\bibinfo{author}{Takeda, K.} \emph{et~al.}
\newblock \bibinfo{title}{Optimized electrical control of a {S}i/{S}i{G}e spin
  qubit in the presence of an induced frequency shift}.
\newblock \emph{\bibinfo{journal}{npj Quantum Inf.}}
  \textbf{\bibinfo{volume}{4}}, \bibinfo{pages}{54} (\bibinfo{year}{2018}).

\bibitem{undseth2023hotter}
\bibinfo{author}{Undseth, B.} \emph{et~al.}
\newblock \bibinfo{title}{Hotter is easier: Unexpected temperature dependence
  of spin qubit frequencies}.
\newblock \emph{\bibinfo{journal}{Phys. Rev. X}} \textbf{\bibinfo{volume}{13}},
  \bibinfo{pages}{041015} (\bibinfo{year}{2023}).

\bibitem{PhysRevApplied.19.044078}
\bibinfo{author}{Undseth, B.} \emph{et~al.}
\newblock \bibinfo{title}{Nonlinear response and crosstalk of electrically
  driven silicon spin qubits}.
\newblock \emph{\bibinfo{journal}{Phys. Rev. Appl.}}
  \textbf{\bibinfo{volume}{19}}, \bibinfo{pages}{044078}
  (\bibinfo{year}{2023}).

\bibitem{ono2002science}
\bibinfo{author}{Ono, K.}, \bibinfo{author}{Austing, D.},
  \bibinfo{author}{Tokura, Y.} \& \bibinfo{author}{Tarucha, S.}
\newblock \bibinfo{title}{Current rectification by pauli exclusion in a weakly
  coupled double quantum dot system}.
\newblock \emph{\bibinfo{journal}{Science}} \textbf{\bibinfo{volume}{297}},
  \bibinfo{pages}{1313--1317} (\bibinfo{year}{2002}).

\bibitem{qiao2021floquet}
\bibinfo{author}{Qiao, H.} \emph{et~al.}
\newblock \bibinfo{title}{Floquet-enhanced spin swaps}.
\newblock \emph{\bibinfo{journal}{Nat. Commun.}} \textbf{\bibinfo{volume}{12}},
  \bibinfo{pages}{2142} (\bibinfo{year}{2021}).

\bibitem{chanrion2020charge}
\bibinfo{author}{Chanrion, E.} \emph{et~al.}
\newblock \bibinfo{title}{Charge detection in an array of {CMOS} quantum dots}.
\newblock \emph{\bibinfo{journal}{Phys. Rev. Appl.}}
  \textbf{\bibinfo{volume}{14}}, \bibinfo{pages}{024066}
  (\bibinfo{year}{2020}).

\bibitem{duan2020remote}
\bibinfo{author}{Duan, J.} \emph{et~al.}
\newblock \bibinfo{title}{Remote capacitive sensing in two-dimensional
  quantum-dot arrays}.
\newblock \emph{\bibinfo{journal}{Nano Lett.}} \textbf{\bibinfo{volume}{20}},
  \bibinfo{pages}{7123--7128} (\bibinfo{year}{2020}).

\bibitem{hsiao2023exciton}
\bibinfo{author}{Hsiao, T.-K.} \emph{et~al.}
\newblock \bibinfo{title}{Exciton transport in a germanium quantum dot ladder}.
\newblock \emph{\bibinfo{journal}{Phys. Rev. X}} \textbf{\bibinfo{volume}{14}},
  \bibinfo{pages}{011048} (\bibinfo{year}{2024}).

\bibitem{borsoi2022shared}
\bibinfo{author}{Borsoi, F.} \emph{et~al.}
\newblock \bibinfo{title}{Shared control of a 16 semiconductor quantum dot
  crossbar array}.
\newblock \emph{\bibinfo{journal}{Nat. Nanotechnol.}} \bibinfo{pages}{1--7}
  (\bibinfo{year}{2023}).

\bibitem{neyens2023probing}
\bibinfo{author}{Neyens, S.} \emph{et~al.}
\newblock \bibinfo{title}{Probing single electrons across 300-mm spin qubit
  wafers}.
\newblock \emph{\bibinfo{journal}{Nature}} \textbf{\bibinfo{volume}{629}},
  \bibinfo{pages}{80--85} (\bibinfo{year}{2024}).

\bibitem{giordano2021germanium}
\bibinfo{author}{Scappucci, G.} \emph{et~al.}
\newblock \bibinfo{title}{The germanium quantum information route}.
\newblock \emph{\bibinfo{journal}{Nat. Rev. Mater.}}
  \textbf{\bibinfo{volume}{6}}, \bibinfo{pages}{926--943}
  (\bibinfo{year}{2021}).

\bibitem{petta2010coherent}
\bibinfo{author}{Petta, J.}, \bibinfo{author}{Lu, H.} \&
  \bibinfo{author}{Gossard, A.}
\newblock \bibinfo{title}{A coherent beam splitter for electronic spin states}.
\newblock \emph{\bibinfo{journal}{Science}} \textbf{\bibinfo{volume}{327}},
  \bibinfo{pages}{669--672} (\bibinfo{year}{2010}).

\bibitem{ribeiro2010harnessing}
\bibinfo{author}{Ribeiro, H.}, \bibinfo{author}{Petta, J.~R.} \&
  \bibinfo{author}{Burkard, G.}
\newblock \bibinfo{title}{Harnessing the {G}a{A}s quantum dot nuclear spin bath
  for quantum control}.
\newblock \emph{\bibinfo{journal}{Phys. Rev. B}} \textbf{\bibinfo{volume}{82}},
  \bibinfo{pages}{115445} (\bibinfo{year}{2010}).

\bibitem{nichol2015quenching}
\bibinfo{author}{Nichol, J.~M.} \emph{et~al.}
\newblock \bibinfo{title}{Quenching of dynamic nuclear polarization by
  spin-orbit coupling in {G}a{A}s quantum dots}.
\newblock \emph{\bibinfo{journal}{Nat. Commun.}} \textbf{\bibinfo{volume}{6}},
  \bibinfo{pages}{7682} (\bibinfo{year}{2015}).

\bibitem{PhysRevLett.128.126803}
\bibinfo{author}{Jirovec, D.} \emph{et~al.}
\newblock \bibinfo{title}{Dynamics of hole singlet-triplet qubits with large
  $g$-factor differences}.
\newblock \emph{\bibinfo{journal}{Phys. Rev. Lett.}}
  \textbf{\bibinfo{volume}{128}}, \bibinfo{pages}{126803}
  (\bibinfo{year}{2022}).

\bibitem{PhysRevB.104.195421}
\bibinfo{author}{Mutter, P.~M.} \& \bibinfo{author}{Burkard, G.}
\newblock \bibinfo{title}{All-electrical control of hole singlet-triplet spin
  qubits at low-leakage points}.
\newblock \emph{\bibinfo{journal}{Phys. Rev. B}}
  \textbf{\bibinfo{volume}{104}}, \bibinfo{pages}{195421}
  (\bibinfo{year}{2021}).

\bibitem{PhysRevApplied.18.054090}
\bibinfo{author}{Fern\'andez-Fern\'andez, D.}, \bibinfo{author}{Ban, Y.} \&
  \bibinfo{author}{Platero, G.}
\newblock \bibinfo{title}{Quantum control of hole spin qubits in double quantum
  dots}.
\newblock \emph{\bibinfo{journal}{Phys. Rev. Appl.}}
  \textbf{\bibinfo{volume}{18}}, \bibinfo{pages}{054090}
  (\bibinfo{year}{2022}).

\bibitem{cai2023coherent}
\bibinfo{author}{Cai, X.}, \bibinfo{author}{Connors, E.~J.},
  \bibinfo{author}{Edge, L.~F.} \& \bibinfo{author}{Nichol, J.~M.}
\newblock \bibinfo{title}{Coherent spin--valley oscillations in silicon}.
\newblock \emph{\bibinfo{journal}{Nature Physics}}
  \textbf{\bibinfo{volume}{19}}, \bibinfo{pages}{386--393}
  (\bibinfo{year}{2023}).

\bibitem{rooney2023gate}
\bibinfo{author}{Rooney, J.} \emph{et~al.}
\newblock \bibinfo{title}{Gate modulation of the hole singlet-triplet qubit
  frequency in germanium}.
\newblock \emph{\bibinfo{journal}{arXiv preprint arXiv:2311.10188}}
  (\bibinfo{year}{2023}).

\bibitem{lodari2021low}
\bibinfo{author}{Lodari, M.} \emph{et~al.}
\newblock \bibinfo{title}{Low percolation density and charge noise with holes
  in germanium}.
\newblock \emph{\bibinfo{journal}{Mater. Quantum Technol.}}
  \textbf{\bibinfo{volume}{1}}, \bibinfo{pages}{011002} (\bibinfo{year}{2021}).

\bibitem{PhysRevLett.115.096801}
\bibinfo{author}{Bertrand, B.} \emph{et~al.}
\newblock \bibinfo{title}{Quantum manipulation of two-electron spin states in
  isolated double quantum dots}.
\newblock \emph{\bibinfo{journal}{Phys. Rev. Lett.}}
  \textbf{\bibinfo{volume}{115}}, \bibinfo{pages}{096801}
  (\bibinfo{year}{2015}).

\bibitem{hendrickx2020single}
\bibinfo{author}{Hendrickx, N.} \emph{et~al.}
\newblock \bibinfo{title}{A single-hole spin qubit}.
\newblock \emph{\bibinfo{journal}{Nat. Commun.}} \textbf{\bibinfo{volume}{11}},
  \bibinfo{pages}{3478} (\bibinfo{year}{2020}).

\bibitem{PhysRevLett.116.116801}
\bibinfo{author}{Martins, F.} \emph{et~al.}
\newblock \bibinfo{title}{Noise suppression using symmetric exchange gates in
  spin qubits}.
\newblock \emph{\bibinfo{journal}{Phys. Rev. Lett.}}
  \textbf{\bibinfo{volume}{116}}, \bibinfo{pages}{116801}
  (\bibinfo{year}{2016}).

\bibitem{PhysRevLett.116.110402}
\bibinfo{author}{Reed, M.~D.} \emph{et~al.}
\newblock \bibinfo{title}{Reduced sensitivity to charge noise in semiconductor
  spin qubits via symmetric operation}.
\newblock \emph{\bibinfo{journal}{Phys. Rev. Lett.}}
  \textbf{\bibinfo{volume}{116}}, \bibinfo{pages}{110402}
  (\bibinfo{year}{2016}).

\bibitem{wang2024operating}
\bibinfo{author}{Wang, C.-A.} \emph{et~al.}
\newblock \bibinfo{title}{Operating semiconductor quantum processors with
  hopping spins}.
\newblock \emph{\bibinfo{journal}{arXiv preprint arXiv:2402.18382}}
  (\bibinfo{year}{2024}).

\bibitem{berritta2024real}
\bibinfo{author}{Berritta, F.} \emph{et~al.}
\newblock \bibinfo{title}{Real-time two-axis control of a spin qubit}.
\newblock \emph{\bibinfo{journal}{Nature Communications}}
  \textbf{\bibinfo{volume}{15}}, \bibinfo{pages}{1676} (\bibinfo{year}{2024}).

\bibitem{ha2021flexible}
\bibinfo{author}{Ha, W.} \emph{et~al.}
\newblock \bibinfo{title}{A flexible design platform for si/sige exchange-only
  qubits with low disorder}.
\newblock \emph{\bibinfo{journal}{Nano Letters}} \textbf{\bibinfo{volume}{22}},
  \bibinfo{pages}{1443--1448} (\bibinfo{year}{2021}).

\bibitem{dagotto1996surprises}
\bibinfo{author}{Dagotto, E.} \& \bibinfo{author}{Rice, T.}
\newblock \bibinfo{title}{Surprises on the way from one-to two-dimensional
  quantum magnets: The ladder materials}.
\newblock \emph{\bibinfo{journal}{Science}} \textbf{\bibinfo{volume}{271}},
  \bibinfo{pages}{618--623} (\bibinfo{year}{1996}).

\bibitem{mkadzik2022precision}
\bibinfo{author}{M{\k{a}}dzik, M.~T.} \emph{et~al.}
\newblock \bibinfo{title}{Precision tomography of a three-qubit donor quantum
  processor in silicon}.
\newblock \emph{\bibinfo{journal}{Nature}} \textbf{\bibinfo{volume}{601}},
  \bibinfo{pages}{348--353} (\bibinfo{year}{2022}).

\end{thebibliography}


\providecommand{\noopsort}[1]{}\providecommand{\singleletter}[1]{#1}%
\begin{thebibliography}{10}
\expandafter\ifx\csname url\endcsname\relax
  \def\url#1{\texttt{#1}}\fi
\expandafter\ifx\csname urlprefix\endcsname\relax\def\urlprefix{URL }\fi
\providecommand{\bibinfo}[2]{#2}
\providecommand{\eprint}[2][]{\url{#2}}

\bibitem{hendrickx2020single}
\bibinfo{author}{Hendrickx, N.} \emph{et~al.}
\newblock \bibinfo{title}{A single-hole spin qubit}.
\newblock \emph{\bibinfo{journal}{Nat. Commun.}} \textbf{\bibinfo{volume}{11}}, \bibinfo{pages}{3478} (\bibinfo{year}{2020}).

\bibitem{hsiao2023exciton}
\bibinfo{author}{Hsiao, T.-K.} \emph{et~al.}
\newblock \bibinfo{title}{Exciton transport in a germanium quantum dot ladder}.
\newblock \emph{\bibinfo{journal}{Phys. Rev. X}} \textbf{\bibinfo{volume}{14}}, \bibinfo{pages}{011048} (\bibinfo{year}{2024}).

\bibitem{wang2023probing}
\bibinfo{author}{Wang, C.-A.} \emph{et~al.}
\newblock \bibinfo{title}{Probing resonating valence bonds on a programmable germanium quantum simulator}.
\newblock \emph{\bibinfo{journal}{npj Quantum Inf.}} \textbf{\bibinfo{volume}{9}}, \bibinfo{pages}{58} (\bibinfo{year}{2023}).

\bibitem{nichol2015quenching}
\bibinfo{author}{Nichol, J.~M.} \emph{et~al.}
\newblock \bibinfo{title}{Quenching of dynamic nuclear polarization by spin-orbit coupling in {G}a{A}s quantum dots}.
\newblock \emph{\bibinfo{journal}{Nat. Commun.}} \textbf{\bibinfo{volume}{6}}, \bibinfo{pages}{7682} (\bibinfo{year}{2015}).

\bibitem{PhysRevLett.128.126803}
\bibinfo{author}{Jirovec, D.} \emph{et~al.}
\newblock \bibinfo{title}{Dynamics of hole singlet-triplet qubits with large $g$-factor differences}.
\newblock \emph{\bibinfo{journal}{Phys. Rev. Lett.}} \textbf{\bibinfo{volume}{128}}, \bibinfo{pages}{126803} (\bibinfo{year}{2022}).

\bibitem{hendrickx2023sweet}
\bibinfo{author}{Hendrickx, N.} \emph{et~al.}
\newblock \bibinfo{title}{Sweet-spot operation of a germanium hole spin qubit with highly anisotropic noise sensitivity}.
\newblock \emph{\bibinfo{journal}{Nature Materials}} \bibinfo{pages}{1--8} (\bibinfo{year}{2024}).

\bibitem{PhysRevLett.96.100501}
\bibinfo{author}{Hu, X.} \& \bibinfo{author}{Das~Sarma, S.}
\newblock \bibinfo{title}{Charge-fluctuation-induced dephasing of exchange-coupled spin qubits}.
\newblock \emph{\bibinfo{journal}{Phys. Rev. Lett.}} \textbf{\bibinfo{volume}{96}}, \bibinfo{pages}{100501} (\bibinfo{year}{2006}).

\bibitem{huang2018spin}
\bibinfo{author}{Huang, P.}, \bibinfo{author}{Zimmerman, N.~M.} \& \bibinfo{author}{Bryant, G.~W.}
\newblock \bibinfo{title}{Spin decoherence in a two-qubit {CPHASE} gate: the critical role of tunneling noise}.
\newblock \emph{\bibinfo{journal}{npj Quantum Inf.}} \textbf{\bibinfo{volume}{4}}, \bibinfo{pages}{62} (\bibinfo{year}{2018}).

\bibitem{shulman2012demonstration}
\bibinfo{author}{Shulman, M.~D.} \emph{et~al.}
\newblock \bibinfo{title}{Demonstration of entanglement of electrostatically coupled singlet-triplet qubits}.
\newblock \emph{\bibinfo{journal}{Science}} \textbf{\bibinfo{volume}{336}}, \bibinfo{pages}{202--205} (\bibinfo{year}{2012}).

\bibitem{nichol2017high}
\bibinfo{author}{Nichol, J.~M.} \emph{et~al.}
\newblock \bibinfo{title}{High-fidelity entangling gate for double-quantum-dot spin qubits}.
\newblock \emph{\bibinfo{journal}{npj Quantum Inf.}} \textbf{\bibinfo{volume}{3}}, \bibinfo{pages}{3} (\bibinfo{year}{2017}).

\bibitem{burkard2023semiconductor}
\bibinfo{author}{Burkard, G.}, \bibinfo{author}{Ladd, T.~D.}, \bibinfo{author}{Pan, A.}, \bibinfo{author}{Nichol, J.~M.} \& \bibinfo{author}{Petta, J.~R.}
\newblock \bibinfo{title}{Semiconductor spin qubits}.
\newblock \emph{\bibinfo{journal}{Rev. Mod. Phys.}} \textbf{\bibinfo{volume}{95}}, \bibinfo{pages}{025003} (\bibinfo{year}{2023}).

\bibitem{DanonPRB80-041301}
\bibinfo{author}{Danon, J.} \& \bibinfo{author}{Nazarov, Y.~V.}
\newblock \bibinfo{title}{Pauli spin blockade in the presence of strong spin-orbit coupling}.
\newblock \emph{\bibinfo{journal}{Phys. Rev. B}} \textbf{\bibinfo{volume}{80}}, \bibinfo{pages}{041301} (\bibinfo{year}{2009}).

\bibitem{geyer2022twoqubit}
\bibinfo{author}{Geyer, S.} \emph{et~al.}
\newblock \bibinfo{title}{Anisotropic exchange interaction of two hole-spin qubits}.
\newblock \emph{\bibinfo{journal}{Nature Physics}} \bibinfo{pages}{1--6} (\bibinfo{year}{2024}).

\bibitem{burkardCoupledQuantumdots}
\bibinfo{author}{Burkard, G.}, \bibinfo{author}{Loss, D.} \& \bibinfo{author}{DiVincenzo, D.~P.}
\newblock \bibinfo{title}{Coupled quantum dots as quantum gates}.
\newblock \emph{\bibinfo{journal}{Phys. Rev. B}} \textbf{\bibinfo{volume}{59}}, \bibinfo{pages}{2070--2078} (\bibinfo{year}{1999}).

\bibitem{jirovec2021singlet}
\bibinfo{author}{Jirovec, D.} \emph{et~al.}
\newblock \bibinfo{title}{A singlet-triplet hole spin qubit in planar {G}e}.
\newblock \emph{\bibinfo{journal}{Nat. Mater.}} \textbf{\bibinfo{volume}{20}}, \bibinfo{pages}{1106--1112} (\bibinfo{year}{2021}).

\bibitem{bravyiSchriefferWolffTransformationQuantum2011}
\bibinfo{author}{Bravyi, S.}, \bibinfo{author}{Divincenzo, D.~P.} \& \bibinfo{author}{Loss, D.}
\newblock \bibinfo{title}{Schrieffer-{{Wolff}} transformation for quantum many-body systems}.
\newblock \emph{\bibinfo{journal}{Annals of Physics}} \textbf{\bibinfo{volume}{326}}, \bibinfo{pages}{2793} (\bibinfo{year}{2011}).

\bibitem{chirolliDecoherenceSolidstateQubits2008}
\bibinfo{author}{Chirolli, L.} \& \bibinfo{author}{Burkard, G.}
\newblock \bibinfo{title}{Decoherence in solid-state qubits}.
\newblock \emph{\bibinfo{journal}{Advances in Physics}} \textbf{\bibinfo{volume}{57}}, \bibinfo{pages}{225--285} (\bibinfo{year}{2008}).

\bibitem{lawrieSimultaneousSinglequbitDriving2023}
\bibinfo{author}{Lawrie, W. I.~L.} \emph{et~al.}
\newblock \bibinfo{title}{Simultaneous single-qubit driving of semiconductor spin qubits at the fault-tolerant threshold}.
\newblock \emph{\bibinfo{journal}{Nature Communications}} \textbf{\bibinfo{volume}{14}}, \bibinfo{pages}{3617} (\bibinfo{year}{2023}).

\bibitem{nielsen2020probing}
\bibinfo{author}{Nielsen, E.} \emph{et~al.}
\newblock \bibinfo{title}{Probing quantum processor performance with pygsti}.
\newblock \emph{\bibinfo{journal}{Quantum science and Technology}} \textbf{\bibinfo{volume}{5}}, \bibinfo{pages}{044002} (\bibinfo{year}{2020}).

\bibitem{mkadzik2022precision}
\bibinfo{author}{M{\k{a}}dzik, M.~T.} \emph{et~al.}
\newblock \bibinfo{title}{Precision tomography of a three-qubit donor quantum processor in silicon}.
\newblock \emph{\bibinfo{journal}{Nature}} \textbf{\bibinfo{volume}{601}}, \bibinfo{pages}{348--353} (\bibinfo{year}{2022}).

\bibitem{PRXQuantum.3.020335}
\bibinfo{author}{Blume-Kohout, R.} \emph{et~al.}
\newblock \bibinfo{title}{A taxonomy of small markovian errors}.
\newblock \emph{\bibinfo{journal}{PRX Quantum}} \textbf{\bibinfo{volume}{3}}, \bibinfo{pages}{020335} (\bibinfo{year}{2022}).

\end{thebibliography}
\bibliographystyle{naturemag}

\section{\label{sec:level5}Methods}
\subsection{\label{sec:level50}Device and setup}
The device is thermally anchored to the mixing chamber of a dilution refrigerator with a base temperature of around 10 mK. All the control electronics are at room temperature, which connect the device via 50 direct current (DC) lines and 24 alternating current (AC) lines in total.  {The DC and AC signals are combined using bias tees on the printed circuit board (PCB) with an RC time constant
of 100 ms to apply both signals to the same gate of the device. For baseband pulses, a compensation pulse to the gate is applied to make the dc offset over the whole measurement cycle equal to zero, which mitigates the charging effects in the bias tees.} DC signals are produced by custom-built battery-powered DC voltage sources and are fed through a matrix module---a breakout box with filters inside---to the Fisher cables of the fridge. AC signals are produced by 6 Keysight M3202A modules which are connected directly to the coaxial lines in the fridge. {The output digital filter of the AWG channels is set to the anti-ringing filter mode to suppress ringing effects in the baseband pulses.} For the filters  {in the lines}, we use common-mode Ferrite chokes at room temperature to filter low-frequency noise (10 kHz - 1 MHz) in the ground of AC lines and use RC filters {(R=100 k$\Omega$, C=47 nF for normal gates, R=470 $\Omega$, C=270 pF for the Ohmic contacts)} as well as copper-powder filters that are mounted on the cold finger attached to the mixing chamber plate to filter high-frequency noise in DC lines.

The sensing dots are measured using radio-frequency (RF) reflectometry with working frequencies of 179 MHz, 190 MHz, 124 MHz and 158 MHz for sensor  $\text{S}_{\text{TL}}$,  $\text{S}_{\text{BL}}$,  $\text{S}_{\text{TR}}$ and  $\text{S}_{\text{BR}}$, respectively. Tank circuits are formed by NbTiN inductors mounted on the PCB and the spurious capacitance of the bonding wires and metal lines on the board and chip. We apply RF signals using custom-built RF generators and combine them into a single coaxial line at room temperature using a power combiner ZFSC-3-1W-S+. The signal is attenuated at each plate in the dilution refrigerator and passes through a directional coupler (ZEDC-15-2B) at the mixing chamber to reach the device. The signal reflected from the device goes through the same directional coupler and is then amplified with a CITLF3 cryogenic amplifier at the 4K plate. At room temperature, the signal is amplified again and demodulated by custom-built IQ mixers. The demodulated signal is sent to the Keysight M3102A module to convert analog readout signals to digital signals. We use DC blocks to reduce low-frequency noise (< 10 MHz) in the RF lines. The DC block inside the refrigerator blocks the DC signal on the inner conductor (PE8210) while the ones at room temperature block that on both the inner and outer conductor (PE8212). To suppress high-frequency noise in the reflected signal, we use a low-pass filter (SBLP-300+) at room temperature.

\subsection{\label{sec:level51}Initialization, control and readout}
In the experiment, we repeatedly perform single-shot readout cycles to obtain singlet or triplet probabilities. The integration time for each single-shot readout is around 10-40 $\mu$s, depending on the signal-to-noise ratio and triplet relaxation time during measurements. To compensate for the drift of the sensor signal, we use a reference readout segment before each measurement sequence~\cite{van2021quantum}. For some datasets, we adjust the single-shot readout threshold by post-processing instead of through a reference segment. In post-processing, we collect a histogram of 500-4000 shots for each data point based on which we set the threshold to analyze those shots. In this way, the sensor drift between data points is mostly filtered out.

{A typical pulse for single-qubit control can include initialization (20 $\mu$s), reference readout (20 $\mu$s), initialization (20-50 $\mu$s), control (30-3000 ns) and readout (20 $\mu$s). A ramp-in time of 20 ns between initialization and control is used to avoid diabatic errors. The position of initialization is in the (0,2) or (2,0) regime but deeper than the PSB readout point to ensure fast relaxation to the ground state. In single-qubit GST measurements, the gate set includes a null operation. In order to avoid the readout immediately following the initialization, a waiting time of 10 ns at a point in the (1,1) regime is added to ensure the data acquisition is done correctly. This may decrease the readout fidelity when the waiting point causes unwanted rotations of the qubit. For this reason, the waiting point was moved to the readout position in the two-qubit GST experiment. For multi-qubit initialization and control, we initialize all the qubits into the singlet simultaneously by ramping from (2,0) or (0,2) to a high detuning point in (1,1), except for the qubit to be subject to single-qubit control, which is pulsed directly to the zero detuning point. We also found that adding a brief pre-control segment after initialization at high detuning in (1,1) for all qubits (wait about 2 ns) can give a better initialization to singlets. This variation is used in some of the experiments on quantum state tomography and gate set tomography.}

{For the qubit operation times we used in the measurements of RB, QST and GST, the typical values are summarized as follows:} 

\begin{itemize}
\item $\sqrt{X}$: 43.5 ns (Q1), 27.5 ns (Q2), 35 ns (Q3), and 25 ns (Q4).
\item $H$: 65 ns (Q1), 40 ns (Q2), 56 ns (Q3), and 40 ns (Q4); 
\item $\sqrt{\text{SWAP}}$: 13 ns (Q1-Q2), 16.5 ns (Q2-Q3), and 11 ns (Q3-Q4). 
\end{itemize}

\begin{comment}
\begin{table}[h]
\centering
\begin{tabular}{ |p{2.5cm}||p{1cm}|p{1cm}|p{1cm}|p{1cm}|  }
 \hline
 Gate time (ns) & Q1 & Q2 & Q3 &Q4 \\
 \hline
 $\sqrt{X}$   &43.5  &27.5  &35   &25\\
 \hline
 $H$   &65 &40 &56 &40\\
 \hline
 $\sqrt{\text{SWAP}}$   &13 &16.5 &11 \\
 \hline
\end{tabular}
\end{table}   
\end{comment}

\subsection{\label{sec:level53}Randomized benchmarking}
{In single-qubit randomized benchmarking (RB), we use the native gates $I$, $\sqrt{X}$ and $\sqrt{Y}$ to compose the sequences of Clifford gates. At the end of each sequence, a rotation is applied to (ideally) bring the qubit back to its initial state, and the final qubit state is measured using PSB. 
Experimentally, the single-qubit $I$ gate is implemented as a pulse segment with zero waiting time. The Clifford gate sequence length varies from 2 to 232, and there are in total 30 random sequences for each sequence length. Single-shot measurement of the tested qubit is repeated 1000 times for each random sequence to obtain the singlet or triplet probability. The measured data is fitted to a function $P_S=Ap_c^N+B$, where $p_c$ is the depolarizing parameter, $A$ and $B$ are the coefficients that absorb the state preparation and
measurement errors, and $N$ is the number of Clifford operations in the sequence. The average Clifford infidelity can then be described as $r_c=(d-1)(1-p_c)/d$, where $d=2^n$ is the dimension of the system and $n$ is the number of qubits. For the single-qubit operations used here, there are on average 3.625 generators per Clifford composition (see Extended Fig.~\ref{RB_cir} for details). Therefore, the average gate fidelity is given by $F_g=1-r_c/3.625$. The uncertainties in the reported numbers represent the standard deviations acquired from the curve fitting.}

\subsection{\label{sec:level52}Quantum state tomography}
{The density matrix of a two-qubit state can be expressed as $\rho=\sum_{i=1}^{16}c_iM_i$ where $M_i$ are 16 linearly independent measurement operators, and the coefficients $c_i$ are calculated from the expectation values $m_i$ of the measurement operators using a maximum-likelihood estimate. In the experiment, we performed 9 combinations of \{$I$, $\sqrt{X}$, $\sqrt{Y}$\} basis-change rotations on the two qubits and obtained the expectation values $m_i$ by determining the joint two-qubit probabilities. To do so, we performed 500 single-shot measurements per sequence, and repeated the whole experiment 3-5 times. After that, the measured probabilities were converted to the estimated actual two-spin probabilities by removing the SPAM errors.} 

{The SPAM matrix was measured by aiming to initialize two qubits into $\ket{SS}$, $\ket{ST_-}$, $\ket{T_-S}$, and $\ket{T_-T_-}$, and repeatedly measuring the two-qubit states in a single-shot manner. Then we use the relationship $\textbf{\textit{P}}_M = M_{\text{SPAM}}\textbf{\textit{P}}$, where $\textbf{\textit{P}}_M$ are the measured two-qubit probabilities, $M_{\text{SPAM}}$ the SPAM matrix, and \textbf{\textit{P}} the actual two-qubit probabilities. We notice this relationship works when the initialization error is negligible compared to the readout error, or it would cause miscorrections in the results.}

{Single-shot readout of two-qubit states was implemented differently for different qubit pairs. For Q1 and Q2, we first measure Q1 with an integration time of 20 $\mu$s while maintaining Q2 in the symmetry condition but with $\delta \text{v}b_{26}=-60$ mV to preserve its state. Next Q2 is measured. This method uses the same sensor for PSB readout of both Q1 and Q2, therefore the two measurements have to be done sequentially. For Q2-Q3 and Q3-Q4, we performed SWAP gates to transfer the qubit information to Q1 and Q4, and the two qubits were measured simultaneously using two sensors on both sides. Also for the characterization of the remote Bell state Q1-Q4, the qubits Q1 and Q4 were measured simultaneously using the two sensors on both sides (after possible single-qubit rotations to change basis).}

{The single-qubit rotations before the final measurement were performed sequentially. Hence, the time between the $\sqrt{\text{SWAP}}$ gate and the single-qubit gate of the second qubit can be dependent on any single-qubit operation being applied to the first qubit. These different times would cause different phase accumulations on the second qubit. To solve this problem, we use a waiting time as long as the longest qubit operation time of the first qubit before performing the basis-change rotation of the second qubit. This ensures the phase of the second qubit is consistent throughout the whole experiment (see Extended Fig.~\ref{Bell state}a).}

{The Bell state fidelity is obtained from the experimentally obtained density matrix $\rho_\text{exp}$ and the ideally expected density matrix, as $\rho_\text{ideal}$, and $F=\text{Tr}(\sqrt{\sqrt{\rho_\text{ideal}}\rho_\text{exp}\sqrt{\rho_\text{ideal}}})$. The phase $\theta$ of the ideal Bell state $\ket{\psi}=\frac{1}{\sqrt{2}}(\ket{ST_-}+e^{i\theta}\ket{T_-S})$ is used as a fitting parameter to incorporate additional (fixed and predictable) single-qubit phase rotations before and after the $\sqrt{\text{SWAP}}$ gate. The fitted $\theta$ for the Bell state Q1-Q2, Q2-Q3, Q3-Q4 and Q1-Q4 are 0.717, -0.614, -2.718 and 2.507, respectively. We note the non-ideal pulse effect between the concatenated single-qubit gate and the $\sqrt{\text{SWAP}}$ gate may also result in other types of single-qubit rotations (see the Extended Data Fig.~\ref{Two qubit GST restuls}), which is not incorporated and can contribute to errors in the Bell state preparation. The uncertainties in the reported numbers are the standard deviations calculated from 2000 bootstrap re-sampling iterations of the single-shot readout data for both the SPAM matrix and $\textbf{\textit{P}}_M$. }

\section{\label{sec:level6}Extended figures}
\setcounter{figure}{0}

\begin{figure*}[htb]
\renewcommand{\figurename}{\textbf{Extended Data Fig}}
\includegraphics{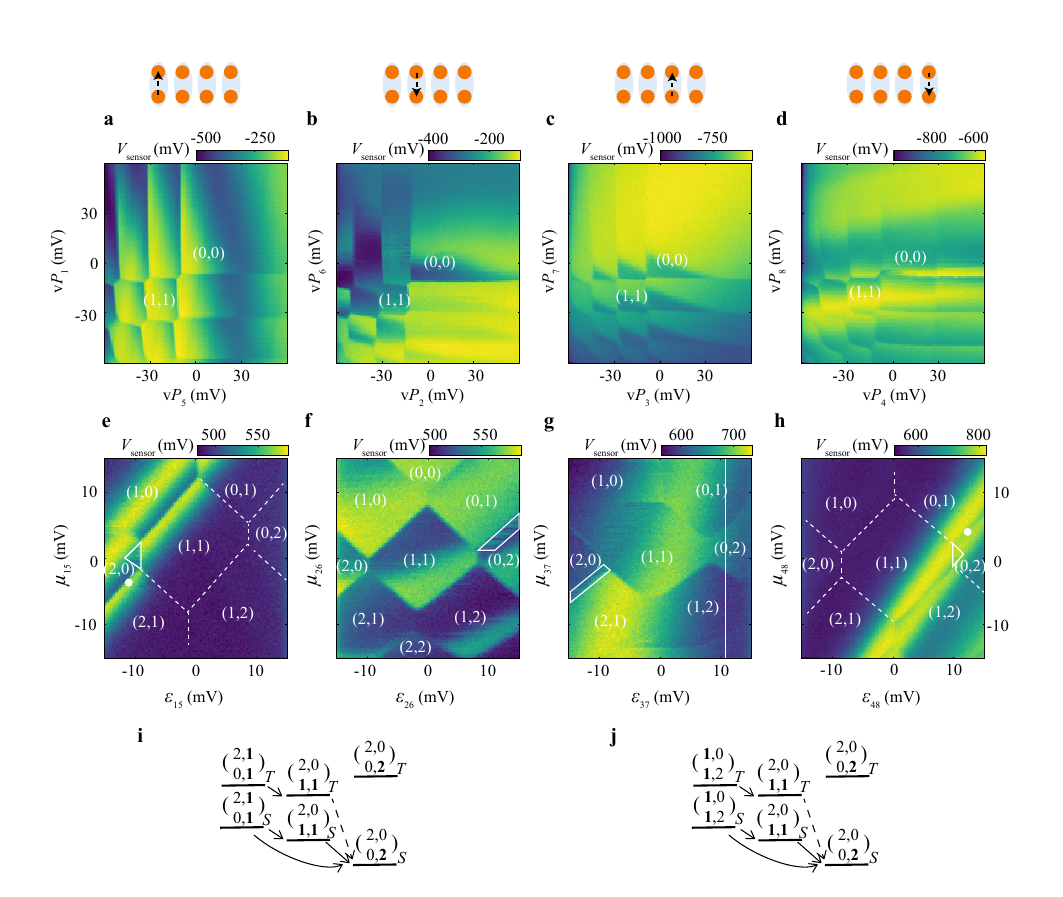}
%\internallinenumbers
\caption{\label{CSD_ED}\textbf{Charge stability diagrams and Pauli spin blockade.} \textbf{a-d}, Charge stability diagrams for DQD 1-5 (\textbf{a}), 2-6 (\textbf{b}), 3-7 (\textbf{c}), and 4-8 (\textbf{d}), respectively. \textbf{a} and \textbf{b} are recorded using the sensor $\text{S}_{\text{BL}}$ while \textbf{c} and \textbf{d} are recorded using the sensor $\text{S}_{\text{BR}}$. Hole numbers inside the relevant charge stability regions are indicated, {showing all the DQDs can be emptied to (0,0). \textbf{e-h}, Charge stability diagrams measured by scanning the detuning $\varepsilon_{ij}$ and the overall chemical potential $\mu_{ij}$ of the DQD. The PSB regions inside the (2, 0) or (0, 2) area }are indicated by solid white triangles and trapezoids. For outer DQD 1-5 and 4-8, we find PSB by pulsing $\varepsilon_{15}$ and $\varepsilon_{48}$ from (1,1) to (2,0) and (0,2), where within a triangular region an electron tunnels between the dots starting from the $S(1, 1)$ but no tunneling occurs (the system is in Pauli spin blockade) from $T_0(1, 1)$, $T_-(1, 1)$ and $T_+(1, 1)$. For inner DQD 2-6 and 3-7, we swap their spin states to those of DQD 5-6 and 7-8 where the sensor signals are stronger. \textbf{i},\textbf{j}, Illustration of PSB using the energy levels in the quadruple quantum dot plaquette for DQD 2-6 (\textbf{i}) and 3-7 (\textbf{j}), respectively. The hole numbers are indicated as $\left( \begin{smallmatrix} n_1, & n_2 \\ n_5, & n_6 \end{smallmatrix} \right)$ for \textbf{i} and $\left( \begin{smallmatrix} n_3, & n_4 \\ n_7, & n_8 \end{smallmatrix} \right)$ for \textbf{j}, and the subscripts $S$ and $T$ show the two-spin states of holes in the quantum dots indicated by bold numbers, respectively. The solid arrows show fast spin-conserving tunneling while the dashed arrows show suppressed tunneling due to PSB. Here we take pair 2-6 as an example to explain the readout process of the inner spin pairs. First, we align DQD 1-5 at the charge stability boundary between (2,0) and (2,1), as shown by the white dot in \textbf{e}, and then pulse $\varepsilon_{26}$ from negative to positive. We subsequently find a shaded region between (0,1) and (0,2) in the diagram for DQD 2-6, which is caused by PSB in DQD 5-6. The mechanism is shown in \textbf{i}: when we pulse DQD 2-6 to the point where $S(0,2)$ is lower in energy than $(0,1)$, the holes in DQD 2-6 moves across to DQD 5-6, irrespective of the spin states. Subsequently, the conventional PSB mechanism in DQD 5-6 allows $S(1,1)$ to transition to $S(0,2)$, while the triplets $T(1,1)$ have to remain in the (1,1) charge state. In this way, we indirectly realize spin-to-charge conversion for the two spins initially in DQD 2-6. Actually, $S(1,1)$ in DQD 2-6 can also directly tunnel to $S(0,2)$ inside the same DQD, as seen by the curved arrow in \textbf{i}. The mechanism to measure DQD 3-7 is analogous.}
\end{figure*}

\begin{figure*}[htb]
\renewcommand{\figurename}{\textbf{Extended Data Fig}}
\includegraphics{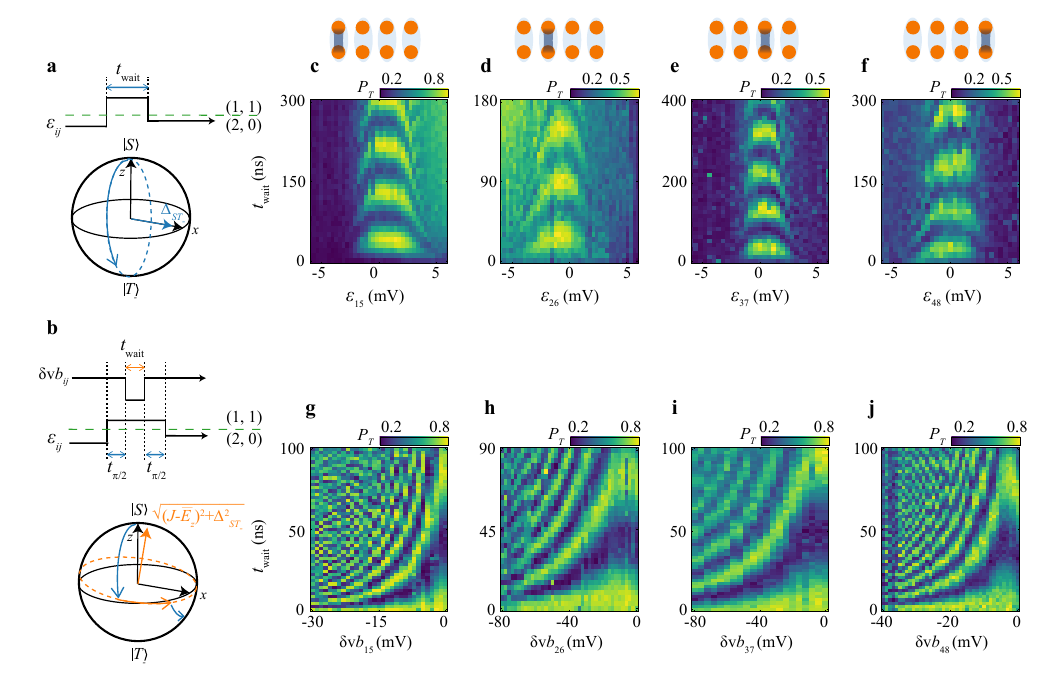}
%\internallinenumbers
\caption{\label{Additional single-qubit results} {\textbf{Data of two-axis qubit control around the $x$- and $z$-axis, measured at $B$ = 10 mT. }\textbf{a},\textbf{b}, Pulse scheme and Bloch sphere illustration of $x$-axis and $z$-axis evolution of $S-T_-$ qubits. The straight blue and orange arrows show the corresponding rotation axis. The $x$-axis rotations are set by the $S-T_-$ coupling, $\Delta_{ST_-}$. For large $J$ such that $J-\overline{E}_z\gg \Delta_{ST_-}$, the rotation axis tilts towards the $z$-axis. The rotation is never exactly around the $z$-axis due to the presence of a finite $\Delta_{ST_-}$, yet, sufficiently orthogonal control is possible when $(J-\overline{E}_z) \gg \Delta_{ST_-}$. In \textbf{b}, we illustrate a Ramsey-like pulse sequence used to demonstrate $z$-axis control. We first initialize the qubit into a singlet, perform a $\pi/2$ rotation around the $x$-axis of duration $t_{\pi/2}$, and then change $J$ diabatically by pulsing the corresponding barrier gate by an amount $\delta\text{v}b_{ij}$ to implement a $z$-axis rotation. Finally, we perform another $\pi/2$ operation around the $x$-axis and project the qubit into the $S-T_-$ basis for spin readout. \textbf{c-f}, Experimental results for $x$-axis rotations of each qubit, showing measured triplet probabilities $P_T$ as a function of $t_\text{wait}$ and the detuning voltage $\varepsilon_{ij}$. \textbf{g-j}, Experimental results for $z$-axis rotations of each qubit, showing $P_T$ as a function of $t_\text{wait}$ and the barrier voltage change $\delta \text{v}b_{ij}$. The oscillation frequency is given by $f_{ST_-}=\sqrt{(J-\overline{E}_z)^2+\Delta^2_{ST_-}}/h$, where $h$ is Planck's constant. We note that the outer two barrier gates $\text{v}b_{15}$ and $\text{v}b_{48}$ have a stronger effect on the corresponding $J_{ij}$ than the inner barrier gates $\text{v}b_{26}$ and $\text{v}b_{37}$. This may be explained by {additional residual resist} below the inner barrier gates, which are fabricated in the last step~\cite{hsiao2023exciton}, and by the different fan-out routing for the outer barrier gates (see Fig.~\ref{Device}a,b in the main text). Within the tuning range of the barrier gate, the highest ratio $(J-\overline{E}_z)/\Delta_{ST_-}$ amounts to around 20 for the outer qubits Q1 and Q4 and about 10 for the inner qubits Q2 and Q3 (see Supplementary Information section IV for more details). }}
\end{figure*}

\begin{figure*}[htb]
\renewcommand{\figurename}{\textbf{Extended Data Fig}}
\includegraphics{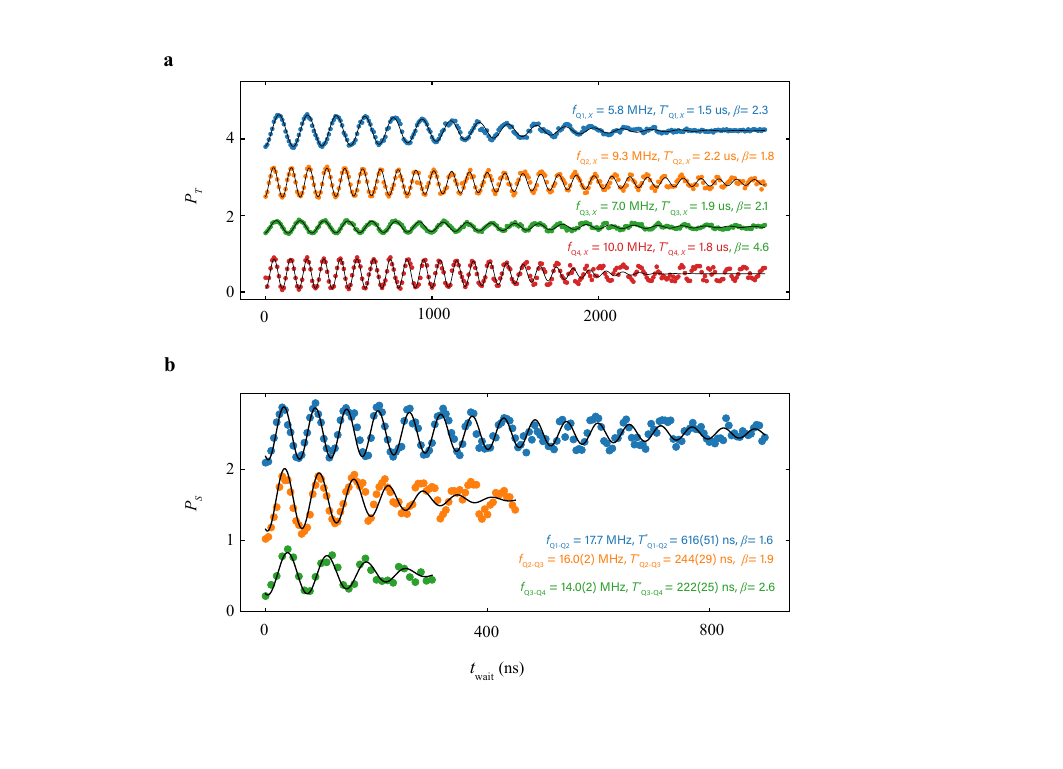}
%\internallinenumbers
\caption{\label{Additional single-qubit results2} {\textbf{Decoherence times of the qubits under control.} \textbf{a}, Measured triplet probabilities $P_T$ of long-time evolutions around the $x$-axis for Q1-Q4 at $B$=5 mT. \textbf{b}, Measured singlet probabilities $P_S$ as a function of the evolution time $t_\text{wait}$ at the center of the chevron patterns of the SWAP oscillations for each pair of qubits at $B$=5 mT. The data of \textbf{a} and \textbf{b} are fitted with a function of the form $P_T=P_0 + A \text{cos}(2\pi ft+\phi)\text{exp}[-(t/T^*)^\beta]$, where $P_0$, $A$, $\beta$, $f$, $T^*$ are fitting parameters. Here $f$ refers to the oscillation frequency, $T^*$ refers to $T_x^*$, the coherence time under $x$-axis rotations, or $T_{Qi-Qj}^*$, the coherence time under SWAP oscillations between adjacent qubits. Furthermore, $\beta$ determines the shape of the decay envelope, and the fitted values are also shown in the inset. $\beta$ gives information on the noise spectrum in the system. If the system is dominated by quasi-static or low-frequency noise, $\beta=2$, which gives a Gaussian decay; if the system is dominated by high-frequency noise, $\beta=1$, which gives an exponential decay. The extracted $\beta$ for the $x$-axis rotations of Q1-Q3 and all the SWAP oscillations are close to 2, indicating the dominance of low-frequency noise in this system. Notice the large value of $\beta$ for the $x$-axis rotations of Q4 may result from the fitting error.}}
\end{figure*}

\begin{figure*}[htb]
\renewcommand{\figurename}{\textbf{Extended Data Fig}}
\includegraphics{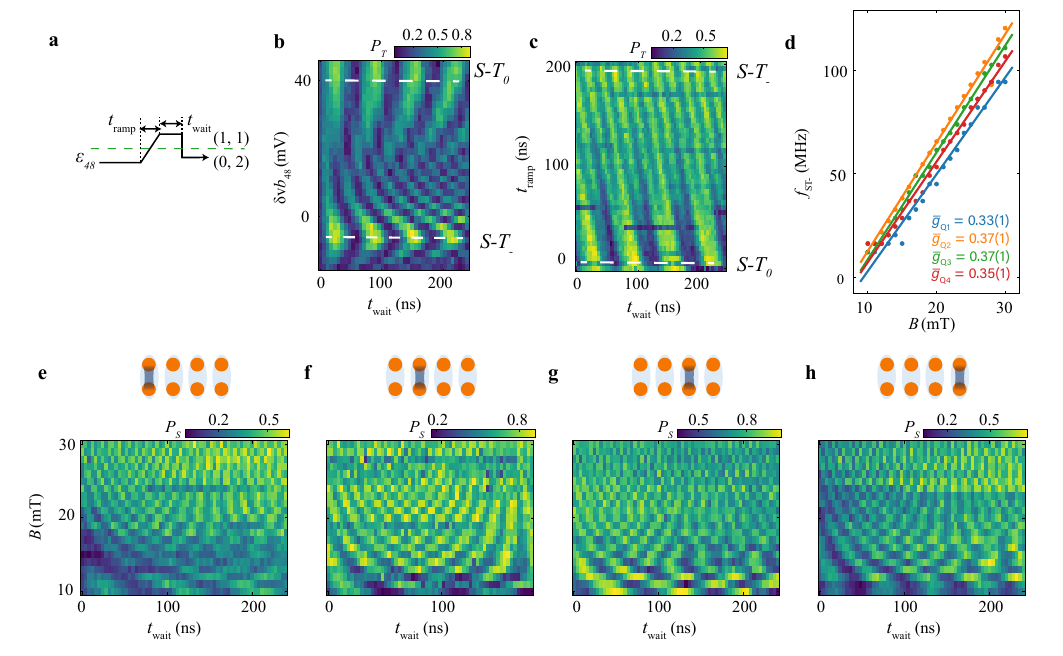}
%\internallinenumbers
\caption{\label{Additional single-qubit results3} {\textbf{Coherent control of singlet-triplet states under different conditions and the average $g$-factor.} \textbf{a}, Pulse scheme to measure $S-T_-$ and $S-T_0$ oscillations. A square pulse is used with a ramp-in time $t_\text{ramp}$ and waiting time $t_\text{wait}$. \textbf{b}, Measured triplet probabilities of DQD 4-8 as a function of $t_\text{wait}$ and the barrier voltage amplitude $\delta \text{v}b_{48}$ with $t_\text{ramp}=20$ ns. As mentioned in the main text, a block pulse along the detuning axis with 20-nanosecond ramp time ($t_\text{ramp}$) from (0,2) to (1,1) is adiabatic with respect to the {tunnel coupling} but diabatic with respect to the $S-T_-$ anticrossing. Therefore, we can drive $x$-rotations of $S-T_-$ qubits when the pulse amplitude reaches zero detuning with $J(\varepsilon_{ij} = 0) \sim \overline{E_z}$ (see Fig.~\ref{Device}d in the main text). However, when we increase the barrier voltage change $\delta \text{v}b_{48}$ until $J(\varepsilon_{ij} = 0) < \overline{E_z}$ ($\delta \text{v}b_{48} \sim 40$ mV), the $S-T_-$ anticrossings appear away from zero detuning (see Fig.~\ref{Device}c in the main text), thus the same pulse does not produce $x$-axis oscillations of the $S-T_-$ qubit. Moreover, under this condition, the $S-T_0$ splitting is reduced and the 20-nanosecond ramp time eventually becomes {diabatic} with respect to the $S-T_0$ splitting. As a result, the singlet state will rotate between the $S$ and $T_0$ states under the Zeeman energy difference between the two dots $\Delta E_z$. \textbf{c}, Measured triplet probabilities of DQD 4-8 as a function of $t_\text{wait}$ and $t_\text{ramp}$ with $\delta \text{v}b_{48}=40$ mV. When $t_\text{ramp}$ is small, the observed oscillations are between $S$ and $T_0$; however, when $t_\text{ramp}$ is increased until the pulse is adiabatic with respect to the $S-T_0$ splitting (over 100 ns), the $S-T_0$ oscillations can no longer be observed. Such a long ramp time can rotate the initial state to a superposition state between $S$ and $T_-$ states, and $z$-axis rotations of the $S-T_-$ qubit become visible~\cite{PhysRevLett.128.126803}. Therefore, we also observe a transition of $S-T_0$ oscillations and $S-T_-$ oscillations as a function of $t_\text{ramp}$ in the figure.  \textbf{d}, The rotation frequency $f_{ST_-}$ of each qubit as a function of the magnetic field strength $B$. When the external magnetic field strength is varied while keeping the gate voltages fixed, the frequency of these $S-T_-$ oscillations increases {nearly} linearly with the field due to the contribution from Zeeman energy in $f_{ST_-}$. From the slope, we extract $\overline{g}_{ij}$ for the four qubits as shown in the inset. The data are acquired using the fast Fourier transform (FFT) of time-domain oscillations in \textbf{e-h}. \textbf{e-h}, Measured singlet probabilities $P_S$ of Q1 (\textbf{i}), Q2 (\textbf{j}), Q3 (\textbf{k}), and Q4 (\textbf{l}) as a function of $t_\text{wait}$ and magnetic field strength $B$. The rotations are induced using the pulse scheme of panel \textbf{a} with $t_\text{ramp}=100$ ns.}}
\end{figure*}

\begin{figure*}[htb]
\renewcommand{\figurename}{\textbf{Extended Data Fig}}
\includegraphics{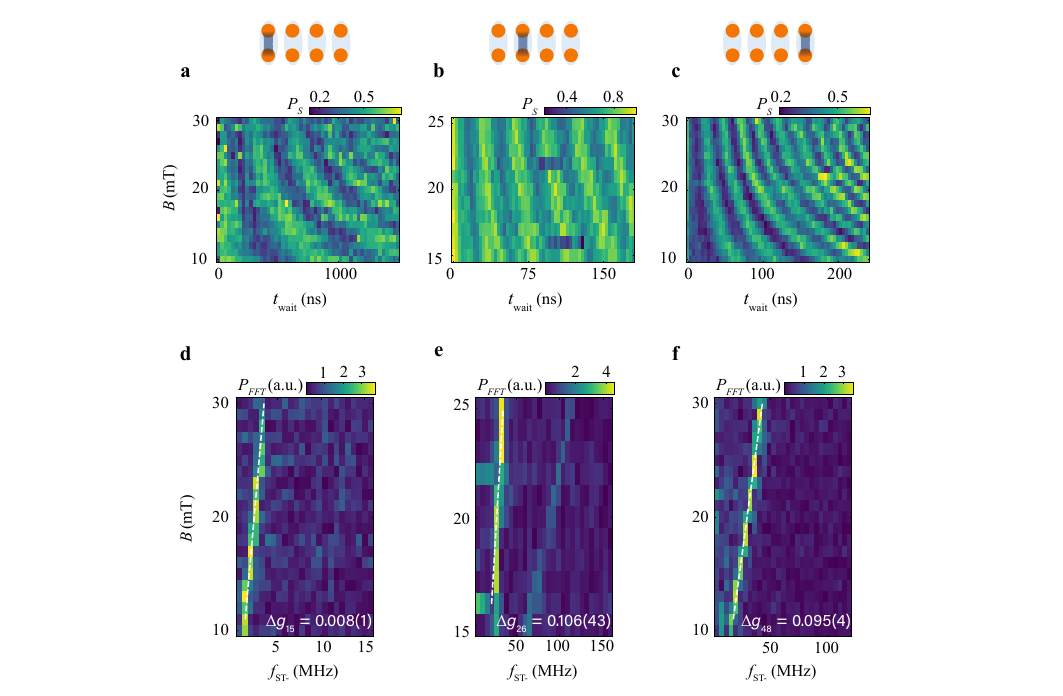}
%\internallinenumbers
\caption{\label{Additional single-qubit results4} {\textbf{Coherent control of $S-T_0$ states and $g$-factor differences.} \textbf{a-c}, Measured singlet probabilities $P_S$ as a function of $t_\text{wait}$ and the magnetic field strength $B$ for Q1 (\textbf{c}), Q2 (\textbf{d}), and Q4 (\textbf{e}) during $S-T_0$ oscillations.  Here $t_\text{ramp}$ is set to zero and the barrier gate voltage is set such that $J(\varepsilon_{ij} = 0) < \overline{E_z}$ to suppress unwanted $S-T_-$ oscillations. \textbf{d-f}, The fast Fourier transforms of the data in \textbf{a-c}, with a signal that can be line-fitted using the $g$-factor difference $\Delta g_{ij}$ (inset) of two dots.  For Q3, we didn't find $S-T_0$ oscillations, which may be because the corresponding $\Delta g$ is too low to detect.}}
\end{figure*}

\begin{figure*}[htb]
\renewcommand{\figurename}{\textbf{Extended Data Fig}}
\begin{tabular}{ |p{1.2cm}|p{4cm}|  }
 \hline
  Clifford & Gate decomposition \\
 \hline
 1   & $I$  \\
 2   & $\sqrt{X}$ \\
 3   & $\sqrt{Y}$  \\
 4   & $\sqrt{X}$$\sqrt{X}$  \\
 5   & $\sqrt{Y}$$\sqrt{Y}$  \\
 6   & $\sqrt{X}$$\sqrt{Y}$ \\
 7   & $\sqrt{Y}$$\sqrt{X}$ \\
 8   & $\sqrt{X}$$\sqrt{X}$$\sqrt{X}$  \\
 9   & $\sqrt{Y}$$\sqrt{Y}$$\sqrt{Y}$   \\
 10   & $\sqrt{X}$$\sqrt{X}$$\sqrt{Y}$   \\
 11   & $\sqrt{Y}$$\sqrt{Y}$$\sqrt{X}$  \\
 12   & $\sqrt{X}$$\sqrt{Y}$$\sqrt{X}$  \\
 13   & $\sqrt{X}$$\sqrt{Y}$$\sqrt{Y}$$\sqrt{Y}$  \\
 14   & $\sqrt{X}$$\sqrt{X}$$\sqrt{X}$$\sqrt{Y}$  \\
 15   & $\sqrt{Y}$$\sqrt{X}$$\sqrt{X}$$\sqrt{X}$  \\
 16   & $\sqrt{Y}$$\sqrt{Y}$$\sqrt{Y}$$\sqrt{X}$  \\
 17   & $\sqrt{Y}$$\sqrt{Y}$$\sqrt{X}$$\sqrt{X}$  \\
 18   & $\sqrt{X}$$\sqrt{X}$$\sqrt{Y}$$\sqrt{Y}$$\sqrt{Y}$  \\ 
 19   & $\sqrt{Y}$$\sqrt{Y}$$\sqrt{X}$$\sqrt{X}$$\sqrt{X}$  \\  
 20   & $\sqrt{X}$$\sqrt{X}$$\sqrt{X}$$\sqrt{Y}$$\sqrt{X}$  \\  
 21   & $\sqrt{X}$$\sqrt{X}$$\sqrt{X}$$\sqrt{Y}$$\sqrt{Y}$$\sqrt{Y}$  \\
 22   & $\sqrt{Y}$$\sqrt{Y}$$\sqrt{Y}$$\sqrt{X}$$\sqrt{X}$$\sqrt{X}$  \\
 23   & $\sqrt{X}$$\sqrt{X}$$\sqrt{X}$$\sqrt{Y}$$\sqrt{Y}$$\sqrt{Y}$$\sqrt{X}$  \\
 24   & $\sqrt{X}$$\sqrt{X}$$\sqrt{X}$$\sqrt{Y}$$\sqrt{X}$$\sqrt{X}$$\sqrt{X}$  \\
 
 \hline
\end{tabular}
%\internallinenumbers
\caption{\label{RB_cir} \textbf{{Single-qubit Clifford gate decomposition.}} }
\end{figure*}

\begin{figure*}[htb]
\renewcommand{\figurename}{\textbf{Extended Data Fig}}
\includegraphics{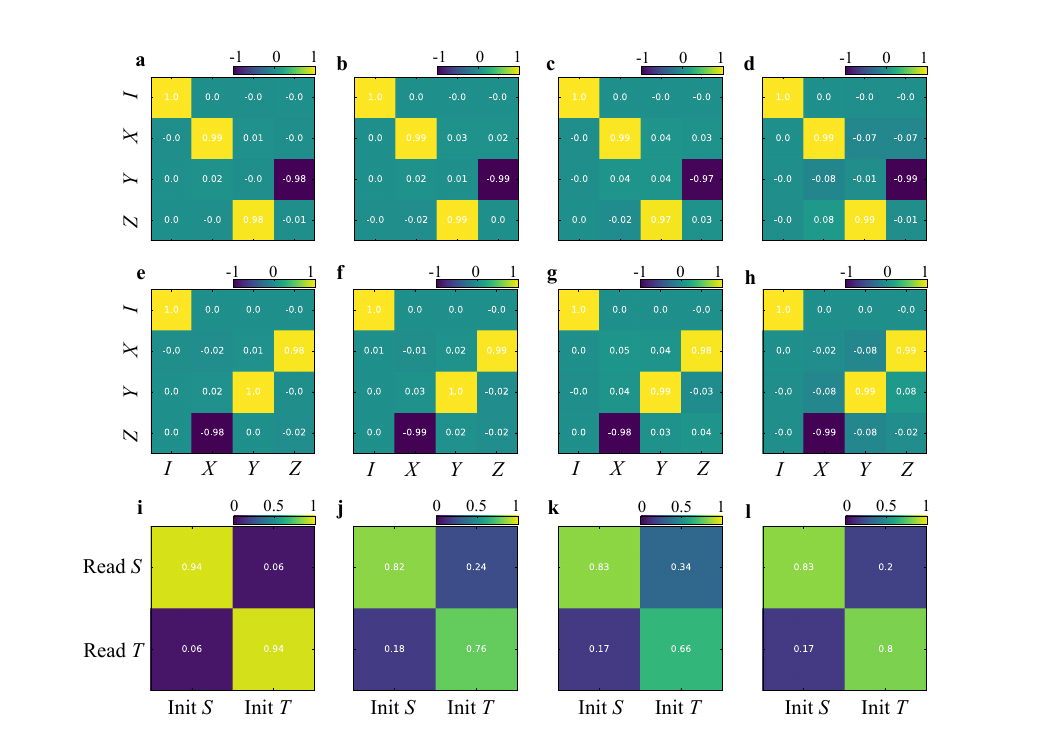}
%\internallinenumbers
\caption{\label{Single qubit GST restuls} {\textbf{Results of single-qubit gate set tomography.} \textbf{a-h} Single-qubit Pauli transfer matrices (PTM) of the $\sqrt{X}$ gate (\textbf{a-d}) and $\sqrt{Y}$ gate (\textbf{e-h}) for Q1-Q4 (from left to right) obtained from gate set tomography. \textbf{i-l}, Estimated state preparation and measurement (SPAM) error probabilities from GST results for Q1-Q4, using the same method as used in ~\cite{mkadzik2022precision}. We find that the SPAM errors of Q2 and Q3 are worse than those of Q1 and Q4. There are two reasons. Firstly, the indirect PSB mechanism is more sensitive to the readout point we chose and the idling point of the other qubit. In particular, the readout fidelity of the triplet is lower when triplet relaxation is faster at the readout point. Secondly, the slower tunneling rate from (0,2) or (2,0) to (1,1) causes an initialization error in some cases. Also, the instability of the charge sensor can contribute to the readout error, which makes the readout visibility vary between different measurements. } }
\end{figure*}

\begin{figure*}[htb]
\renewcommand{\figurename}{\textbf{Extended Data Fig}}
\includegraphics{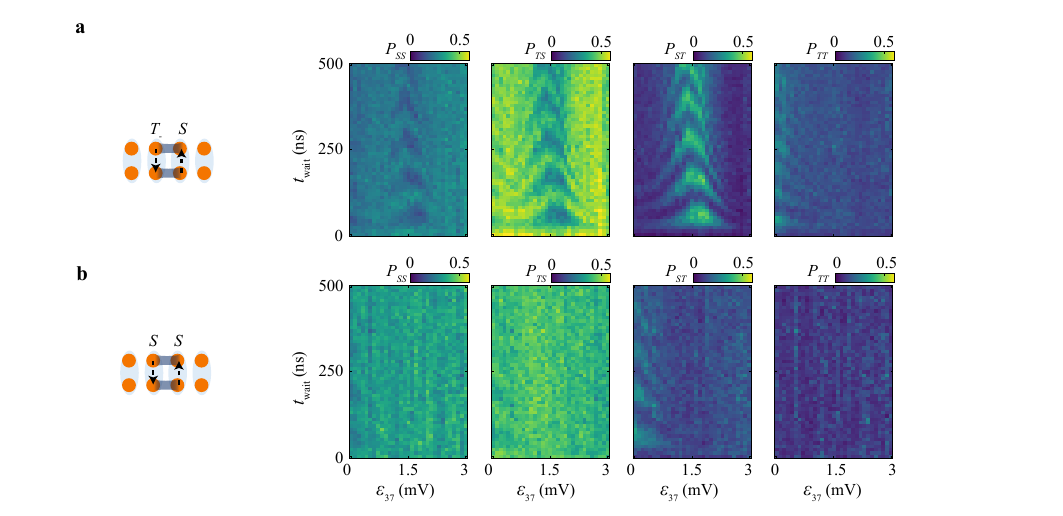}
%\internallinenumbers
\caption{\label{Additional two-qubit results} {\textbf{Sequential readout and joint probabilities of two qubits under SWAP oscillations.} \textbf{a},\textbf{b}, Sequentially measured probabilities $P_{SS}$, $P_{TS}$, $P_{ST}$ and $P_{TT}$ of Q2 and Q3 as a function of $t_\text{wait}$ and the detuning of Q3, $\varepsilon_{37}$, after initializing Q2-Q3 into $\ket{T_-S}$ (\textbf{a}) and $\ket{SS}$ (\textbf{b}). The data is acquired at $B=5$ mT. In \textbf{a}, the out-of-phase signals in $P_{TS}$ and $P_{ST}$ observed around $\varepsilon_{37}=1.5$ mV are the result of the SWAP oscillations between these two qubits. A similar signal to $P_{TS}$ but with lower visibility is observed in $P_{SS}$, which can be explained by the higher triplet readout error for Q2 than for Q3. The sequential readout is achieved by pulsing the barrier gate $\delta\text{v}b_\text{26}$ to -60 mV, where we measure Q3 first for a duration of 20 $\mu$s and simultaneously keep Q2 in the center of (1,1) with sufficiently large $J$, where the Hamiltonian eigenbasis corresponds to the qubit readout basis~\cite{shulman2012demonstration, van2021quantum, wang2023probing}. In the next step of the sequence, Q2 is measured. In panel \textbf{b}, we do not observe any apparent leakage to $\ket{TT}$ but only see signs of single-qubit rotations at low detunings. This is expected given that, in this regime, the states $\ket{TT}$ are far away in energy from the other states (see Fig.~\ref{two qubit gate}b in the main text).}}
\end{figure*}

\begin{figure*}[htb]
\renewcommand{\figurename}{\textbf{Extended Data Fig}}
\includegraphics{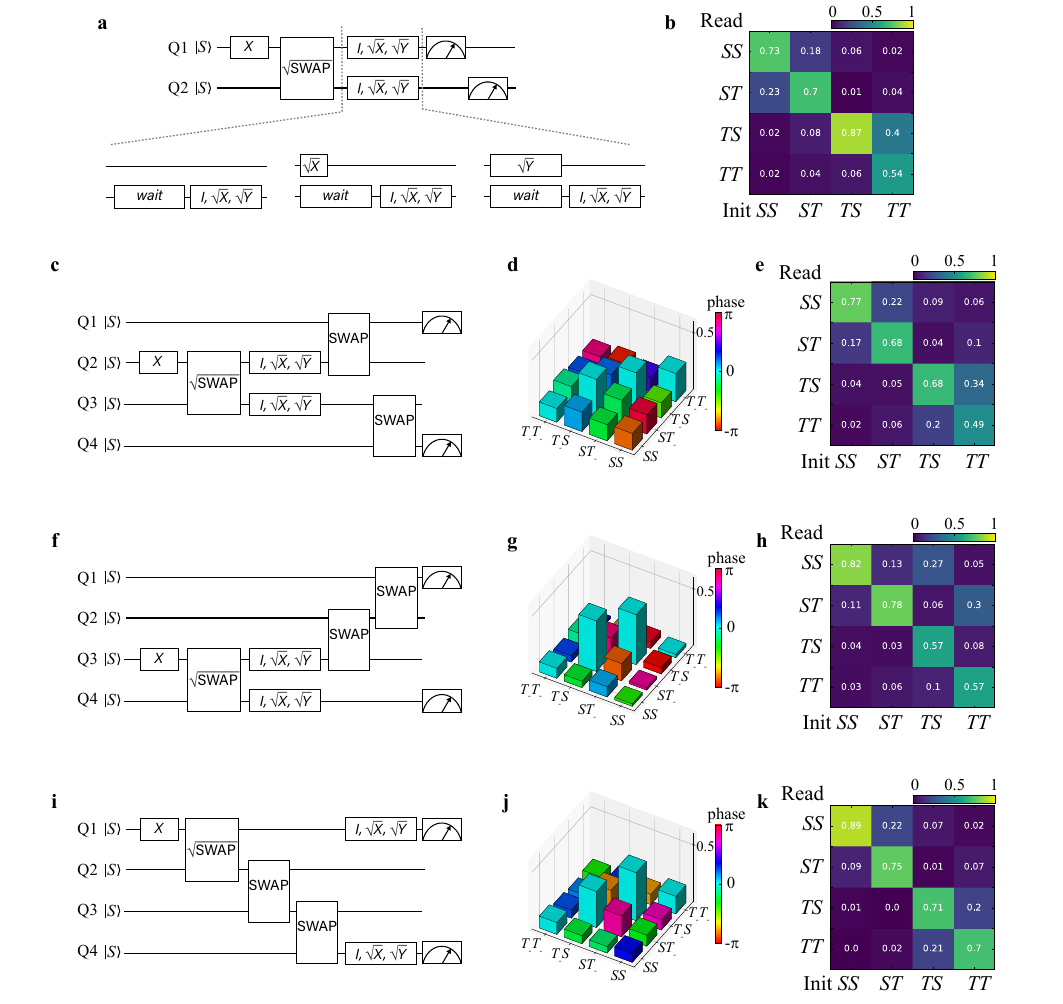}
%\internallinenumbers
\caption{\label{Bell state} {\textbf{Quantum gate circuit and results for quantum state tomography of the Bell states.} \textbf{a},\textbf{c},\textbf{f},\textbf{i}, Quantum circuit used to prepare and characterize a Bell state for different qubit pairs. In \textbf{a}, we plot the details of the single-qubit basis changes after the generation of the Bell state, where we apply a fixed wait time before performing gates on Q2 to keep its phase consistent through all the experiments. In \textbf{c}, SWAP gates are used to transfer the state of Q2 to Q1 and that of Q3 to Q4. Next Q1 and Q4 are measured simultaneously using two sensors. In \textbf{f}, two consecutive SWAP gates are used to transfer Q3 to Q1. Next Q1 and Q4 are measured simultaneously using two sensors as well. In \textbf{i}, the quantum information is transferred from Q2 to Q4 before performing the single-qubit gates for basis changes. This allows us to quantify the entanglement between Q1 and Q4 after state transfer. \textbf{d},\textbf{g},\textbf{j}, Two-qubit density matrices obtained from the corresponding quantum circuit after removal of measurement errors and using MLE for Q2-Q3 (\textbf{d}), Q3-Q4 (\textbf{g}) and Q1-Q4 (\textbf{j}) (Fig.~\ref{two qubit gate}h shows the density matrix for Q1-Q2). Measurement errors were removed based on the SPAM matrices. These matrices include not only measurement errors but also initialization errors, hence we are overcorrecting. The fact that initialization errors for most qubits were much smaller than measurement errors combined with the fact that MLE forces the resulting density matrix to be physical, helps ensure a reliable outcome. If we don't attempt to remove readout errors, the density matrices show state fidelities and concurrences of 71.3(6)\% and 9(2)\% for Q1-Q2, 64.2(6)\% and 10(2)\% for Q2-Q3,  64.6(7)\% and 0(0)\% for Q3-Q4 and 64.4(9)\% and 0(0)\% for Q1-Q4.    \textbf{b},\textbf{e},\textbf{h},\textbf{k}, SPAM matrices used in the quantum state tomography analysis of Q1-Q2 (\textbf{b}), Q2-Q3 (\textbf{e}), Q3-Q4 (\textbf{h}) and Q1-Q4 (\textbf{k}). The SPAM matrices of Q1-Q2 and Q1-Q4 were measured directly by initializing them to the indicated states and measuring the corresponding pair in a single-shot manner. For Q2-Q3 or Q3-Q4, we initialized the qubits to the indicated states and measured the state of Q1-Q4 after the SWAP operations. These SPAM matrices do include errors from the SWAP operations.}}
\end{figure*}

\begin{figure*}[htb]
\renewcommand{\figurename}{\textbf{Extended Data Fig}}
\includegraphics{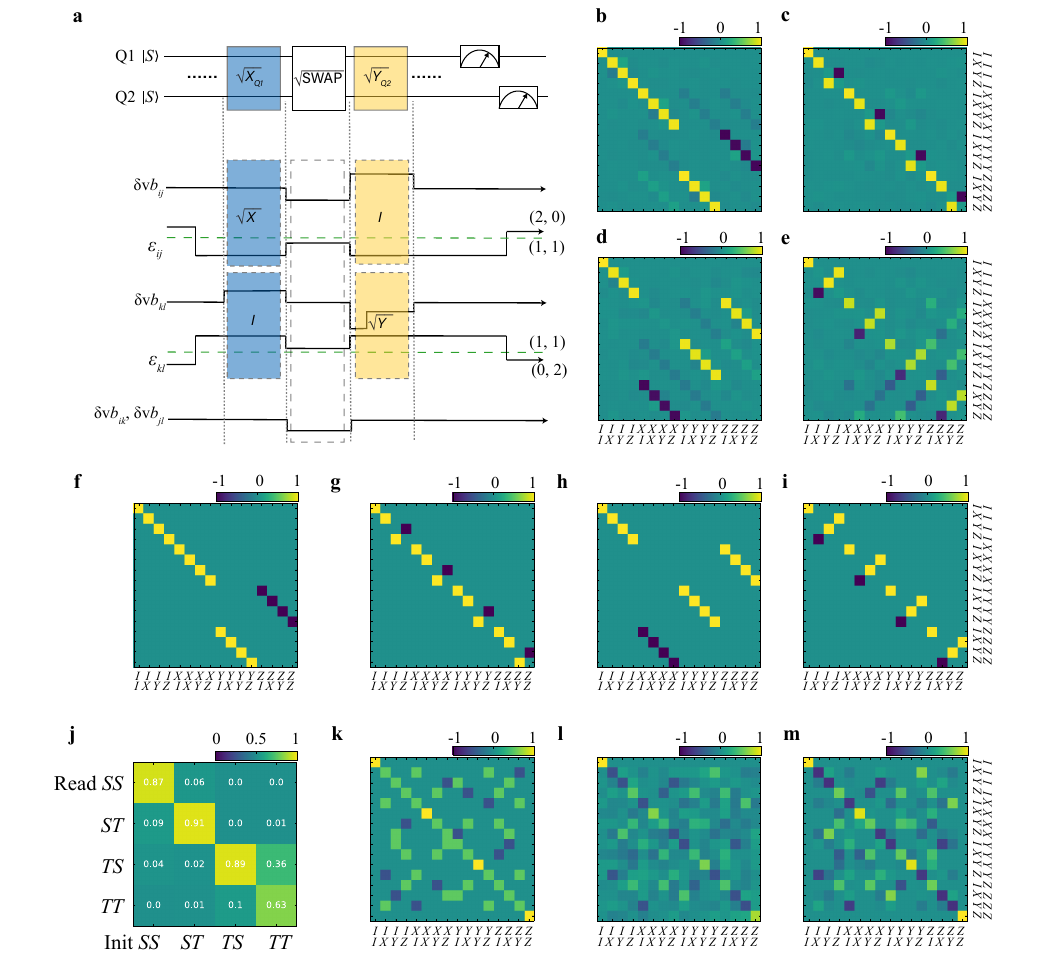}
%\internallinenumbers
\caption{\label{Two qubit GST restuls} {\textbf{Measurement sequence and results of two-qubit gate set tomography.} \textbf{a}, Illustration of the gate voltage pulses for a two-qubit circuit. Performing a single-qubit gate in the two-qubit space is nontrivial since during the time one qubit is undergoing an operation, the idling qubit could suffer from unwanted rotations and crosstalk. To solve this problem, we pulse the idle qubit to an operating point where it completes a 2$\pi$ rotation during the time needed to operate on the other qubit (see Supplementary Information section VIII for more details). \textbf{b-e}, Measured PTMs obtained from GST for single-qubit gates in the two-qubit space, including $\sqrt{X}_{Q1}$ (\textbf{b}), $\sqrt{X}_{Q2}$ (\textbf{c}), $\sqrt{Y}_{Q1}$ (\textbf{d}) and $\sqrt{Y}_{Q2}$ (\textbf{e}). \textbf{f-i}, The ideal PTMs from GST for single-qubit gates in the two-qubit space, including $\sqrt{X}_{Q1}$ (\textbf{f}), $\sqrt{X}_{Q2}$ (\textbf{g}), $\sqrt{Y}_{Q1}$ (\textbf{h}) and $\sqrt{Y}_{Q2}$ (\textbf{u}). \textbf{j}, SPAM error matrix of the measured two qubits estimated from GST, using the same method as used in ~\cite{mkadzik2022precision}. \textbf{k}, The PTM of the standard $\sqrt{\text{SWAP}}$ based on an isotropic Heisenberg exchange Hamiltonian. \textbf{l}, The experimentally measured PTM, $M_\text{exp}$, obtained from GST. \textbf{m}, The theoretical PTM, $M_\text{the}$, of the $\sqrt{\text{SWAP}}$-style gate obtained by fitting the experimentally measured PTM, $M_\text{exp}$, with a PTM generated by Eq. (28) in Supplementary Information section VII (the fitted parameters are given there). The Hamiltonian Eq. (28) includes effects of spin-orbit coupling that are left out in the  Hamiltonian of Eq.~\ref{SWAP} of the main text. The fidelity of the $\sqrt{\text{SWAP}}$-style gate is obtained from $F=\frac{1}{d+1}(\text{Tr}[M_\text{the}^{-1}M_\text{exp}]/d+1)$, where $d=2^N$ is the dimension of the Hilbert space, and $N$ refers to the number of qubits.}}
\end{figure*}

% \onecolumn
\end{document}

%\linenumbers\relax % Commence numbering lines
\title{Supplementary Information: Universal control of four singlet-triplet qubits}

\author{Xin Zhang}
\email{These authors contributed equally to this work}
\affiliation{QuTech, Delft University of Technology, Delft, The Netherlands.}%
\affiliation{Kavli Institute of Nanoscience, Delft University of Technology, Delft, The Netherlands.}%

\author{Elizaveta Morozova}
\email{These authors contributed equally to this work}
\affiliation{QuTech, Delft University of Technology, Delft, The Netherlands.}%
\affiliation{Kavli Institute of Nanoscience, Delft University of Technology, Delft, The Netherlands.}%

\author{Maximilian Rimbach-Russ}
\affiliation{QuTech, Delft University of Technology, Delft, The Netherlands.}%
\affiliation{Kavli Institute of Nanoscience, Delft University of Technology, Delft, The Netherlands.}%

\author{Daniel Jirovec}
\affiliation{QuTech, Delft University of Technology, Delft, The Netherlands.}%
\affiliation{Kavli Institute of Nanoscience, Delft University of Technology, Delft, The Netherlands.}%

\author{Tzu-Kan Hsiao}
\affiliation{QuTech, Delft University of Technology, Delft, The Netherlands.}%
\affiliation{Kavli Institute of Nanoscience, Delft University of Technology, Delft, The Netherlands.}%

\author{Pablo Cova Fari$\tilde{\text{n}}$a}
\affiliation{QuTech, Delft University of Technology, Delft, The Netherlands.}%
\affiliation{Kavli Institute of Nanoscience, Delft University of Technology, Delft, The Netherlands.}%

\author{Chien-An Wang}
\affiliation{QuTech, Delft University of Technology, Delft, The Netherlands.}%
\affiliation{Kavli Institute of Nanoscience, Delft University of Technology, Delft, The Netherlands.}%

\author{Stefan D. Oosterhout}
\affiliation{QuTech, Delft University of Technology, Delft, The Netherlands.}%
\affiliation{Netherlands Organisation for Applied Scientific Research (TNO), Delft, The Netherlands.}%

\author{Amir Sammak}
\affiliation{QuTech, Delft University of Technology, Delft, The Netherlands.}%
\affiliation{Netherlands Organisation for Applied Scientific Research (TNO), Delft, The Netherlands.}%

\author{Giordano Scappucci}
\affiliation{QuTech, Delft University of Technology, Delft, The Netherlands.}%
\affiliation{Kavli Institute of Nanoscience, Delft University of Technology, Delft, The Netherlands.}%

\author{Menno Veldhorst}
\affiliation{QuTech, Delft University of Technology, Delft, The Netherlands.}%
\affiliation{Kavli Institute of Nanoscience, Delft University of Technology, Delft, The Netherlands.}%

\author{Lieven M. K. Vandersypen}
\email{L.M.K.Vandersypen@tudelft.nl}
\affiliation{QuTech, Delft University of Technology, Delft, The Netherlands.}%
\affiliation{Kavli Institute of Nanoscience, Delft University of Technology, Delft, The Netherlands.}%

\date{\today}

\keywords{Suggested keywords}
\maketitle
This supplementary information includes:
\begin{itemize}
\item Supplementary Note ~\ref{sec:level1} Experimental setup
%\item Supplementary Note ~\ref{sec:level2} Charge stability diagrams and spin readout
\item Supplementary Note ~\ref{sec:level3} Virtual gate matrix
\item Supplementary Note ~\ref{sec:level4} Asymmetry in measured qubit energy spectrum
\item Supplementary Note ~\ref{sec:level5} Additional data of $S-T_-$ oscillations
\item Supplementary Note ~\ref{sec:level6} Additional data of SWAP operations
\item Supplementary Note ~\ref{sec:level7} Pulse scheme and calibration for quantum state transfer
\item Supplementary Note ~\ref{sec:level8} A theoretical model for the SWAP operation
\item Supplementary Note ~\ref{sec:level12} Gate set tomography of the single- and two-qubit gate
\end{itemize}

\newpage
\section{\label{sec:level1} Experimental setup}
\begin{figure*}[htb]
\renewcommand{\figurename}{\textbf{Supplementary Fig}}
\includegraphics{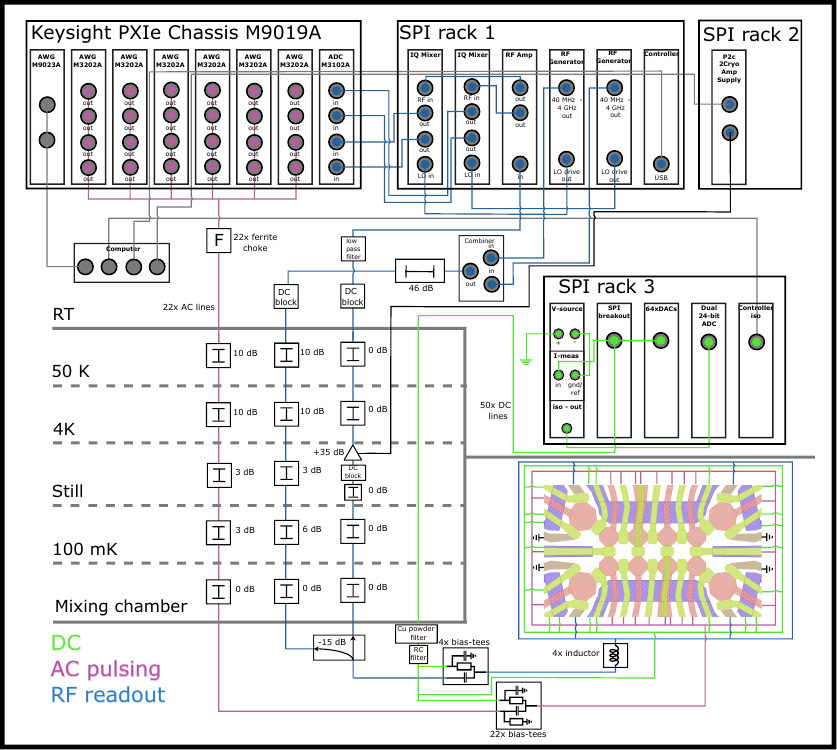}
%\internallinenumbers
\caption{\label{circuit} Measurement circuit for the device. The DC and AC control lines as well as radio-frequency (RF) readout lines are fed from room-temperature instruments to the device at a base temperature of the dilution refrigerator through the cables and various electronic components shown in the figure. The room-temperature instruments include three custom-built SPI racks for supplying DC voltages and RF readout signals, and one Keysight PXIe Chassis for AC pulsing (arbitrary waveform generator, AWG) and data acquisition (analog-to-digital converter, ADC). {The on-board bias-tees used for combining RF readout signals with DC voltages have an R=5 k$\Omega$ resistor, a C=100 pF capacitor to ground  and a C=100 pF capacitor at the AC input; the one used for combining voltage pulses with DC signals has an R=1 M$\Omega$ resistor and a C=100 pF capacitor to connect the DC signal, and an R=100 k$\Omega$ resistor and a C=100 nF capacitor to connect the AC signal.} }
\end{figure*}

\newpage
\section{\label{sec:level3} Virtual gate matrix}
As mentioned in the main text, we use virtualized gates to independently control the chemical potential in each quantum dot. An example of the virtual gate matrix we used is as follows:

\begin{align*}
\begin{pmatrix}
\text{vP}_{1}\\  \text{vP}_{2}\\  \text{vP}_{3}\\ \text{vP}_{4}\\  
\text{vP}_{5}\\  \text{vP}_{6}\\  \text{vP}_{7}\\ \text{vP}_{8}\\ 
\text{vb}_{12}\\  \text{vb}_{23}\\  \text{vb}_{34}\\ \text{vb}_{56}\\ \text{vb}_{67}\\   \text{vb}_{78}\\
   \text{vb}_{15}\\  \text{vb}_{26}\\   \text{vb}_{37}\\  \text{vb}_{48}\\  
\end{pmatrix}=
\left(\begin{array}{cccccccccccccccccc}
   1     & 0.3   & 0.1  & 0.05& 0.35  & 0.2 & 0.05 & 0.02  & 0.4  & 0.08  & 0    & 0.05 & 0.05 & 0    & 0.4  & 0.15 & 0    & 0  \\
   0.15  & 1     & 0.3  & 0.05& 0.05  & 0.25& 0.05 & 0     & 0.3  & 0.2   & 0    & 0.05 & 0.1  & 0    & 0.1  & 0.2  & 0    & 0 \\
   0.05  & 0.25  & 1    & 0.25& 0     & 0.1 & 0.35 & 0.05  & 0    & 0.15  & 0.06 & 0    & 0.15 & 0.12 & 0    & 0    & 0.08 & 0.07 \\
   0.02  & 0.05  & 0.35 & 1   & 0.02  & 0   & 0.15 & 0.35  & 0    & 0.05  & 0.06 & 0    & 0    & 0.11 & 0    & 0    & 0.03 & 0.28 \\
   0.15  & 0.15  & 0.03 & 0.02& 1     & 0.25& 0.05 & 0.03  & 0.1  & 0.05  & 0    & 0.2  & 0.05 & 0.03 & 0.3  & 0    & 0.02 & 0  \\
   0.1   & 0.2   & 0.1  & 0.05& 0.2   & 1   & 0.2  & 0.05  & 0.1  & 0.1   & 0.05 & 0.1  & 0.25 & 0    & 0.1  & 0.1  & 0    & 0 \\
   0.05  & 0.15  & 0.25 & 0.1 & 0.08  & 0.22& 1    & 0.15  & 0    & 0.1   & 0.02 & 0.05 & 0.25 & 0.28 & 0.05 & 0    & 0.05 & 0.05 \\
   0     & 0.03  & 0.2  & 0.25& 0.03  & 0.05& 0.45 & 1     & 0    & 0.02  & 0.025& 0.02 & 0.05 & 0.4  & 0    & 0    & 0.03 & 0.22 \\
   0     & 0     & 0    & 0   & 0     & 0   & 0    & 0     & 1    & 0     & 0    & 0    & 0    & 0    & 0    & 0    & 0    & 0  \\
   0     & 0     & 0    & 0   & 0     & 0   & 0    & 0     & 0    & 1     & 0    & 0    & 0    & 0    & 0    & 0    & 0    & 0 \\
   0     & 0     & 0    & 0   & 0     & 0   & 0    & 0     & 0    & 0     & 1    & 0    & 0    & 0    & 0    & 0    & 0    & 0 \\
   0     & 0     & 0    & 0   & 0     & 0   & 0    & 0     & 0    & 0     & 0    & 1    & 0    & 0    & 0    & 0    & 0    & 0 \\
   0     & 0     & 0    & 0   & 0     & 0   & 0    & 0     & 0    & 0     & 0    & 0    & 1    & 0    & 0    & 0    & 0    & 0  \\
   0     & 0     & 0    & 0   & 0     & 0   & 0    & 0     & 0    & 0     & 0    & 0    & 0    & 1    & 0    & 0    & 0    & 0 \\
   0     & 0     & 0    & 0   & 0     & 0   & 0    & 0     & 0    & 0     & 0    & 0    & 0    & 0    & 1    & 0    & 0    & 0 \\
   0     & 0     & 0    & 0   & 0     & 0   & 0    & 0     & 0    & 0     & 0    & 0    & 0    & 0    & 0    & 1    & 0    & 0 \\
   0     & 0     & 0    & 0   & 0     & 0   & 0    & 0     & 0    & 0     & 0    & 0    & 0    & 0    & 0    & 0    & 1    & 0 \\
   0     & 0     & 0    & 0   & 0     & 0   & 0    & 0     & 0    & 0     & 0    & 0    & 0    & 0    & 0    & 0    & 0    & 1 \\
\end{array}\right)
\begin{pmatrix}
    \text{P}_{1}\\  \text{P}_{2}\\  \text{P}_{3}\\ \text{P}_{4}\\  
    \text{P}_{5}\\  \text{P}_{6}\\  \text{P}_{7}\\ \text{P}_{8}\\ 
   \text{b}_{12}\\  \text{b}_{23}\\  \text{b}_{34}\\ \text{b}_{56}\\ \text{b}_{67}\\   \text{b}_{78}\\
   \text{b}_{15}\\  \text{b}_{26}\\   \text{b}_{37}\\  \text{b}_{48}\\  
\end{pmatrix}
\end{align*}

\begin{comment}
\begin{align*}
\begin{pmatrix}
\text{vb}_{15}\\  \text{vb}_{26}\\  \text{vb}_{37}\\ \text{vb}_{48}\\  
\end{pmatrix}=
\begin{pmatrix}
   1     & 0.035 & 0    & 0   & -0.13 & 0   & 0    & 0.03  & 0    & 0 \\
   0.3   & 1     & 0.05 & 0   & -2    & 0.65& 0    & -0.3  & 1.4  & 0 \\
   0     & 0.15  & 1    &0.45 & 0     & 0.21& 0    & 0     & 0.4  & -0.2\\
   0     & 0     &0.04  &1    & 0     & 0   & 0    & 0     & 0    & 0.25 \\
\end{pmatrix}
\begin{pmatrix}
   \text{b}_{15}\\  \text{b}_{26}\\   \text{b}_{37}\\  \text{b}_{48}\\  \text{b}_{12}\\  \text{b}_{23}\\  \text{b}_{34}\\ \text{b}_{56}\\ \text{b}_{67}\\   \text{b}_{78}
\end{pmatrix}
\end{align*}
\end{comment}

\newpage
\section{\label{sec:level4} Asymmetry in measured qubit energy spectrum}
\begin{figure*}[htb]
\renewcommand{\figurename}{\textbf{Supplementary Fig}}
\includegraphics{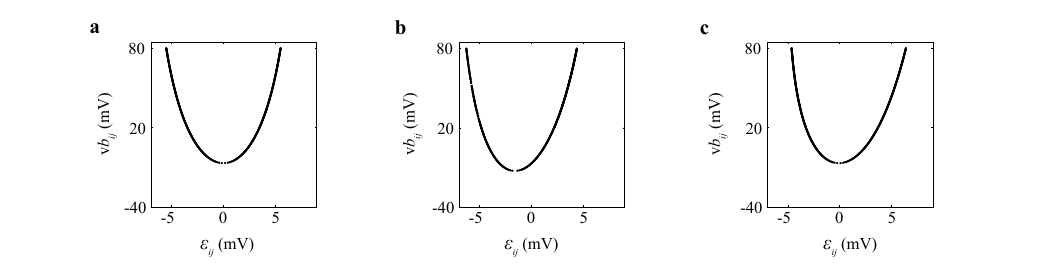}
%\internallinenumbers
\caption{\label{energy spectroscopy} \textbf{a}, Simulated parabola-like curve of the energy spectroscopy as a function of detuning $\varepsilon_{ij}$ and barrier voltage $\text{v}b_{ij}$ with a standard single-qubit Hamiltonian. \textbf{b}, Simulated asymmetrical curve by considering a detuning dependent $g$-factor. \textbf{c}, Simulated asymmetrical curve by considering a linear change of detuning offset as a function of $\text{v}b_{ij}$.}
\end{figure*}
In Fig. 1 of the main text, we show energy spectrums of the $S-T_-$ qubits and mention that the asymmetry of the parabola-like curve may be caused by the detuning-dependent $g$-factor or an imperfect virtualization of the barrier gate. Here we do a numerical simulation to determine their effects. The results are shown in Supplementary Fig.~\ref{energy spectroscopy}a-c, which compares the standard energy spectrum (a) with the one with a detuning dependent $g$-factor (b) and that with imperfect virtualization (c).  To numerically simulate these curves, we use the single-qubit Hamiltonian ~\eqref{ST-} and a relationship of $J = 2t^2U/(U^2-\varepsilon^2)$. For the parameters, we use $U$ = 295 GHz, $\overline{g}=$ 0.33, $B$ = 10 mT and $\Delta_{ST_-}$ = 16 MHz. For the effect of $g$-factor change, we use a linear change of $\overline{g}=$ 0.19-0.47 as a function of detuning, while for the effect of imperfect virtualization, we use a linear variation of detuning offset $\varepsilon_0$ as a function of the barrier gate voltage $\text{v}b_{ij}$, which changes in the range of -0.88 - 0.44 mV when $ \text{v}b_{ij}$ is changed from 80 to -40 mV. The variation of $g$-factor is a bit far away from the values reported in a similar device, where they report a change of 0.21-0.29 as a function of detuning~\cite{hendrickx2020single}. But the $1\%$ variation of $\varepsilon_0$ is plausible considering the imperfect virtualization. Still, the observed asymmetrical curve in the main text may result from their combined effects.

%\begin{figure*}[htb]
%\includegraphics{Fig_g_factor_2 copy.pdf}
%\caption{\label{g_factor_2} }
%\end{figure*}
%\begin{figure*}[htb]
%\includegraphics{Fig_qubit_control_FFT copy.pdf}
%\caption{\label{z control} z control.}
%\end{figure*}

\newpage
\section{\label{sec:level5}  Additional data of \texorpdfstring{$S-T_-$ } ooscillations}

\begin{figure*}[htb]
\renewcommand{\figurename}{\textbf{Supplementary Fig}}
\includegraphics{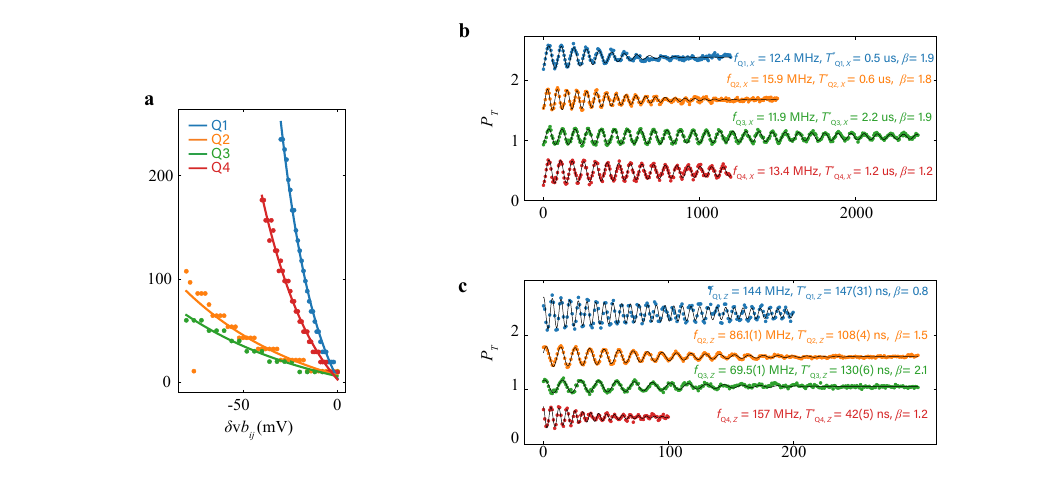}
%\internallinenumbers
\caption{\label{g_factor_2} {\textbf{a}, Rotation frequency $f_{ST_-}$ of each qubit as a function of $\delta \text{v}b_{ij}$, extracted from Extended Data Fig.2\textbf{g-j} using a FFT. The data points for Q2 and Q3 show discrete steps as the barrier gate voltage changes, which are caused by the FFT precision. The solid lines are fits of the data with an exponential function. \textbf{b,c}, Measured triplet probabilities $P_T$ of long-time evolutions around the $x$- (\textbf{b}) and $z$-axis (\textbf{c}) for Q1-Q4 at $B$=10 mT. The data is vertically shifted for clarity. In \textbf{b}, the detuning is around 0 mV. In \textbf{c}, we take the data with the same detuning but with different barrier voltage pulses: $\delta \text{v}b_{15}=-20$ mV, $\delta \text{v}b_{26}=-80$ mV, $\delta \text{v}b_{37}=-80$ mV, and $\delta \text{v}b_{48}=-30$ mV. The barrier voltages are chosen to achieve the condition $J-\overline{E}_z\gg \Delta_{ST_-}$ depending on the gate tunability.}}
\end{figure*}

The additional data here is collected at $B$= 10 mT, under similar conditions as the data shown in Extended Data Fig.2. 

Supplementary Fig.~\ref{g_factor_2}a summarizes how $f_{ST_-}$ can be tuned via $\delta\text{v}b_{ij}$. As discussed in the main text, the outer two barrier gates $\text{v}b_{15}$ and $\text{v}b_{48}$ have a stronger effect on the corresponding $J_{ij}$ than the inner barrier gates $\text{v}b_{26}$ and $\text{v}b_{37}$. This may {possibly be explained by resist residues} below the inner barrier gates, which are fabricated in the last step~\cite{hsiao2023exciton}, and by the different fan-out routing for the outer barrier gates (see Fig.1a,b in the main text). 

The dephasing time under free evolution, which is traditionally termed $T_2^*$, is an important metric for assessing the qubit quality. Since the qubit rotations around both the $x$-axis and the $z$-axis are the result of free evolution, we introduce $T_x^*$ and $T_z^*$ to describe the corresponding dephasing times. Supplementary Fig.~\ref{g_factor_2}b,c show the measured damped oscillations of the qubits under $x$-axis and $z$-axis control. From the fits, we obtain a $T_x^*$ of 0.5 - 2.1 $\mu$s and a $T_z^*$ of 42(5) - 147(31) ns. 

{The measured values of $T_x^*$ are slightly lower than previously reported values measured at $B = 1$ mT~\cite{wang2023probing} under the same condition that $B$ is parallel to the hole movement direction. This can be partly attributed to the larger magnetic field in panel b: $\Delta_{ST_-}$ scales with $B$~\cite{nichol2015quenching, PhysRevLett.128.126803} and not only sets $f_{ST_-}$ but also constitutes a proportionality factor between noise and $f_{ST_-}$ fluctuations (see also Supplementary Information section VIII). The extracted $x$-axis rotation frequencies in Supplementary Fig.~\ref{g_factor_2}b reflect the values of $\Delta_{ST_-}$ for each qubit, which are around 11.9-15.9 MHz, much larger than the results reported at $B = 1$ mT~\cite{wang2023probing}. This confirms that $\Delta_{ST_-}$ is stronger in the present experiment than in the previous work at 1 mT. The small variation in $\Delta_{ST_-}$ and in the measured average $g$-factors suggests a fairly homogeneous spin-orbit coupling in this device. The extracted parameter $\beta$ also has a big variation among qubits, and considering the variations in $T_x^*$ are also large, especially for the data at $B=10$ mT, these qubits may suffer from spatially dependent charge noise or inhomogeneous hyperfine-induced dephasing due to dot size differences~\cite{hendrickx2023sweet}. }

We also observe that $T_z^*$ is {roughly} an order of magnitude smaller than $T_x^*$ in Supplementary Fig.~\ref{g_factor_2}b. Possibly this is due to the fact that fluctuations in the tunnel barrier translate to fluctuations in $J$, which couple in directly during $z$-axis evolution but only to second order for $x$-axis evolution~\cite{PhysRevLett.96.100501, huang2018spin}. Additionally, the increased curvature of the singlet branch for larger $J$ implies a higher sensitivity to detuning noise. The variations in $J$ thus may contribute to the spread of the $T_z^*$ values we obtained in Supplementary Fig.~\ref{g_factor_2}c.

\newpage
\section{\label{sec:level6} Additional data of SWAP operations}
\begin{figure*}[htb]
\renewcommand{\figurename}{\textbf{Supplementary Fig}}
\includegraphics{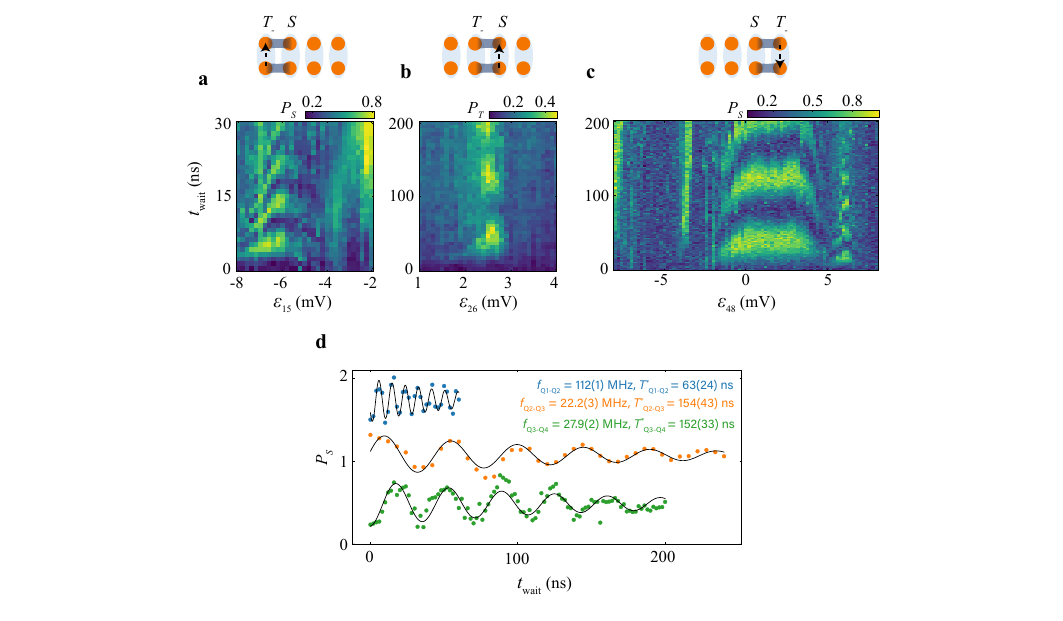}
%\internallinenumbers
\caption{\label{SWAP_full} {\textbf{a-c}, The experimental results of SWAP oscillations measured at $B=10$ mT, showing triplet probabilities $P_T$ or singlet probabilities $P_S$ as a function of operation time $t_\text{wait}$ and the detuning voltage for Q1-Q2 (\textbf{a}), Q2-Q3 (\textbf{b}) and Q3-Q4 (\textbf{c}). The initial states of two qubits (before the SWAP oscillations) are denoted on the top, and the qubit pair that is read out is indicated by the dashed arrow showing the readout pulse direction. \textbf{d}, Measured singlet probabilities $P_S$ at $B=10$ mT as a function of the evolution time $t_\text{wait}$ at the center of the chevron patterns of the SWAP oscillations for each pair of qubits. Notice the conditions are slightly different from the two-dimensional scanned data due to device tuning.}}
\end{figure*}

The additional data here is collected at $B$= 10 mT, under similar conditions as the data shown in Fig.4b of the main text. 

{Supplementary Fig.~\ref{SWAP_full}a-c shows SWAP oscillations of Q1-Q2, Q2-Q3, and Q3-Q4.} For Q3-Q4, we scanned over a wide range of detuning and there are three sets of oscillations visible from left to right, where the rightmost corresponds to SWAP oscillations of Q3-Q4, the middle oscillations correspond to a single-qubit operation of Q4, and the leftmost is also a set of SWAP oscillations but with a very slow oscillation speed, which corresponds to a smaller interqubit coupling than the rightmost one (see the leftmost anticrossing in Fig. 3b of the main text). These features can all be qualitatively understood using Fig. 3b in the main text. However, to quantitatively model the oscillations, the contribution of anisotropic exchange couplings may be needed.

The oscillation frequencies in the middle of the chevron patterns are in the range of 22.2(3) - 112(1) MHz, corresponding to a $\sqrt{\text{SWAP}}$ (entangling) gate with durations of just 2.2 ns to 11.3 ns, more than an order of magnitude faster than the entangling gate based on capacitive coupling~\cite{shulman2012demonstration, nichol2017high}. To determine the dephasing times of the SWAP oscillations, we collect data in the middle of the chevron patterns, as shown in Supplementary Fig.~\ref{SWAP_full}d, and fit them with the same function as used for single-qubit oscillations. The extracted dephasing times are between 63(24) and 154(43) ns. {An increased quality of SWAP oscillations can be observed in the Extended Fig. 3b, of which the data was collected at $B=5$ mT in a separate cool-down of the device.} The dephasing times may be further increased by executing the SWAP oscillations at the symmetry points of the detuning of each qubit, which requires a stronger tunability of the exchange interactions using the barrier gates. 

\newpage
\section{\label{sec:level7} Pulse scheme and calibration for quantum state transfer}
\begin{figure*}[htb]
\renewcommand{\figurename}{\textbf{Supplementary Fig}}
\includegraphics{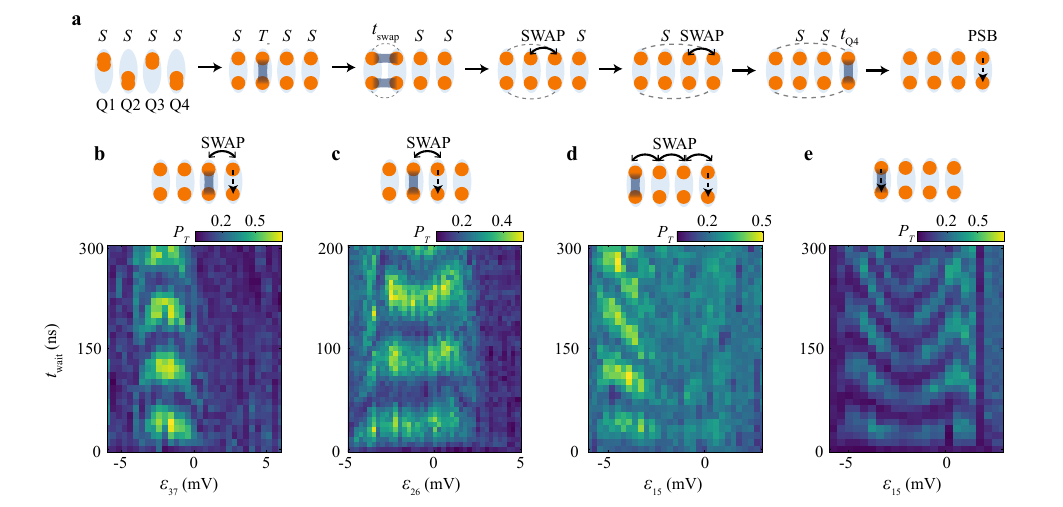}
%\internallinenumbers
\caption{\label{state_trasnfer_supple} \textbf{a}, Schematic representation of the different steps in the quantum-state-transfer experiment of Fig. 4 of the main text. The gray dashed curves indicate potential entanglement between two qubits. The black curved double-arrow refers to a SWAP gate that is intended to transfer information from one qubit to the next. \textbf{b}, Measured triplet probabilities of Q4 as a function of waiting time $t_\text{wait}$ and detuning of Q3, $\varepsilon_{37}$. The inset illustrates that we control Q3 to perform an $x$-rotation with a time $t_\text{wait}$ and subsequently perform a SWAP gate of Q3-Q4 and measure Q4 using PSB. \textbf{c}, Measured triplet probabilities of Q3 as a function of $t_\text{wait}$ and detuning of Q2, $\varepsilon_{26}$. The inset illustration is similar to that in \textbf{b} but with single-qubit control of Q2, a subsequent SWAP gate of Q2-Q3 and PSB measurement of Q3. \textbf{d}, Measured triplet probabilities of Q4 as a function of $t_\text{wait}$ and the detuning of Q1, $\varepsilon_{15}$. The inset shows that we perform an $x$-rotation of Q1 with a time $t_\text{wait}$ followed by three consecutive SWAP gates to transfer the state of Q1 to Q2, Q3 and Q4, after which we measure Q4 using PSB. \textbf{d} has lower visibility on the right-hand side of the figure compared to its left-hand side, which can be attributed to a possible parameter shift of the consecutive SWAP gates by low-frequency charge noise after the left part was scanned. \textbf{e}, Direct readout of Q1 with the same single-qubit control as in \textbf{d}. }
\end{figure*}
The detailed pulse scheme for collecting the data of quantum state transfer in Fig.4b of the main text is shown in Supplementary Fig.~\ref{state_trasnfer_supple}a. We initialize Q1-Q4 in the $\left( \begin{smallmatrix} 2 & 0 & 2 & 0 \\ 0 & 2 & 0 & 2 \end{smallmatrix} \right)$ regime and then diabatically pulse to the $\left( \begin{smallmatrix} 1 & 1 & 1 & 1 \\ 1 & 1 & 1 & 1 \end{smallmatrix} \right)$ regime, preserving four singlets. At this point, the detuning of Q1 and Q3 is set close to the (1,1)-(2,0) transition and that of Q4 close to the (1,1)-(0,2) transition in order to stay away from the $S-T_-$ anticrossing. This ensures that these three qubits remain in the singlet state. The detuning of Q2 is set to zero such that Q2 rotates around its $x$ axis to $\ket{T_-}$. Then we pulse the detunings and barrier gates of Q1-Q2 to initiate a SWAP interaction for a duration $t_\text{wait}$, after which we transfer the state of Q2 to Q4 by performing sequential SWAP operations of Q2-Q3 and Q3-Q4. Finally, we pulse Q4 to zero detuning to kickstart single-qubit evolution around the $x$-axis for a duration $t_\text{Q4}$. For readout, we pulse Q1-Q4 into $\left( \begin{smallmatrix} 1 & 1 & 1 & 0 \\ 1 & 1 & 1 & 2 \end{smallmatrix} \right)$ where we perform PSB readout of Q4. The detunings of Q1-Q3 are set to zero while Q4 is read out, but this will not affect the readout since Q4 is protected by a very large $J$ in the readout window. 

Before the quantum state transfer experiment, we confirm the operation of the individual SWAP gates by performing a single-qubit rotation of one qubit and measuring its state using the other after a SWAP gate. Some results are shown in Supplementary Fig.~\ref{state_trasnfer_supple}b and c, where we perform this test for Q3-Q4 and Q2-Q3, respectively. Supplementary Fig.~\ref{state_trasnfer_supple}d shows a result where Q1 is rotated and the measured qubit is Q4 following three consecutive SWAP gates from Q1 to Q4. A comparison of the direct readout of Q1 is shown in Supplementary Fig.~\ref{state_trasnfer_supple}e, with similar oscillations to that of Supplementary Fig.~\ref{state_trasnfer_supple}d at similar detuning voltages. These results demonstrate the SWAP gates are of sufficient quality to allow quantum state transfer across the entire array.

\section{\label{sec:level8} A theoretical model for the SWAP operation}
Holes confined in quantum dots are well described by the conventional Fermi-Hubbard model~\cite{burkard2023semiconductor}. In the presence of a finite magnetic field and spin-orbit interaction, the Fermi-Hubbard model can be written as:

\begin{align}
	H_\text{FH}=&\sum_{i=1} \left[ \mu_in_i + \mu_B \boldsymbol{B}\cdot \tilde{\mathcal{G}}_i\boldsymbol{S}_i\right] + \sum_{i=1} \frac{\tilde{U}_{i}}{2}n_i(n_i-1) + \sum_{\braket{i,j}} V_{ij} n_{i}n_{j}\nonumber\\ &+ \sum_{\sigma=\uparrow,\downarrow}\left(\tilde{t}_{ij}c_{i,\sigma}^\dagger c_{j,\sigma}+\text{h.c.}\right)
	+ \sum_{\sigma=\uparrow,\downarrow}\left(\tilde{t}_{\text{SO},ij}c_{i,\sigma}^\dagger c_{j,\bar{\sigma}}+\text{h.c.}\right).
	\label{eq:FH}
\end{align}
Here, h.c. denotes hermitian conjugate, $\bar{\sigma}$ the opposite spin state of $\sigma$, the operator $c_{i,\sigma}^{\dagger}$ ($c_{i,\sigma}$) creates (annihilates) a hole in QD $i$ with spin $\sigma=\uparrow,\downarrow$, $n_i\equiv\sum_\sigma c_{i,\sigma}^\dagger c_{i,\sigma}$ is the charge number operator, and $\boldsymbol{S}=(S_{x,i},S_{y,i},S_{z,i})^T$ is the vector consisting of the spin matrices

\begin{align}
	S_{x,i} &= \frac{\hbar}{2} (c_{i,\uparrow}^\dagger c_{i,\downarrow} + c_{i,\downarrow}^\dagger c_{i,\uparrow})\\
	S_{y,i} &= -i\frac{\hbar}{2} (c_{i,\uparrow}^\dagger c_{i,\downarrow} - c_{i,\downarrow}^\dagger c_{i,\uparrow})\\
	S_{z,i} &= \frac{\hbar}{2} (c_{i,\uparrow}^\dagger c_{i,\uparrow} - c_{i,\downarrow}^\dagger c_{i,\downarrow}).
\end{align}
The first line in Hamiltonian~\eqref{eq:FH} describes the hole energies, i.e., $\mu_{i}$ denotes the chemical potential in QD~$i$, $\tilde{U}_{i}$ is the Coulomb repulsion for doubly occupying the $i$-th QD, $V_{ij}$ is the Coulomb repulsion of two holes occupying neighboring QDs $i$ and $j$, $\mu_B$ is Bohr's magneton, and $\tilde{\mathcal{G}}_i$ is the quantum dot g-tensor. The second line describes the hopping terms between neighboring sites denoted by the spin-conserving tunnel matrix elements $\tilde{t}_{ij}$ and spin-non-conserving tunnel matrix elements $\tilde{t}_{\text{SO},ij}$. Note, that this notation is consistent with Refs.~\cite{DanonPRB80-041301,geyer2022twoqubit} as $\tilde{t}_{ij}$ and $\tilde{t}_{\text{SO},ij}$ can be complex in general.

\subsection*{Single-qubit Hamiltonian}

Considering only two neighboring quantum dots, the Hamiltonian can be simplified by transforming into the spin-orbit frame~\cite{geyer2022twoqubit}. Intuitively, since the spin-orbit interaction only rotates the spin during tunneling, local spin rotations can always ``unwind'' the rotation. As a consequence of the spin-orbit interaction, the g-tensor of quantum dot $i$ is rotated as $\tilde{\mathcal{G}}_i\rightarrow R_i \tilde{\mathcal{G}}_i\equiv \mathcal{G}_i$, and the tunnel coupling is renormalized as $t = \sqrt{|\tilde{t}|^2+|\tilde{t}_\text{SO}|^2}$~\cite{geyer2022twoqubit}. 
\paragraph*{Hamiltonian} The respective Hamiltonian in the basis $\lbrace S(2,0)$, $S(0,2)$, $S(1,1)$, $T_-(1,1)$, $T_0(1,1)$, $T_+(1,1)\rbrace$ of the spin-orbit frame reads

%\begin{align}
%    H_\text{DQD}=\begin{bmatrix}
%   U-\varepsilon_{ij}  & \sqrt{2}t & 0 & 0 & 0 & 0 \\
%   \sqrt{2}t & 0 & \Delta E_z &  t_\text{SO} &  t_\text{SO} & \sqrt{2}t \\
%   0  & \Delta E_z & 0 & 0 & 0 & 0 \\
%   0  & t^\star_\text{SO}  &  0 & \overline{E_z} & 0 & 0 \\
%   0  & t^\star_\text{SO}  &  0 & 0 & -\overline{E_z} & 0  \\
%   0  & \sqrt{2}t & 0 & 0 & 0 & U+\varepsilon_{ij}
%\end{bmatrix}
%\end{align}
\begin{align}
	H_\text{DQD}=\begin{bmatrix}
		U+\varepsilon  & 0 & \sqrt{2}t & 0 & 0 & 0 \\
		0 & U-\varepsilon  & \sqrt{2}t & 0 & 0 & 0 \\
		\sqrt{2}t & \sqrt{2}t & 0 & \frac{\Delta E_x - i \Delta E_y}{\sqrt{2}} & \Delta E_z & -\frac{\Delta E_x + i \Delta E_y}{\sqrt{2}} \\
		0  & 0 & \frac{\Delta E_x + i \Delta E_y}{\sqrt{2}} & - \bar{E}_z & \frac{\bar{E}_x + i \bar{E}_y}{\sqrt{2}} & 0 \\
		0  & 0 &  \Delta E_z & \frac{\bar{E}_x - i \bar{E}_y}{\sqrt{2}} & 0 & \frac{\bar{E}_x + i \bar{E}_y}{\sqrt{2}} \\
		0  & 0 & \frac{-\Delta E_x + i \Delta E_y}{\sqrt{2}} & 0 & \frac{\bar{E}_x - i \bar{E}_y}{\sqrt{2}} & \bar{E}_z
	\end{bmatrix}.
\label{Ham:DQD}
\end{align}
Here, we introduce the energy detuning $\varepsilon = \mu_1 -\mu_2$, and the average and difference in Zeeman energy for a fixed magnetic field direction, $2\hat{E}_\xi = \mu_B \boldsymbol{B}(\mathcal{G}_1+\mathcal{G}_2) \hat{\xi}$ and $2\Delta E_\xi = \mu_B \boldsymbol{B}(\mathcal{G}_1-\mathcal{G}_2) \hat{\xi}$, where $\hat{\xi}$ is the unit vector along axis $\xi=x,y,z$. We note that in our experiment we only have access to the energies and not the individual g-tensors due to the lack of access to a vector magnet.

\paragraph*{Simulations}
To compute the single-qubit energy spectra in the main text (Fig. 1c-e), we use Hamiltonian~\eqref{eq:FH} assuming only real tunnel matrix elements. Explicitly, we set the spin-conserving tunneling as $\widetilde{t}=$ 1.2, 1.86, 2.4 GHz, the spin-non-conserving tunneling as $\widetilde{t}_\text{SO}=$ 10 MHz, and $U=$ 300 GHz, $B=|\boldsymbol{B}|=$ 10 mT, $\overline{g}=|\mathcal{G}_1+\mathcal{G}_2|/2=$ 0.33, $\Delta g=|\mathcal{G}_1-\mathcal{G}_2|/2=$ 0.008, and with all other components of the g-tensors set to zero.

\paragraph*{Effective qubit Hamiltonian}
To separate the spin dynamics from the charge dynamics, we now restrict ourselves to the subspace spanned by the spin states $\lbrace S(1,1)$, $T_-(1,1)$, $T_0(1,1)$, $T_+(1,1)\rbrace$. This can be done either by block-diagonalizing via Schrieffer-Wolff perturbation theory~\cite{burkardCoupledQuantumdots} or by diagonalizing the singlet sector spanned by $\lbrace S(2,0)$, $S(0,2)$, $S(1,1)\rbrace$ with a subsequent projection or block-diagonalization~\cite{geyer2022twoqubit}. Regardless of the chosen method, the resulting Hamiltonian has the form of a generalized Heisenberg Hamiltonian

\begin{align}
	H_\text{DQD}=\begin{bmatrix}
		J_0 & \frac{\Delta E_x - i \Delta E_y}{\sqrt{2}} & \Delta E_z & -\frac{\Delta E_x + i \Delta E_y}{\sqrt{2}} \\
		 \frac{\Delta E_x + i \Delta E_y}{\sqrt{2}} & - \bar{E}_z & \frac{\bar{E}_x + i \bar{E}_y}{\sqrt{2}} & 0 \\
		  \Delta E_z^\ast & \frac{\bar{E}_x - i \bar{E}_y}{\sqrt{2}} & 0 & \frac{\bar{E}_x + i \bar{E}_y}{\sqrt{2}} \\
		 \frac{-\Delta E_x + i \Delta E_y}{\sqrt{2}} & 0 & \frac{\bar{E}_x - i \bar{E}_y}{\sqrt{2}} & \bar{E}_z
	\end{bmatrix},
	\label{Ham:DQD4}
\end{align}
where $J_0$ is the exchange interaction. In the regime of single dot occupation, $|t|\ll|U\pm \epsilon|$, the exchange interaction can be approximated by~\cite{burkardCoupledQuantumdots}

\begin{align}
    J=\frac{2t^2 U}{U^2-\varepsilon^2}.
    \label{eq:J_heisenberg}
\end{align}
{We observe that during the transformation $\Delta E_{x,y,z}$ is renormalized by the charge-state admixture as $\Delta E^\prime_{x,y,z}$~\cite{jirovec2021singlet,geyer2022twoqubit}. For simplicity and without impacting practical outcomes, we henceforth omit the primed notation.} 

Starting from Hamiltonian~\eqref{Ham:DQD4} we can now construct the qubit Hamiltonian. The $\ket{0}\equiv\ket{S}=S(1,1)$ qubit state is defined via the  PSB readout mechanism. We find the second qubit state by diagonalizing the triplet sector spanned by $\lbrace T_-(1,1)$, $T_0(1,1)$, $T_+(1,1)\rbrace$. Fortunately, the transformation can be parameterized as $e^{-i \boldsymbol{v}\cdot \boldsymbol{S}^0/\hbar}$, where $\boldsymbol{v}$ is a real vector and $\boldsymbol{S}^0$ is a vector containing the spin-0 matrices

\begin{align}
    S^0_{x}&=\frac{1}{2}(S_{x,1}+S_{x,2})\\
    S^0_{y}&=\frac{1}{2}(S_{y,1}+S_{y,2})\\
    S^0_{z}&=\frac{1}{2}(S_{z,1}+S_{z,2}).
\end{align}
The $\ket{1}\equiv\ket{T_-}$ qubit state is then consequently given by the lowest-energy state. The final qubit Hamiltonian reads

\begin{align}
	\tilde{H}_{ST_-}
	&=\begin{bmatrix}
		-J && \frac{\Delta_{ST_-}e^{i\phi_{ST_-}}}{2} \\
		\frac{\Delta_{ST_-}e^{-i\phi_{ST_-}}}{2} && -\overline{E}_{z}\\
	\end{bmatrix}.
\end{align}
In lowest-order perturbation theory the exchange energy $J = J_0$, the average Zeeman splitting $\overline{E}_{z} = \sqrt{\bar{E}_x^2+\bar{E}_y^2+\bar{E}_z^2}$, and the spin-orbit coupling reads

\begin{align}
	\Delta_{\text{SO}}e^{i\phi_{ST_-}} = \sqrt{2}\frac{\Delta E_z \left(\bar{E}_x^2+\bar{E}_y^2\right)- \bar{E}_z (\Delta E_x \bar{E}_x+\Delta E_y \bar{E}_y)}{\overline{E}_{z} (\hat{E}_x-i \bar{E}_y)} + \sqrt{2}i\frac{\Delta E_x \bar{E}_y-\Delta E_y \bar{E}_x}{ (\hat{E}_x-i \bar{E}_y)}.
\end{align}
To arrive at the single-qubit Hamiltonian in the main text, we apply the local phase transformation $U_{\phi}=\exp(-i \phi_{ST_-} (\ket{0}\bra{0}-\ket{1}\bra{1}) /2)$
\begin{align} \label{ST-}
	H_{ST_-}=U_{\phi}\tilde{H}_{ST_-}U_{\phi}^\dagger
	&=\begin{bmatrix}
		-J && \frac{\Delta_{ST_-}}{2} \\
		\frac{\Delta_{ST_-}}{2} && -\overline{E}_{z}\\
	\end{bmatrix},
\end{align}
We remark that a more accurate Hamiltonian can be computed recursively using perturbation theory for a sufficiently large spectral gap~\cite{bravyiSchriefferWolffTransformationQuantum2011}.
We also remark that leakage outside the qubit subspace can be strongly suppressed using optimal control theory and Hamiltonian engineering.

\subsection*{Decoherence times}
Charge noise and nuclear spin noise are ubiquitous in germanium semiconductor devices. Due to the low-frequency nature of both types of noise, they can be approximately modeled as quasistatic fluctuations of input parameters, giving rise to pure dephasing. Furthermore, we ignore in our analysis any energy relaxation mechanism, which is a good approximation for spin qubits~\cite{burkard2023semiconductor}. Hamiltonian~\eqref{ST-} under weak low-frequency noise thus becomes~\cite{chirolliDecoherenceSolidstateQubits2008}

\begin{align}
    H_{ST_-} &=\begin{bmatrix}
		-J -\delta J && \frac{\Delta_{ST_-}+\delta \Delta_x-i\delta \Delta_y}{2} \\
		\frac{\Delta_{ST_-}+\delta \Delta_x+i\delta \Delta_y}{2} && -\overline{E}_{z} - \delta \overline{E}_{z}\\
	\end{bmatrix},
\end{align}
where $\delta$ denotes fluctuations. Note that there are two off-diagonal contributions $\delta \Delta_x$ and $\delta \Delta_y$ that arise from the complex quantity $\Delta_{ST_-}e^{i\phi_{ST_-}}$. Assuming quasi-static noise (or low-frequency noise within the adiabatic approximation) the qubit resonance frequency is modulated by noise as follows

\begin{align}
    \hbar \omega_q = \sqrt{4(J-\overline{E}_{z})^2+\Delta_{ST_-}^2}\rightarrow \sqrt{4(J-\overline{E}_{z}+\delta J- \delta \overline{E}_{z})^2+(\Delta_{ST_-}+\delta \Delta_x)^2+\delta \Delta_y^2}.
    \label{eq:fluctuations}
\end{align}
The $S-T_-$ qubit is operated in two regimes which we separately discuss.
\paragraph*{Dephasing during rotations around the $x$-axis}
For single-qubit $x$-rotations, we operate at $J=\overline{E}_{z}$ and we can expand Eq.~\eqref{eq:fluctuations} up to first order

\begin{align}
    \hbar \omega_q^x = \Delta_{ST_-} + \delta\Delta_{x} + \mathcal{O}(\delta^2)
\end{align}
to find the dominating contributions. Assuming quasi-static Gaussian distributed noise, the decoherence time in the $x$-axis rotation regime is then given by

\begin{align}
    T_x^\ast = \frac{1}{\sqrt{2} \sigma_{\Delta_{x}}},
\end{align}
where $\sigma_{\Delta_{x}}^2=\langle \delta\Delta_{x}^2 \rangle - \langle\delta\Delta_{x}\rangle^2$ is the standard deviation of the $\delta\Delta_{x}$ noise. Since $\Delta E_\xi\propto B$ and $\overline{E}_{\xi}\propto B$ for $\xi=x,y,z$, we expect $T_x^\ast\propto B^{-1}$ for an in-plane magnetic field~\cite{lawrieSimultaneousSinglequbitDriving2023,hendrickx2023sweet}. Therefore, operating at small magnetic fields is beneficial for $x$-gates.

\paragraph*{Dephasing during rotations around the $z$-axis}
For single-qubit $z$-rotations, we operate at $|J-\overline{E}_{z}|\gg \Delta_{ST_-}$ and we can expand Eq.~\eqref{eq:fluctuations} up to first order

\begin{align}
    \hbar \omega_q^z = |J-\overline{E}_{z} +\delta J- \delta \overline{E}_{z}| + \mathcal{O}(\delta^2)
\end{align}
to find the dominating contributions. Assuming Gaussian distributed noise the decoherence time in the $z$-axis rotation regime is then given by

\begin{align}
    T_z^\ast = \frac{1}{\sqrt{2} \sigma_{J,\overline{E}_{z}}(1+c)},
\end{align}
where $\sigma_{\sigma_{J,\overline{E}_{z}}}^2=\langle (\delta J- \delta \overline{E}_{z})^2 \rangle - \langle(\delta J- \delta \overline{E}_{z})\rangle^2$ is the standard deviation of the noise difference, $c = \text{cov}(J,\overline{E}_{z})/(\sigma_J \sigma_{\overline{E}_{z}})$ the noise correlation factor, $\text{cov}(J,\overline{E}_{z})$ the covariance, and $\sigma_J$ ($\sigma_{\overline{E}_{z}}$) the standard deviation of the individual fluctuations $\delta J$ and $ \delta \overline{E}_{z}$. Since $\overline{E}_{z}\propto B$, operating at small magnetic fields can suppress fluctuations of the Zeeman splitting, giving rise to $T_z^\ast = \frac{1}{\sqrt{2} \sigma_{J}}$. In contrast to $T^\ast_x$, thus, $T^\ast_z$ is not suppressed at small magnetic fields. This is consistent with the observations in the main text that $T^\ast_z < T^\ast_x $.

\subsection*{Two-qubit Hamiltonian}
The derivation of the Hamiltonian of two coupled $S-T_-$ qubits is analogous to that of the single-qubit Hamiltonian. The first step is separating the spin and charge degree of freedom, resulting in a generalized Heisenberg Hamiltonian. 
We give explicit expressions for the two-qubit Hamiltonian for two different architectures, a simplified architecture with single connectivity and the design from the experiment.

\paragraph*{Linear chain}
with a single inter-qubit tunnel (exchange) coupling and isotropic g-tensor up to a scaling. In this case, the multi-qubit Hamiltonian is given by 

\begin{align}
	H_\text{FH}=&\sum_{i=1} \mu_B  g_i B S_{z,i} + \sum_{i=1}  \frac{\Delta_{ST_-,i}}{\hbar} S_x + \sum_{\braket{i,j}} + J_{ij} \left(\frac{\boldsymbol{S}_i\cdot\boldsymbol{S}_{j}}{\hbar^2} - \frac{1}{4}\right),
	\label{eq:Heisenberg}
\end{align}
where $\braket{i,j}$ denotes neighboring quantum dots.
Projected on the two-qubit basis $\lbrace \ket{S,S},\ket{S,T_-},\ket{T_-,S},\ket{T_-,T_-}\rbrace$ the two-qubit Hamiltonian reads

\begin{align}
	H_\text{2Q}=H_\text{Q1} + H_\text{Q2} + H_\text{coup,iso}
\label{2q}
\end{align}
with

\begin{align} 
H_\text{Q1} + H_\text{Q2} &=\begin{bmatrix}
   -J_{ij} - J_{kl} && \frac{\Delta_{\text{SO}, kl}}{2} && \frac{\Delta_{\text{SO}, ij}}{2} && 0 \\
   \frac{\Delta_{\text{SO}, kl}}{2}  && -J_{ij}-\overline{E}_{z, kl} && 0 && \frac{\Delta_{\text{SO}, ij}}{2} \\
   \frac{\Delta_{\text{SO}, ij}}{2}  && 0  &&  -\overline{E}_{z, ij} - J_{kl} && \frac{\Delta_{\text{SO}, kl}}{2} \\
   0  && \frac{\Delta_{\text{SO}, ij}}{2} && \frac{\Delta_{\text{SO}, kl}}{2} && -\overline{E}_{z, ij}-\overline{E}_{z, kl} \\
\end{bmatrix},\label{SWAP1}\\
% H_\text{coup,iso}=&\frac{1}{2}\begin{bmatrix}
%    -J_{\text{coup}}  && 0 && 0 && 0  \\
%    0  && -J_{\text{coup}}  && J_{\text{coup}}  && 0 \\
%    0  && J_{\text{coup}}   &&  -J_{\text{coup}}  && 0 \\
%    0  && 0 && 0 && 0 \\
%\end{bmatrix},
H_\text{coup,iso}=&\frac{1}{2}\begin{bmatrix}
    0  && 0 && 0 && 0  \\
    0  && 0  && J_{\text{coup},ij,kl}  && 0 \\
    0  && J_{\text{coup},ij,kl}   &&  0  && 0 \\
    0  && 0 && 0 && J_{\text{coup},ij,kl} \\
\end{bmatrix},
\label{SWAP2}
% H_\text{coup,SOI}=&\frac{1}{2}\begin{bmatrix}
%    0  && 0 && 0 && J^{\text{SO}}_{\perp2,ij,kl} \\
%    0  && 0 && J^{\text{SO}}_{\perp1,ij,kl} && J^{\text{SO}}_{a,ij,kl} \\
%    0  && J^{\text{SO},\star}_{\perp1,ij,kl}  &&  0 && J^{\text{SO},\star}_{a,kl,ij} \\
%    J^{\text{SO},\star}_{\perp2,ij,kl}  && J^{\text{SO},\star}_{a,ij,kl} && J^{\text{SO}}_{a,kl,jk} && -J^{\text{SO}}_{||,ij,kl} \\
% \end{bmatrix}.
% \label{SWAP3}
\end{align}
Here the subscripts indicate the sites $i$, $j$, $k$ and $l$, respectively. Note that this Hamiltonian is identical to the Hamiltonian used in the main text up to an energy offset that corresponds to a global phase shift.

From a quantum information point of view, the isotropic exchange Hamiltonian~\eqref{SWAP2} generates a universal two-qubit gate. More precisely, the Hamiltonian generates a SWAP + iSWAP gate. This can be visualized through the symmetry of the system: since the isotropic exchange interaction is given by the projector on the singlet subspace, the $\ket{T_-T_-}\bra{T_-T_-}$ matrix element of the two-qubit interaction has to be zero.

% The last term, the anisotropic exchange Hamiltonian~\eqref{SWAP3} arises from the spin-orbit interaction that allows for differences between the g-tensors in DQD~1 and DQD~2. For a fixed magnetic field direction, the resulting exchange interaction can be written as a rotation matrix. Explicetly, the two-qubit interaction can be computed as
% \begin{align}
%     H_\text{coup,SOI} = P_\text{2Q} e^{i \boldsymbol{v}_{ij}\cdot \boldsymbol{S}^0_{ij}/\hbar} \, e^{i \boldsymbol{v}_{kl}\cdot \boldsymbol{S}_{kl}^0/\hbar}
% \left[J_{ik}  \left(\frac{\boldsymbol{S}_i\cdot\boldsymbol{S}_{k}}{\hbar^2} - \frac{1}{4}\right)\right]
%  e^{-i \boldsymbol{v}_{ij}\cdot \boldsymbol{S}^0_{ij}/\hbar} \, e^{-i \boldsymbol{v}_{kl}\cdot \boldsymbol{S}_{kl}^0/\hbar}P_\text{2Q},\\
%  = [J_{ik}  \left(\frac{\boldsymbol{S}_i\cdot\boldsymbol{S}_{k}}{\hbar^2} - \frac{1}{4}\right)
% \end{align}
% where $P_\text{2Q}$ is the projector on the two-qubit subspace.

\paragraph*{Ladder}
with two inter-qubit tunnel (exchange) couplings and anisotropic g-tensors. This is the architecture used in the present device. Consequently, the Hamiltonian cannot be cast into Eq.~\eqref{eq:Heisenberg} and an additional rotation of the spin has to be taken into account~\cite{geyer2022twoqubit}. Fortunately, the ``additional'' rotation from the exchange interaction can be absorbed into the rotation caused by the differences in g-tensor for a fixed magnetic field setting. In this case, the total two-qubit Hamiltonian is given by

\begin{align}
	H_\text{2Q}=H_\text{Q1} + H_\text{Q2} + H_\text{coup,iso} +  H_\text{coup,SOI} \;,
 \label{eq:ham2Qtot}
\end{align}
where the first three terms are given in Eqs.~\eqref{SWAP1}-\eqref{SWAP2}. The last term is caused by the spin-orbit interaction and is given by

\begin{align}
    H_\text{coup,SOI}&= P_\text{2Q} U_{ij,kl}
\left[ \left(\frac{\boldsymbol{S}_i\cdot\mathcal{J}_{ik}\boldsymbol{S}_{k}}{\hbar^2} - \frac{1}{4}\right) + \left(\frac{\boldsymbol{S}_j\cdot \mathcal{J}_{jl}\boldsymbol{S}_{l}}{\hbar^2} - \frac{1}{4}\right)\right]
 U_{ij,kl}^\dagger P_\text{2Q} - H_\text{coup,iso},\\
&=\frac{1}{2}\begin{bmatrix}
   0  && 0 && 0 && J^{\text{SO}}_{\perp2,ij,kl} \\
   0  && 0 && J^{\text{SO}}_{\perp1,ij,kl} && J^{\text{SO}}_{a2,ij,kl} \\
   0  && J^{\text{SO},\ast}_{\perp1,ij,kl}  &&  0 && J^{\text{SO}}_{a1,kl,ij} \\
   J^{\text{SO},\ast}_{\perp2,ij,kl}  && J^{\text{SO},\ast}_{a2,ij,kl} && J^{\text{SO},\ast}_{a1,kl,jk} && -J^{\text{SO}}_{||,ij,kl} \\
\end{bmatrix},
\label{SWAP3}
\end{align}
where $P_\text{2Q}$ is the projector on the two-qubit subspace,

\begin{align}
    U_{ij,kl} = U_{\phi_{ij}}U_{\phi_{kl}} e^{i \boldsymbol{v}_{ij}\cdot \boldsymbol{S}^0_{ij}/\hbar} \, e^{i \boldsymbol{v}_{kl}\cdot \boldsymbol{S}_{kl}^0/\hbar},
\end{align}
is the combination of the single $S-T_-$ qubit basis transformation from the previous section, and $\mathcal{J}_{ik(jl)}$ is the exchange tensor between the spins in quantum dot pair $ik$ ($jl$) in the spin-orbit basis~\cite{geyer2022twoqubit} of quantum dot pair $ij$ and $kl$.
We also note that the full characterization of all elements cannot be resolved with our current operation regime and measurement setup, as high-fidelity sequential spin readout is required. For example, the $J^{\text{SO}}_{\perp2,ij,kl}$ term could be characterized in the regime where the energies of the $\ket{SS}$ and $\ket{T_-T_-}$ states are identical but energetically separated from the other states. Therefore, we consider for simplicity an isotropic model to describe the swapping dynamics, which fits well to the observations.

\paragraph*{Simulation of the two-qubit energy spectrum}
To compute the two-qubit energy spectrum between Q3 and Q4 in the main text (Fig. 3b), we use the Hamiltonian of ~\eqref{2q} and Eq.~\eqref{eq:J_heisenberg} and the estimated single-qubit parameters at $B=$ 10 mT. Explicitly, we use $U_{Q3}=$ 290 GHz, $U_{Q4}=$ 326 GHz, $t_{Q3}=$ 3 GHz, $t_{Q4}=$ 2.8 GHz, $\varepsilon_{Q3}=$ -174 GHz, $\overline{g}_{Q3}=$ 0.37, $\overline{g}_{Q4}=$ 0.35, and assume for simplicity a homogeneous $\Delta_{\text{SO}, ij}=\Delta_{\text{SO}, kl}=$ 10 MHz. For the two-qubit interaction between Q3 and Q4, we use $H_\text{coup,iso}$ with $J_\text{coup}=(J_{ik}+J_{jl})/2$. We have estimated the inter-qubit exchange using $t_{ik}=$ 2 GHz and $t_{jl}=$ 0.4 GHz and $\varepsilon_{ik}=(\varepsilon_{ij}-\varepsilon_{kl}+\mu_0)/2$ and $\varepsilon_{jl}=(\varepsilon_{kl}-\varepsilon_{ij}+\mu_0)/2$ by assuming the difference between the chemical potential offsets of two qubits $\mu_0=\mu_{ij}-\mu_{kl} = 0$.
\paragraph*{Simulation of the state transfer}
For the simulation of the state transfer in the main text (Fig. 4b), we use the Hamiltonian of Eq.~\eqref{SWAP2} for the SWAP operations with $J_\text{coup}=95$ MHz and a single-qubit $x$-axis rotation frequency of Q4 of 13 MHz.

\paragraph*{Fitting Hamiltonian}
{Since our two-qubit characterization was performed in a regime where the single-qubit energies are larger than the inter-qubit coupling energies, the upper Hamiltonian can be further simplified using another block-diagonalization (SW approximation) step to ease fitting. In particular, we assume $|J^{\text{SO},\ast}_{\perp 2,kl,jk}|,|J^{\text{SO},\ast}_{a1,kl,jk}|,|J^{\text{SO},\ast}_{a2,kl,jk}|\ll |J_{ij}-\overline{E}_{z,ij}|,|J_{kl}-\overline{E}_{z,kl}|$. The effective Hamiltonian can then be written as follows:
\begin{equation} \label{SWAP_model}
\begin{split}
H_\text{2Q, eff}
 =&\frac{J_\text{trans}}{4}(\sigma_x^{ij}\sigma_x^{kl} + \sigma_y^{ij}\sigma_y^{kl}) + \frac{J_\text{cross}}{4}(\sigma_y^{ij}\sigma_x^{kl} - \sigma_x^{ij}\sigma_y^{kl}) + \frac{J_\text{long}}{4}\sigma_z^{ij}\sigma_z^{kl}+
 \frac{\varepsilon}{4}(\sigma_z^{ij}-\sigma_z^{kl})+\frac{\Sigma}{4}(\sigma_z^{ij}+\sigma_z^{kl})
\end{split}
\end{equation}
with the parameters 
\begin{align}
    J_\text{trans} &= J_{\text{coup},ij,kl} + \text{Re}\left(J^{\text{SO}}_{\perp1,ij,kl} + \frac{1}{2}  \frac{J^{\text{SO},\ast}_{a2,ij,kl} J^{\text{SO}}_{a1,ij,kl}}{\overline{E}_{z, ij}+\overline{E}_{z, kl}}\right),\\
    J_\text{cross} &= -\text{Im}\left(J^{\text{SO}}_{\perp1,ij,kl} + \frac{1}{2}  \frac{J^{\text{SO},\ast}_{a2,ij,kl} J^{\text{SO}}_{a1,ij,kl}}{\overline{E}_{z, ij}+\overline{E}_{z, kl}}\right)\\
    J_\text{long} &= \frac{J_{\text{coup},ij,kl}-J^{\text{SO}}_{||,ij,kl}}{2} - \frac{1}{2}  \frac{|J^{\text{SO}}_{a1,ij,kl}|^2 +  |J^{\text{SO}}_{a2,ij,kl}|^2}{\overline{E}_{z, ij}+\overline{E}_{z, kl}},\\
    \varepsilon &=  \overline{E}_{z, ij}-\overline{E}_{z, kl}-J_{ij} + J_{kl} + \frac{1}{4}  \frac{|J^{\text{SO}}_{a2,ij,kl}|^2 -  |J^{\text{SO}}_{a1,ij,kl}|^2}{\overline{E}_{z, ij}+\overline{E}_{z, kl} },\\
    \Sigma &= \overline{E}_{z, ij}+\overline{E}_{z, kl} -J_{ij} - J_{kl} -\frac{J_{\text{coup},ij,kl}-J^{\text{SO}}_{||,ij,kl}}{2} + \frac{1}{4}  \left(\frac{|J^{\text{SO}}_{a1,ij,kl}|^2 +  |J^{\text{SO}}_{a2,ij,kl}|^2}{\overline{E}_{z, ij}+\overline{E}_{z, kl}}+ \frac{2|J^{\text{SO}}_{\perp 2,ij,kl}|^2}{\overline{E}_{z, ij}+\overline{E}_{z, kl} -J_{ij} - J_{kl} }\right).
\end{align}
Here, $J_\text{trans}$ denotes the real part and $J_\text{cross}$ the imaginary part of the center off-diagonal terms of the Hamiltonian~\eqref{eq:ham2Qtot}; $J_\text{long}$ mainly refers to the exchange interaction of $S-T_-$ qubits plus the spin-orbit coupling induced anisotropic exchange interactions; $\varepsilon$ includes mainly the detuning of qubit energies in the experiment, and $\Sigma$ is the total qubit energy, with a correction of the anisotropic exchange couplings. We remark, that the upper Hamiltonian also considers the effect of single-qubit phase rotations that may happen before and after the $\sqrt{\text{SWAP}}$ gate in the GST experiment, e.g., through errors in the idling gate before and after the exchange pulse or crosstalk. Such single-qubit phase rotations shift the $\sqrt{\text{SWAP}}$ rotation axis $\theta \equiv \text{arg}(J_\text{trans}+i J_\text{cross})\rightarrow \theta +\phi$, where $\text{arg}$ denotes the argument of the complex number. It can also be noted that this realistic Hamiltonian is similar in form to the isotropic Hamiltonian in the main text up to single-qubit phases and controlled phase rotations, thus we used only the isotropic model to describe the SWAP dynamics in the main text.}

\section{\label{sec:level12} Gate set tomography of the single- and two-qubit gate}

\subsection*{Gate set tomography of the single-qubit gate}
{The gate set tomography experiments are all carried out using the python package pyGSTi \cite{nielsen2020probing}.}

{For single-qubit GST, we use the model smq1Q\_XY with a gateset including only $\sqrt{X}$ and $\sqrt{Y}$. The fiducials for state preparation and measurements are \{null, $\sqrt{X}$, $\sqrt{Y}$, $\sqrt{X}$$\sqrt{X}$, $\sqrt{X}$$\sqrt{X}$$\sqrt{X}$, $\sqrt{Y}$$\sqrt{Y}$$\sqrt{Y}$\}, where null is an idle gate with zero waiting time, and the germs for amplifying qubit errors are \{$\sqrt{X}$$\sqrt{Y}$, $\sqrt{X}$$\sqrt{X}$$\sqrt{Y}$\}. Unless indicated otherwise, the implemented circuit lengths are powers of two from 1 up to 32, resulting in 568 circuits in total. In every sequence, the singlet or triplet probability of the involved qubit is acquired by averaging over 1000 single-shot cycles. The data was analyzed using pyGSTi with the CPTP (completely positive trace-preserving) model, from which we obtained Pauli transfer matrices (PTM) $M_\text{exp}$ for each gate operation. By comparing them to the ideal matrices, $M_\text{ideal}$, we can derive an entanglement fidelity as $F_p=\text{Tr}[M_\text{ideal}^{-1}M_\text{exp}]/d^2$, and thus a gate infidelity $1-F_g=\frac{d}{d+1}(1-F_p)$, where $d=2^N$ is the dimension of the Hilbert space, and $N$ refers to the qubit number. Two datasets of the gate infidelities of each qubit are shown in Table~\ref{GST_1Q_tab}.} 

{It is worth mentioning that in some cases, the $\sqrt{Y}$ gate shows a smaller error than the $\sqrt{X}$ gate, which may be an artefact caused by the GST protocol, i.e., biased attribution of relational errors~\cite{mkadzik2022precision}. Table~\ref{GST_1Q_tab} also includes the non-unitary gate infidelities, $\frac{d-1}{d}(1-\sqrt{u(M^{-1}_\text{exp}M_\text{ideal})}$, where $u(M)=\text{Tr}(J_\alpha (M)^2)$ and $J_\alpha (M)$ is the Jamio\l kowski isomorphism map between the matrix $M$ and the corresponding Choi Matrix. Almost all the data show a larger error of the gate infidelity than the non-unitary error, but only by a small amount. Therefore, we can expect slightly higher control fidelities with improved pulse control.} 

\begin{table}[h]
\centering
\begin{tabular}{ |p{3.5cm}||p{3cm}|p{3cm}|p{3cm}|p{3cm}|  }
 \hline
 Gate & Q1 & Q2 & Q3 &Q4 \\
 \hline
 $\sqrt{X}$   & 0.0084$\pm$0.0004  &0.0039$\pm$0.0004  &0.0102$\pm$0.0009   &0.0044$\pm$0.0004\\
 $\sqrt{X}$, non-unitary   & 0.0083$\pm$0.0004  &0.0038$\pm$0.0004  &0.0097$\pm$0.0009   &0.0025$\pm$0.0004\\
 $\sqrt{Y}$   &0.0062$\pm$0.0004  &0.0051$\pm$0.0005 &0.0068$\pm$0.0008   &0.0055$\pm$0.0004\\
 $\sqrt{Y}$, non-unitary   &0.0061$\pm$0.0004  &0.0049$\pm$0.0005 &0.0060$\pm$0.0008   &0.0034$\pm$0.0004\\
 \hline
 $\sqrt{X}$
   & 0.0079$\pm$0.0004  &0.0031$\pm$0.0006  &0.0100$\pm$0.0009   &0.0054$\pm$0.0015\\
$\sqrt{X}$, non-unitary   & 0.0072$\pm$0.0004  &0.0002$\pm$0.0006  &0.0093$\pm$0.0009   &0.0029$\pm$0.0015\\
 $\sqrt{Y}$   
    &0.0084$\pm$0.0004  &0.0114$\pm$0.0007 &0.0112$\pm$0.0009   &0.0035$\pm$0.0014\\
$\sqrt{Y}$, non-unitary   &0.0075$\pm$0.0004  &0.0046$\pm$0.0007 &0.0105$\pm$0.0009   &0.0010$\pm$0.0014\\
 \hline
\end{tabular}
\caption{\label{GST_1Q_tab} {Summary of the gate infidelities of single-qubit gates of Q1-Q4 from two different datasets. The uncertainty represents the 95\% confidence interval. Notice the second dataset for Q4 show higher error bars, which was obtained with circuit length L=8, others were obtained with L=32.}}
\end{table}

\subsection*{Gate set tomography of the two-qubit gate}

{The two-qubit GST are performed on Q1 and Q2. We derive a model named smq2Q\_XYSQRT and the default gateset includes \{$\sqrt{X}_{Q1}$, $\sqrt{X}_{Q2}$, $\sqrt{Y}_{Q1}$, $\sqrt{Y}_{Q2}$, $\sqrt{\text{SWAP}}$\}, where the subscript in the single-qubit gate refers to the corresponding qubit. The fiducials and germs are shown in Table~\ref{GST_gate_set_germ}. The circuit lengths are powers of two from 1 up to 2, resulting in 1317 circuits in total. In every sequence, the joint probabilities of each qubit are acquired by averaging over 500 single-shot cycles.}

\begin{table}[h]
\centering
\begin{tabular}{ |p{0.5cm}||p{4cm}|p{3cm}|p{7cm}|  }
 \hline
  & fiducial: preparation & fiducial: measurement & germ \\
 \hline
 1   &null  &null &$\sqrt{X}_{Q1}$   \\
 2   & $\sqrt{X}_{Q2}$ &$\sqrt{X}_{Q2}$ &$\sqrt{X}_{Q2}$   \\
 3   & $\sqrt{Y}_{Q2}$  &$\sqrt{Y}_{Q2}$  &$\sqrt{Y}_{Q1}$   \\
 4   & $\sqrt{X}_{Q2}$$\sqrt{X}_{Q2}$ &$\sqrt{X}_{Q2}$$\sqrt{X}_{Q2}$  &$\sqrt{Y}_{Q2}$   \\
 5   & $\sqrt{X}_{Q1}$ &$\sqrt{X}_{Q1}$  &$\sqrt{\text{SWAP}}$   \\
 6   & $\sqrt{X}_{Q1}$$\sqrt{X}_{Q2}$ &$\sqrt{Y}_{Q1}$ &$\sqrt{\text{SWAP}}$$\sqrt{X}_{Q2}$$\sqrt{X}_{Q2}$$\sqrt{\text{SWAP}}$$\sqrt{X}_{Q2}$$\sqrt{X}_{Q1}$   \\
 7   & $\sqrt{X}_{Q1}$$\sqrt{Y}_{Q2}$ &$\sqrt{X}_{Q1}$$\sqrt{X}_{Q1}$  &$\sqrt{X}_{Q1}$$\sqrt{Y}_{Q2}$$\sqrt{Y}_{Q1}$$\sqrt{Y}_{Q2}$$\sqrt{Y}_{Q1}$   \\
 8   & $\sqrt{X}_{Q1}$$\sqrt{X}_{Q2}$$\sqrt{X}_{Q2}$ &$\sqrt{X}_{Q1}$$\sqrt{X}_{Q2}$  &$\sqrt{X}_{Q2}$$\sqrt{Y}_{Q2}$$\sqrt{X}_{Q1}$$\sqrt{Y}_{Q1}$  \\
 9   & $\sqrt{Y}_{Q1}$ &$\sqrt{X}_{Q1}$$\sqrt{Y}_{Q2}$  &$\sqrt{\text{SWAP}}$$\sqrt{\text{SWAP}}$$\sqrt{Y}_{Q2}$$\sqrt{X}_{Q1}$$\sqrt{Y}_{Q1}$   \\
 10   & $\sqrt{Y}_{Q1}$$\sqrt{X}_{Q2}$ &$\sqrt{Y}_{Q1}$$\sqrt{X}_{Q2}$ & $\sqrt{X}_{Q2}$$\sqrt{Y}_{Q2}$$\sqrt{Y}_{Q2}$ \\
 11   & $\sqrt{Y}_{Q1}$$\sqrt{Y}_{Q2}$ &$\sqrt{Y}_{Q1}$$\sqrt{Y}_{Q2}$  & $\sqrt{Y}_{Q2}$$\sqrt{X}_{Q1}$$\sqrt{\text{SWAP}}$$\sqrt{X}_{Q1}$ \\
 12   & $\sqrt{Y}_{Q1}$$\sqrt{X}_{Q2}$$\sqrt{X}_{Q2}$ & & $\sqrt{\text{SWAP}}$$\sqrt{X}_{Q1}$$\sqrt{Y}_{Q1}$$\sqrt{X}_{Q2}$$\sqrt{Y}_{Q1}$$\sqrt{\text{SWAP}}$  \\
 13   & $\sqrt{X}_{Q1}$$\sqrt{X}_{Q1}$ & & $\sqrt{Y}_{Q2}$$\sqrt{Y}_{Q2}$$\sqrt{Y}_{Q1}$$\sqrt{Y}_{Q1}$$\sqrt{X}_{Q2}$$\sqrt{X}_{Q2}$$\sqrt{X}_{Q1}$  \\
 14   & $\sqrt{X}_{Q1}$$\sqrt{X}_{Q1}$$\sqrt{X}_{Q2}$ &  &   \\
 15   & $\sqrt{X}_{Q1}$$\sqrt{X}_{Q1}$$\sqrt{X}_{Q2}$ &  &   \\
 16   & $\sqrt{X}_{Q1}$$\sqrt{X}_{Q1}$$\sqrt{X}_{Q2}$$\sqrt{X}_{Q2}$ &  &   \\
 \hline
\end{tabular}
\caption{\label{GST_gate_set_germ} {Fiducials and germs for the two-qubit GST circuits}}
\end{table}

\begin{figure*}[htb]
\renewcommand{\figurename}{\textbf{Supplementary Fig}}
\includegraphics{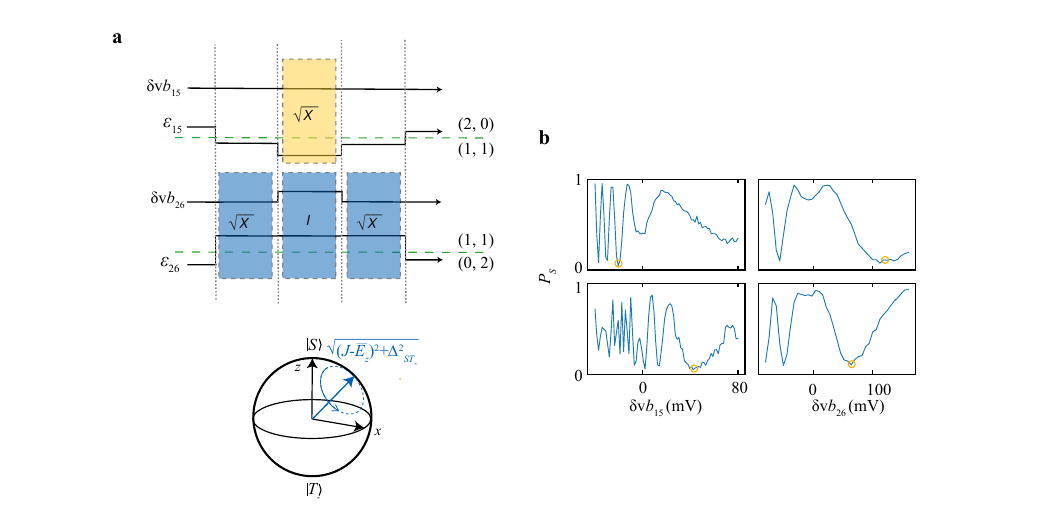}
%\internallinenumbers
\caption{\label{cali_GST} {\textbf{a}, Illustration of the gate voltage pulses for the idle gate calibration of Q2 with Q1 doing an $\sqrt{X}$ gate operation. In this Ramsey-like sequence, a perfect idle gate will result in a zero singlet probability of Q2. The Bloch sphere below shows the qubit rotation with a barrier gate pulse. \textbf{b}, Measured singlet probabilities $P_S$ as a function of the barrier voltage change with a fixed duration that equals the operation time on the other qubit. They are the results for the calibration of $\sqrt{X}_{Q2}$ (top left), $\sqrt{X}_{Q1}$ (top right), $\sqrt{Y}_{Q2}$ (bottom left) and $\sqrt{Y}_{Q1}$ (bottom right). The orange circles show the expected position of the barrier gate voltage for the idle operation. }}
\end{figure*}

{As discussed in the Extended Fig. 10 in the main text, the single-qubit gate in the two-qubit space requires an idle gate. Our strategy is to pulse the idle qubit to an operating point where it completes a 2$\pi$ rotation during the time needed to operate on the other qubit, with a rotation speed of $f_{ST_-}=\sqrt{(J-\overline{E}_z)^2+\Delta^2_{ST_-}}/h$ by pulsing the barrier gate. To calibrate the barrier pulse, we use a Ramsey-like pulse sequence as shown in Supplementary Fig.~\ref{cali_GST}a. One example of the barrier pulse calibration for the single-qubit gates of Q1 and Q2 is shown in Supplementary Fig.~\ref{cali_GST}b. Here, we keep the idle qubit at the sweet point of the detuning and try to pulse the barrier gate positively to reduce the effect of charge noise from strong exchange coupling. For $\sqrt{X}_{Q1}$, $\sqrt{Y}_{Q1}$, $\sqrt{Y}_{Q2}$ the idling qubit is pulsed positively, while for $\sqrt{X}_{Q2}$, the idling qubit is pulsed negatively because the short $\sqrt{X}$ gate duration of Q2 otherwise does not allow a full rotation on the idling qubit (Q1) given the accessible values of $|\overline{E}_z-J|$.}

{The raw data of running GST circuits was analyzed using the CPTP model, and the gate infidelities of three datasets are shown in Table.~\ref{GST_2Q_tab}. The single-qubit gate errors in the two-qubit space are quite large compared to those measured in the single-qubit space. To get a better understanding, we calculate the Jamio\l kowski probability $\epsilon_J(\mathcal{L})=-\text{Tr}[\rho_J(\mathcal{L})\ket{\Psi}\bra{\Psi}]$ and the Jamio\l kowski amplitude $\theta_J(\mathcal{L})=||(1-\ket{\Psi}\bra{\Psi})\rho_J(\mathcal{L})\ket{\Psi}||$, which approximately describe the amount of incoherent and coherent Hamiltonian errors of the quantum process, respectively \cite{PRXQuantum.3.020335}. Here $\rho_J(\mathcal{L})=(\mathcal{L}\otimes1_{d^2})\ket{\Psi}\bra{\Psi}$ is the Jamio\l kowski state, $\mathcal{L}=\text{log}(M_\text{exp}M_\text{ideal}^{-1})$ is the error generator of the process, and $\ket{\Psi}$ is a maximally entangled state that originates from the relation of quantum processes to states in a Hilbert space twice the dimension via the Choi-Jamio\l kowski isomorphism. In Table.~\ref{GST_2Q_tab}, we can see that the coherent errors are all larger than the incoherent errors, suggesting the qubits suffer more from calibration errors in the gate pulses rather than decoherence. These calibration errors affect not only the qubit we aim to rotate but also the idling qubit, which is ideally undergoing a 2$\pi$ rotation. The total error they contribute to the gate operation can be approximately calculated as $1-F_J=\frac{d}{d+1}[\epsilon_J(\mathcal{L})+\theta_J(\mathcal{L})^2]$ when errors are small. We can observe that they are similar to the values $1-F_g$ calculated before.}

\begin{table}[h]
\centering
\begin{tabular}{ |p{2cm}||p{3cm}|p{3cm}|p{3cm}|p{3cm}|  }
 \hline
 Gate & Gate infidelity (1-$F_g$) & Jamio\l kowski probability $\epsilon_J(\mathcal{L})$ & Jamio\l kowski amplitude $\theta_J(\mathcal{L})$ & Total error (1-$F_J$)\\
 \hline
 $\sqrt{X}_{Q1}$   & 0.02826    &0.01827 & 0.1337 &0.02891 \\
 $\sqrt{X}_{Q2}$ &   0.03796  & 0.04513   &0.06360 &0.03934\\
 $\sqrt{Y}_{Q1}$  &0.04175 & 0.03593 &  0.13510 &0.04335\\
 $\sqrt{Y}_{Q2}$    &0.13356 & 0.10504 &  0.28204 &0.14767	\\
 $\sqrt{\text{SWAP}}$ &0.20320 &   0.25753  &  0.22752 &0.24743  \\
 \hline
 $\sqrt{X}_{Q1}$   & 0.02535    &0.02029 & 0.11011 &0.02593 \\
 $\sqrt{X}_{Q2}$ &   0.04334  & 0.05166   &0.06701 &0.04492\\
 $\sqrt{Y}_{Q1}$  &0.04078 & 0.04056 &  0.11085 &0.04228\\
 $\sqrt{Y}_{Q2}$    &0.13500 & 0.10932 &  0.27864 &0.14957	\\
 $\sqrt{\text{SWAP}}$ &0.20390 &0.25132     &0.23363  & 0.24472\\
 \hline
$\sqrt{X}_{Q1}$   & 0.03027    &0.02342 & 0.12412 &0.03106 \\
 $\sqrt{X}_{Q2}$ &   0.04032  & 0.04576   &0.08005 &0.04173\\
 $\sqrt{Y}_{Q1}$  &0.04232 & 0.04027 &  0.12052 &0.04384\\
 $\sqrt{Y}_{Q2}$    &0.13230 & 0.10893 &  0.27269 &0.14663	\\
 $\sqrt{\text{SWAP}}$ &0.19560 &0.24216     &0.22064  &0.23267 \\
 \hline
\end{tabular}
\caption{\label{GST_2Q_tab} {Summary of the gate infidelities of the two-qubit GST measurement from three different datasets. Notice the fidelities for the $\sqrt{\text{SWAP}}$ is analyzed by fitting the theoretical model.}}
\end{table}

{For the two-qubit gate, the GST model we used for data analysis is a standard $\sqrt{\text{SWAP}}$. From the GST result, we obtain a PTM and fit it to our theoretical model~\eqref{SWAP_model}. The fitted parameters are shown in Table~\ref{GST_SWAP_fit}. The fact that $J_\text{long}$ differs from  $J_\text{trans}$, as well as the non-zero value of $J_\text{cross}$, arise from spin-orbit coupling. The detuning of two qubit splittings $\varepsilon$ can originate from imperfect calibration of the $\sqrt{\text{SWAP}}$ gate operation point, whose strength compared to the exchange coupling is calculated and shown as $\varepsilon/\sqrt{J_\text{trans}^2+J_\text{cross}^2}$. $\Sigma$ is a single-qubit z rotation, see Eq.~\eqref{SWAP_model}.  With those fitted parameters, we can rebuild the Pauli transfer matrix $M_\text{ideal}$ of the $\sqrt{\text{SWAP}}$ gate, and by comparing it to the experimental result $M_\text{exp}$, we obtain the gate fidelities as shown in the Table~\ref{GST_SWAP_fit}. These gate fidelities are lower than the Bell state fidelity, which possibly may be explained by the fact that the gate fidelity is a measure that averages over all possible input states, while the Bell state fidelity considers only one specific input state. Moreover, the pulse schemes for measuring the $\sqrt{\text{SWAP}}$ gate fidelity and the Bell state are different, which can result in different contributions from single-qubit errors. These fitted parameters also enable us to rebuild the two-qubit gate unitary matrix as shown below (using data set 3 in Table.~\ref{GST_SWAP_fit})}:
\begin{align}
    \sqrt{\text{SWAP}}_{ST} &=\begin{pmatrix}
		1 && 0 && 0 && 0 \\
		0 && 0.748e^{3.131i} && 0.664e^{-1.45i} && 0\\
            0 && 0.664e^{-1.95i} && 0.748e^{2.894i} && 0\\
            0 && 0 && 0 && 1e^{0.266i}\\
	\end{pmatrix},
\end{align}

\begin{table}[h]
\centering
\begin{tabular}{ |p{3cm}||p{3cm}|p{3cm}|p{3cm}|  }
 \hline
 Parameter & Data set 1  & Data set 2 & Data set 3\\
 \hline
 $J_\text{trans}t$ (rad)  & 4.313 & 4.528 &4.624\\
 $J_\text{cross}t$ (rad)  &-1.187 & -1.285 &-1.181 \\
 $J_\text{long}t$ (rad) & -0.465 & -0.531 &-0.524\\
$\varepsilon t$ (rad)  & 1.540 &1.148 &0.635 \\
 $\sum t$ (rad) &  0.190  & 0.218  &0.266 \\
$\varepsilon/\sqrt{J_\text{trans}^2+J_\text{cross}^2}$ & 0.344 & 0.244 & 0.133 \\
Gate Fidelity &79.7\% &79.6\% &80.4\%\\ 
 \hline
\end{tabular}
\caption{\label{GST_SWAP_fit} {Summary of the fitting parameters for the $\sqrt{\text{SWAP}}_{ST}$ gate. In the table, $t$ refers to the operation time of the $\sqrt{\text{SWAP}}_{ST}$ gate. The values of the parameters with a radiation unit are modulo $2\pi$.}}
\end{table}

\begin{comment}
\subsection*{\label{sec:level9}Supplementary Note 9. Spin volume}

\centering
\begin{tabular}{ |p{4cm}||p{1.5cm}|p{2cm}|p{1.5cm}|  }
 \hline
 \multicolumn{4}{|c|}{Spin Volume} \\
 \hline
 Paper & Spin number & Spin connection &Spin quality\\
 \hline
 1*4 GaAs {~\cite{kandel2019coherent,  van2021quantum}} & 4  & 3   & \\
 2*2 GaAs {~\cite{dehollain2020nagaoka}}   & 4  & 4 & \\
 2*4 GaAs {~\cite{fedele2021simultaneous}} & 8  &4   & \\
 3*3 GaAs {~\cite{mortemousque2021coherent}} & 2 & 4 &\\
 1*3 Si {~\cite{andrews2019quantifying}} &3 & 2 &  \\
 1*6 Si {~\cite{philips2022universal, weinstein2023universal}} &6 & 5 &  \\
 2*2 Ge {~\cite{hendrickx2021four, wang2023probing}}   & 4  & 4 & \\
 2*4 Ge (This article)   & 8  & 10 & \\
 \hline
\end{tabular}    
\end{comment}

\bibliography{supple}
\bibliographystyle{naturemag}